\documentclass[a4paper,epsfig,oneside,12pt]{book}
\usepackage{psfig}
\begin{document}
\setlength{\baselineskip}{1.5\baselineskip}
\thispagestyle{empty}
\begin{titlepage}
\begin{center}
{\LARGE\bf D-Branes and Noncommutative String Theory\\ }
\end{center}

\vskip 1.5in

\centerline{\bf A thesis submitted to the}
\centerline{\bf Utkal University}
\centerline{\bf for the}
\centerline{\bf Ph.D. Degree in Physics}

\vskip 1.5in

\centerline{\bf by}
\centerline{\bf Kamal Lochan Panigrahi}
\centerline{\bf Institute of Physics,}
\centerline{\bf Bhubaneswar, 751 005, India.}

\vskip 1.5in


\end{titlepage}

\chapter*{Abstract}
\thispagestyle{empty}
In this thesis we discuss some nonperturbative and
noncommutative aspects of string theory. We present
low-energy background field solutions corresponding to 
various D-branes (and their bound states) and intersecting
branes in flat and pp-wave spacetime. A class of D-brane
bound states are constructed from charged macroscopic strings and they are 
shown to satisfy the mass-charge relationship of 1/2 BPS bound states.
Another class of D-branes and intersecting branes 
are constructed, by solving the type II field equations explicitly,
in pp-wave background with constant and non-constant flux. 
The worldsheet construction of these D-branes and their spacetime 
supersymmetry properties have also been analyzed in some detail. 

Further, we study an example of open strings ending on
D-branes with mixed boundary condition. In this context, we analyze 
various open and mixed sector tree-level amplitudes of N=2 strings 
in the presence of constant NS-NS antisymmetric tensor (B) field.

\chapter*{Acknowledgements}
\thispagestyle{empty}

I take this opportunity to express my sincere gratitude to my 
advisor, Alok Kumar, for his invaluable
guidance throughout my research period at Institute of Physics. 
He has been the constant source of inspiration 
all along the way. I have learnt a lot
not only about physics but also about life in general from him. 
I heartily thank him and his family.
I am indebted to Sudipta Mukherji for 
various useful discussions and fruitful collaborations. 

I would like to thank all the teachers of my pre-doctoral period at the 
institute for their excellent teaching. I have learnt quite a lot from
all of them. 

I thank my collaborators Anindya Biswas, Aalok Misra, Nobuyoshi Ohta
and Sanjay for the fruitful and enjoyable collaborations. I would
also like to thank all the members of `Theoretical High Energy Physics
Group' of Instutute of Physics for helping me in understanding various aspects
of Physics in general. I thank Koushik Ray for making me comfortable 
with diverse topics in string theory. I would like to thank the Abdus 
Salam ICTP, Trieste, Italy for giving me financial support for 
participating in the `Spring School on Superstring theroy' in 2002 and 
the `School on Mathematics in String and Field theory' in 2003
which helped me a lot in understanding some interesting topics of string
theory. I also take this opportunity to thank Augusto Sagnotti for 
his kind invitation for visiting Universita' di Roma, ``Tor Vergata'',
ITALY. I sincerely thank the string theory group members at ``Tor Vergata''
for various useful discussions.         
\newpage

\thispagestyle{empty}
I thank all the academic and nonacademic staffs of the institute for their
cooperation during my whole stay. 
Finally, my deep sense of gratitude to my parents and my sisters for 
giving me moral support throughout. I dedicate this thesis to my parents 
and to my Mitul.
\newpage

\pagestyle{plain}
\pagenumbering{roman}

\tableofcontents
\pagenumbering{arabic}
\chapter{Introduction}
\pagenumbering{arabic}
\newcommand{\sect}[1]{\setcounter{equation}{0}\section{#1}}
\newcommand{\subsect}[1]{\subsection{#1}}
\newcommand{\subsubsect}[1]{\subsubsection{#1}}
\renewcommand{\theequation}{\thesection.\arabic{equation}}
\def\sxn#1{\sect{#1}}
\def\subsxn#1{\subsect{#1}}
\def\subsubsxn#1{\subsubsect{#1}}
\font\mybb=msbm10 at 12pt
\def\bb#1{\hbox{\mybb#1}}
\newcommand{\be}{\begin{equation}}
\newcommand{\ee}{\end{equation}}
\newcommand{\bea}{\begin{eqnarray}}
\newcommand{\eea}{\end{eqnarray}}
\newcommand{\up}{\uparrow}
\newcommand{\down}{\downarrow}
\newcommand{\lr}{\leftrightarrow}
\newcommand{\la}{\langle}
\newcommand{\ra}{\rangle}
\newcommand{\da}{\dagger}
\newcommand{\p}{\partial}
\newcommand{\varep}{\varepsilo}
\def\del{\nabla}
\def\half{\frac{1}{2}}
\def\CC{{C \n{11} C}}
\def\ab {(a)}
\def\pa{\partial}
\def\dmu{\partial_\mu}
\def\ten{(10)}
\def\K {\bb{K}}
\def\S {\bb{S}}
\def\Z {\bb{Z}}
\def\T {\bb{T}}
\def\ab {(a)}
\def \cf {{\cal {F}}}
\def \ca {{\cal {A}}}
\def \cH {{\cal{H}}}
\def\ten{(10)}
\def\lsim{\mathrel{
\def\arraystretch{.4}
\begin{array}{c}
$$<$$\\
$$\sim$$
\end{array}}}
\def\gsim{\mathrel{
\def\arraystretch{.4}
\begin{array}{c}
>$$\\
$$\sim$$
\end{array}}}
\markright{Chapter 1. Introduction}
\vspace{-1cm}
\sect{\bf String theory: A First Look}
Unification of all the existing forces of nature still remains
a big challenge for Particle physicists. String theory is the
consistent quantum theory of gravity which includes graviton,
the exchange particle for the gravitational interaction, in its spectrum.
Truely `the theory of strings' is a bold step towards the unification of 
gravity, quantum mechanics and particle physics. 
It fits nicely into the pre-existing picture of what physics
beyond standard model would look like. Besides gravity, string theory
necessarily incorporates a number of previously known unifying ideas like
grand unification, Kaluza-Klein compactifications and supersymmetry 
etc.. Moreover, it unifies these ideas in an efficient way and
resolves some of the difficulties which arose previously like the 
renormalizability problem of Kaluza-Klein theory etc..
In fact, some of 
the simplest string theories give rise to the gauge group 
and matter representations which appeared previously in grand
unification. String theory is based on the simple idea that 
elementary objects which appear
to be point-like to the present day experimentalists, are actually
different vibrational modes of `strings'.
The energy per unit length of the string, known as string tension,
is parametrized by $(2 \pi \alpha^{\prime})^{-1}$, where
$\alpha^{\prime}$ has the dimension of $(length)^2$. 
The typical size of the string is $10^{-33}~cm$ - a distance which can't 
be resolved by the present day accelerators. So
there is no direct way of testing 
string theory apart from its theoretical consistency.
The propagation of strings in space-time is governed by the dynamics
of the one dimensional objects. Strings moving in space-time sweep out
a two dimensional surface known as the string worldsheet. 
The dynamics of such an object is governed by an action, popularly
known as Nambu-Goto action:
\begin{eqnarray}
S &=& - {1\over 2\pi\alpha^{\prime}} {\rm (Area~of~the~Worldsheet)}\cr
& \cr
&=& - {1\over 2\pi\alpha^{\prime}}\int d^2 \sigma
\sqrt{-\rm det\partial_{\alpha}{X}^{\mu}\partial_{\beta}X_{\mu}},
\label{nambu}
\end{eqnarray}
which is nothing but the generalization of relativistic action for a point
particle. The string worldsheet is parametrized by 
$X^{\mu}(\sigma,\tau)$, but the action (\ref{nambu}) is invariant 
under the choice of 
reparametrization (world-sheet coordinate invariant). It is useful to
write down the above action in the following form as well, by adding a
world-sheet metric $(g_{\alpha \beta})$:
\begin{eqnarray}
S_P = - {1\over 2\pi\alpha^{\prime}}\int d^2 \sigma \sqrt{g}g^{\alpha
  \beta}\partial_{\alpha}X^{\mu}\partial_{\beta}X_{\mu},
\label{poly}
\end{eqnarray}     
where $g = det g_{\alpha \beta}$. 
In addition to the {\it diffeomorphism} invariance, the action (\ref{poly})
has another local symmetry known as {\it Weyl} invariance:
\begin{eqnarray}
g^{\prime}_{\alpha \beta} (\sigma) = e^{2 w \sigma} g_{\alpha \beta}
(\sigma).
\end{eqnarray}
The action (\ref{poly}) is commonly known as Polyakov action\cite{polyakov}.

The spectrum of string theory consists of a set of massless states and
an infinite tower of massive states. The massive string states have
the mass of the order of $10^{19}$ Gev and is well beyond the reach 
of present day accelerators. There are five consistent string
theories, known as {\it Type IIA}, {\it Type IIB}, {\it Type I}, 
{\it $SO(32)$ heterotic} and {\it $E_8 \times E_8$ heterotic}. 
In Type IIA and Type IIB, strings are closed and  oriented.
In Type I, strings are open or closed; Open Type 
I strings have electric charges at the end points of the strings.
Heterotic $SO(32)$ and $E_8 \times E_8$ strings are closed and 
oriented.

The above five different string theories are 
related by certain symmetries of string theory known as 
`duality symmetry'\cite{WITTEND,DABH,HULLOPEN,POLCWIT,HULLTOWN,
DUFFSS,SSSD,HARSTRSSD,sen,schwarzi,polchinski,dine}. In particular,
the orientifold construction, namely the fact that type I string
theories can be regarded as the quotient of type IIB theory by the
worldsheet parity transformation was first discovered in
\cite{sagnotti87} (see also {\cite{ps89,bps92}).
Below, we discuss briefly various duality symmetries of string theory.\\ 

\noindent
\begin{itemize}
\item {\bf T-Duality:}
\end{itemize}
\noindent
Target space duality or T-duality in short, is the best understood
duality symmetry of string theory. They hold order by order in string
perturbation theory to all order even though they have
not been established for the full non-perturbative string theory. This
duality transformation maps the weak coupling region of one theory
to the weak coupling region of another theory or of the same theory. 
(for a review of this subject see\cite{poratti})
For example:\\
\noindent 
(i) Bosonic string theory compactified on radius $R$ and on radius
$1/R$ are equivalent.\\ 
\noindent
(ii) Type IIA string theory compactified on a circle of radius $R$ is
dual to Type IIB string theory compactified on a circle of radius
$1\over R$\cite{dine89,pol89,lei89,pol94}.\\
\noindent
(iii) $SO(32)$ heterotic string theory compactified on circle of
radius $R$, with gauge group broken down to $SO(16)\times SO(16)$
by wilson line is dual to $E_8 \times E_8$ heterotic string theory
compactified on circle of radius $1\over R$, with gauge group broken
down to  $SO(16)\times SO(16)$ by wilson
line\cite{narain86,narain87,gins87}. 

\noindent
\begin{itemize}
\item{\bf S-Duality:}\\
\end{itemize}  
\noindent
S-duality is a transformation of the coupling constant of the theory:
a theory with coupling constant $g$ is S-dual to 
some other theory with coupling constant $1/g$. This is a
generalization of Electric- Magnetic duality of Maxwell's theory of
electrodynamics. Below we give few theoretical examples of 
S(elf)-duality in string theory.\\
\noindent
(i) In ten dimensions type I theory with coupling constant $g$ is dual
to SO(32) heterotic string theory with coupling constant $1/g$.\\
\noindent
(ii) Type IIB string theory in D=10 has been conjectured to have an
$SL(2,Z)$ self duality group\cite{HULLTOWN}.
\noindent
\begin{itemize}
\item{\bf U-Duality:}\\
\end{itemize}  
\noindent
U-duality can be thought of as a nontrivial combination of S-duality
and T-duality. Naively speaking, this duality relates small volume
limit of some theory with large coupling limit of some other
theory. This symmetry is nonperturbative by nature.
The U-duality group arises from the toroidal compactification 
of type II and heterotic string theories. 

\sect{\bf $D$-Branes in String Theory} 

Besides the vibrational modes of strings, which is the
basic quanta of string theory, multiplets of string duality includes:
solitons, black holes and some extended objects, known as
`$D$-branes'\cite{harmin,ishi,ps89,sagnotti,horava,pol89,polchinskid}.
(see \cite{pol-tasi,JCH,john-d} for a detailed review of D-branes.
See \cite{Open-Sag} for a review of Open strings.)
$D$-branes are nonperturbative and extended objects on which open
strings can end (like the ones as shown in picture 1.1), almost in 
the same way as QCD strings can end on quarks. In contrast to the
quarks of QCD, $D$-branes are, however,
intrinsic excitations of fundamental theory: their existence is
required for the consistency of string theory. 
They play an important role in understanding various duality
conjectures in string theory. As described earlier they, along with
the strings and other solitons, essentially fill up the multiplets
of string dualities. Besides this, $D$-branes play an
important role in understanding the microscopic structure of quantum
gravity. The $D$-brane model of black holes has been proved to be very 
useful in
understanding the black hole thermodynamics, such as Hawking temperature
and entropy. 
It also has revealed a surprising relationship between supersymmetric
gauge theory and geometry. In particular, the counting of microscopic
BPS sates of black hole can be mapped into a familiar problem of 
studying the moduli space of 
supersymmetric gauge theories. In the context of AdS/CFT duality, 
$D$-branes corresponds to nonperturbative objects in the gauge theory 
side, or to defects on which the lower dimensional field theory lives.
In order to know the field theory dynamics in the presence of such
extended objects, it is often useful to have access to
the supergravity solutions of $D$-branes as well. 


The open string worldsheet is parametrized by two coordinates
$(\sigma, \tau)$, with the spatial coordinate $\sigma$ runs
$0\leq \sigma \leq \pi$. In conformal gauge the action is given by:
\begin{eqnarray}
S = {1 \over {4\pi\alpha^{\prime}}}\int_{\cal M} d^2 \sigma 
\partial_a X^{\mu}\partial_a X_{\mu}.
\end{eqnarray}       
Varying with respect to $X^\mu$ and integrating by parts, one gets:
\bea
\delta S = - {1 \over {2\pi\alpha^{\prime}}}\int_{\cal M} d^2 \sigma 
\delta X^{\mu}\partial^2 X_{\mu} + {1 \over {2\pi\alpha^{\prime}}}
\int_{\partial M} d \sigma \delta X^{\mu}\partial_n X_{\mu},
\label{var}
\eea 
where $\partial_n$ is the derivative normal to the boundary. Vanishing
of the first term in (\ref{var}) gives the Laplace equation. 
At the boundary, the only Poincare invariant condition is:
\be
\partial_n X^{\mu} = 0.
\ee   
This is called  `Neumann' boundary condition. The `Dirichlet' boundary
condition $\delta X^{\mu} = 0 \Rightarrow X^{\mu} = \rm constant$ is 
also consistent with the equations of motion, which describe space-time
defects. A static defect extending over $p$ spatial directions is
described by the boundary conditions:
\be
\partial_{\perp} X^{\alpha = 0,...p} = X^{m = p+1, ..., 9} = 0,
\ee
which constrain the open string to move on a $p+1$ dimensional
hypersurface. Since open strings do not propagate in the bulk, their
presence is intimated by the existence of these defects, which are
known as $Dp$-branes. Below we summarize the $Dp$-branes of various
string theories.

\begin{table}[h]
\begin{center}
\begin{tabular}{|c|c|}
\hline
Type IIA & p = 0, 2, 4, 6, 8 \\
\hline
Type IIB & p = -1, 1, 3, 5, 7, 9 \\
\hline
Type I & p = 1, 5, 9 \\
\hline
\end{tabular}
\caption{Dp-branes of various string theories}
\end{center}
\end{table}

The $D(p)$-brane is characterized by a tension
$T_{(p)}$, charge density under Ramond-Ramond $(p + 1)$-form,
$\mu_{(p)}$, defined through the effective worldvolume action
\cite{bachasd}:
\be
S_{Worldvol} = T_{(p)}\int d^{p +1}\xi e^{\phi \over 2}
\sqrt{|det {\hat {G_{\alpha \beta}}}|} + \mu_{(p)}\int d^{p +1} 
\xi C_{(p + 1)}, 
\ee  
where 
\be
\hat {G_{\alpha \beta}} = G^{\mu \nu} \partial_{\alpha}X_{\mu}
\partial_{\beta}X_{\nu},
\ee 
is the induced metric on the worldsheet and $\phi$ is the dilaton.
The cases $p= -1, 7, 8$ are somewhat special. The $D(-1)$ brane sits
at a particular space-time point and should be treated as a (Euclidean) 
instanton with the action:
\be
S_{D(-1)} = T_{(-1)} e^{-\phi} + i \rho_{(-1)} C^{(0)}|_{position}.
\ee
On the other hand, the massless closed strings coupling to 
$D$-branes themselves have a bulk action:
\be
S_{bulk} = - {1\over {2\kappa^2_{10}}}\int d^{10}x\sqrt{-G}\Big[
e^{-\phi} \Big(R - d\phi^2 + {1\over 12} dB^2\Big) + \sum{1\over
  {2 k!}} F^2_{(k)}\Big].
\ee 
The bosonic part of the low energy action for type IIB string theory
in ten dimension is given by:
\bea
S_{IIB} &=& {1\over 2\kappa^2}\int d^{10}x \sqrt{-G} \Big \{e^{-2\phi_b}
\Big[R + 4 {(\nabla\phi_b)^2} - {1\over 12} {(H)}^2\Big] - {1\over 2}
{(\p \chi)}^2 \cr
& \cr
&-& {1\over 12}{(F^{(3)} + \chi H)}^2 
- {1\over 480}{(F^5)}^2\Big\} + {1\over {4\kappa^2}}\int A^{(4)}F^{(3)}H,
\eea
Where $G_{\mu \nu}$ is the string frame metric, $H = dB$ is the field
strength of the Kalb-Ramond field, $F^{(3)} = d A^{(2)}$ and  
$F^{(5)} = d A^{(4)} - {1\over 2}(B F^{(3)} - A^{(2)} H)$ are the
Ramond-Ramond (RR)  field strengths, while $\chi = A^{(0)}$ is the RR
scalar and $\phi$ is the dilaton. 

$D$-brane can be described by a combination of {\it Neumann} and 
{\it Dirichlet} boundary conditions on the string worldsheet
boundaries. A $Dp$-brane is specified by a 
$(p+1)$ dimensional hypersurface, on which open strings can end and 
Neumann boundary condition is imposed on them: 
$\partial_{normal} X^{\mu} = 0$.
The fields associated with the remaining $(9-p)$ coordinates satisfy
Dirichlet boundary condition: $X^{\mu} = constant$. $T$-duality
changes the Neumann boundary condition to a Dirichlet-type boundary
condition: $\partial_{tangent} X^{\mu} = 0$. Hence implementation of
$T$-duality along the worldvolume direction of a $Dp$-brane, replaces
a Neumann condition by a Dirichlet-like condition and one gets the
`delocalized' $D(p-1)$-brane. If one considers a $Dp$-brane which is
oriented at an angle with respect to the orthogonal axis: tilted in
$(X^1, X^2)$ plane, it will be
specified by a combination of both Neumann and Dirichlet boundary
conditions:
\bea
&&\p_{n}(X^1 + \tan \phi X^2) = 0, \\   
&&\p_{t}(X^1 - \cot\phi X^2) = 0.
\eea
Applying $T$-duality along $X^2$ direction, one gets the following
{\it mixed} boundary conditions:
\bea
&&\p_n X^1 + i\tan\phi\partial_t X^2 = 0,\\
&&\p_n X^2 - i\tan\phi\partial_t X^1 = 0,
\eea 
and one notices that $T$ duality has induced a worldvolume gauge field 
$F_{12} = -\tan\phi$\cite{myersd}. The above analysis can also be lifted to
supergravity analysis and we will be discussing that aspect in the
next chapter.
\sect{\bf Noncommutative Geometry and String Theory}

Noncommutativity is an old idea  in Mathematics and Physics.
It appears in string theory\cite{douglas,douglas2,seiwit} 
when the $D$-brane is placed
in a constant antisymmetric tensor background ($B_{\mu \nu}$).
This makes the $D$-brane worldvolume coordinates noncommutative:
\be
[x^{\mu}, x^{\nu}] = i\theta^{\mu \nu},
\ee
and leads to an uncertainty relation between the space and time:
\[
\nabla x^\mu \nabla x^\nu \ge {1\over 2}|\theta^{\mu \nu}|,
\]
Perhaps the main reason for not taking the noncommutative theories
seriously is that the uncertainty relation between position
measurements will {\it a priori} lead to a nonlocal theory.
Also noncommutativity of the
space-time coordinates generally conflicts with Lorentz
invariance.  Although it is not implausible
that a theory defined using such coordinates could be effectively
local on length scales larger than that of $\theta$, it is not obvious
to believe that the breaking of Lorentz invariance would not be
observed at these scales.
Gauge theory on noncommutative spaces are relevant for
the quantization of $D$-branes in background $B_{\mu \nu}$
field\cite{douglas}. Supersymmetric Yang-Mills theory in
noncommutative geometry arises by Fourier transforming the winding
modes of $D$-branes living on a transverse torus in the presence of
NS-NS two form background\cite{douglas2}. 
The $D$-string suspended between a pair of $D3$-branes in the presence
of an antisymmetric $B_{\mu \nu}$ field represents a monopole in the non
commutative Yang Mills theory on the worldvolume of $D3$-brane. 
Also a $(p,q)$ string represents a dyon similar to its
commutative counter part.


The structure of the noncommutative theory is similar to that of the
ordinary gauge theory with the usual product of fields being replaced by the 
{\it `star product' (*)} defined by:
\be
\phi * \chi = \phi(X)~ exp \Big\{- i\theta^{\mu \nu}{\partial\over
    \partial X^{\mu}}{\partial \over \partial Y^{\mu}}\Big\}~\chi(Y),
\ee   
where $\theta^{\mu \nu}$ is a constant antisymmetric tensor. The
effect of such a modification is reflected in the momentum space
vertices of the theory by a factor of the form:
\be
exp\big[i\theta^{\mu \nu} p_{\mu}q_{\nu}\big] \equiv e^{i p\wedge q}
\ee 
The dynamics of the open strings is determined in terms of closed
string metric $(g_{\mu \nu})$, the background two form field $(B_{\mu
  \nu})$ and the closed string coupling constant $(g_s)$. The
propagator of the open string worldsheet coordinates between two
boundary points is given by:
\bea
<X^\mu (\tau) X^\nu (\tau')> = - \alpha' G^{\mu \nu} \log(\tau - \tau') +
{i\over 2} \theta^{\mu \nu}sgn(\tau -\tau'), 
\eea
where the open string metric $(G^{\mu \nu})$ and the noncommutative
parameter $(\theta^{\mu \nu})$ is given by:
\bea
G^{\mu \nu} &=& \Bigg({1\over {g +2\pi\alpha'B}}\Bigg)^{\mu \nu}_S =
\Bigg({1\over {g +2\pi\alpha'B}} g {1\over {g -2\pi\alpha'B}}\Bigg),\cr
& \cr
\theta^{\mu \nu} &=& 2\pi\alpha'\Bigg({1\over {g
    +2\pi\alpha'B}}\Bigg)^{\mu \nu}_A = -(2\pi\alpha')^2
\Bigg({1\over {g +2\pi\alpha'B}} B {1\over {g -2\pi\alpha'B}}\Bigg).
\eea
The effective open string coupling is given by:
\bea
G_o = g_s \Big({{det(g + 2\pi\alpha' B)}\over{det ~g}}\Big)^{1\over 2}.
\eea 

\sect{\bf AdS/CFT Correspondence and String Theory 
in PP-wave Background}

It has been believed since long that 't Hooft limit\cite{'thooft} of 
large $N$ gauge theories is related to a string theory. 
In\cite{malda97}, Maldacena proposed a particular string theory for the 't
Hooft limit of  ${\cal{N}}=4$ super Yang-Mills theory in $3+1$
dimensions. The 't Hooft limit is defined by taking
$N\rightarrow\infty$ while keeping $g^2_{YM} N$ fixed. In this limit
one gets string theory on $AdS_5 \times S^5$ with the radius of five
sphere and the curvature radius of anti-de Sitter are proportional to 
$(g^2_{YM} N)^{1/4}$ in string units. Maldacena in \cite{malda97} was led 
to the conjecture that $U(N)$ super-Yang-Mills theory in (3+1)-dimensions
is the same as (or dual to) type IIB superstring theory on 
$AdS_5 \times S^5$. The super
Yang-Mills coupling is given by the (complex) type IIB string coupling. 
The supersymmetry group of $AdS_5 \times S^5$ is known to be the same as
the superconformal group in (3+1) spacetime dimensions, so the 
supersymmetry of both theories are the same. This new
form of duality (commonly known as AdS/CFT duality) is the duality 
between the large N field theory and a particular string theory 
on certain background. This is a strong-weak coupling duality:
the strong coupling regime of quantum field theory corresponds to
the weak coupling regime of string theory and vice versa.
Despite much effort, string theory in AdS background is still poorly
understood. At low energies, the string theory is well approximated by
supergravity. On the gauge theory side the low energy limit corresponds 
to large 't Hooft limit, $\lambda=g^2_{YM} N >>1$. The spectrum of strings 
on $AdS_5 \times S^5$ corresponds to the spectrum of single-trace 
operators  in the Yang-Mills theory. Moreover, by 
using the recipe, for the correspondence between ten
dimensional type IIB supergravity and the (3+1)-dimensional $SU(N)$
Yang-Mills theories, various two point and three point functions have
been calculated\cite{gubser,ewitten,freedman}. Despite much effort,
the main obstacle in proving the conjectured AdS/CFT duality lies in the 
difficulty to quantize strings in AdS spaces. Only
recently, Berenstein, Maldacena and Nastase (BMN)\cite{BMN} took advantage 
of {\it Penrose's} idea that `every space-time has plane-wave as a limit' 
and have argued that a particular sector of ${\cal N} = 4$
super Yang-Mills theory containing  the operators with large $R$
charge ($J$), is dual to Type IIB string theory in pp-wave background
with $RR$ flux. The beauty of the proposal is that it suggests a simple
class of supergravity backgrounds where one can explicitly quantize 
strings. 

The fact that string theory in pp-wave background
is exactly solvable has opened up the possibility of understanding 
the duality beyond the supergravity limit.
Penrose's idea was incorporated by G\"uven into supergravity
backgrounds in ten and eleven dimensions. This was achieved by
extending the limiting procedure to the other fields present in the
supergravity background. In particular, it has been found out that
the eleven dimensional Minkowski space-time and the maximally
supersymmetric $H_{pp}$ wave background are the Penrose limit of
$AdS_7 \times S^4$ solutions of M-theory and $AdS_5 \times S^5$
solution of Type IIB supergravity respectively. The limiting procedure 
depends crucially on the local symmetries (diffeomorphism and local
gauge invariance) of the supergravity background and the homogeneity
of the supergravity action under a constant rescaling of the fields.
The novelty of this background is that string theory simplifies
dramatically due to the existence of the lightcone gauge and  
is also exactly solvable in some cases. In short, the plane wave background
admits a set of {\it covariantly constant} and {\it null} killing vectors.

The plane wave is a geometry of general form:
\be
ds^2 = -4 dx^+ dx^- + H(x^+, y){(dx^+)}^2 + dy^i dy^i,
\ee
where the function $H$ is independent of $x^-$, and 
$x^{\pm} = {1\over 2}(t \pm x)$. The wave structure is due to the
$x^+$ dependence of the function $H$ appearing in the
metric. The particular case of interest, when  $H$ is quadratic in
$x^i$ and is supported by a five form Ramond-Ramond field strength,
has been analyzed in a great detail in recent past.
The metric and the constant 5-form field strengths of such a background is
given by: 
\bea
ds^2 &=& -4dx^+ dx^- - \mu^2 (y^i) {(dx^+)}^2 + dy^i dy^i,
\cr
F &=& dx^+ \varphi,\>\>\>\> \varphi= \mu(dy^1dy^2dy^3dy^4 +
dy^5dy^6dy^7dy^8).
\label{pp}
\eea

The Green-Schwaz action after taking the lightcone gauge\cite{gsw} for 
the superstring, by choosing $x^+ = \tau$ and
$\Gamma_{-}\theta^{\alpha} =0,~~\alpha=1,2$ is given by:\\
\bea
S = {1\over 2\pi\alpha'}\int dt\int^{\pi \alpha'|p_{-}|}_0 d\sigma\Big[
{1\over 2} \dot{z}^2 - {1\over 2} {z'}^2 - {1\over 2} \mu^2 z^2 +
i\bar{S}(\not \partial + \mu I)S\Big], 
\eea
where $I= \Gamma^{1234}$ and $S$ is the Majorana spinor on the
worldsheet and a positive chirality $SO(8)$ spinor under rotations in
the eight transverse directions. The lightcone hamiltonian is given
by:
\be
2p_- = -p_{+} = H_{lc} = \sum^{\infty}_{n = -\infty} N_n \sqrt{\mu^2 +
{n^2 \over (\alpha'|p_{-}|/2)^2}},
\ee
with $n$ denoting the level of the Fourier mode and $N_n$ denoting total
occupation number of that mode. In the limit,
$\mu \alpha' |p_{-}| << 1$, one recovers the flat space limit,
which can also be seen in the metric (\ref{pp}), with $\mu = 
0$. It has also been demonstrated that plane wave admits the giant
gravitons, as in the case of $AdS_5\times S^5$. These giants are the 
$D3$-branes classically sitting at $x^-$ and wrapping the $S^3$ of
the first four directions or $S^3$ of the second four directions.

The maximally supersymmetric plane wave of type IIB can be obtained
from the Penrose limit of $AdS_5 \times S^5$. The energy and the 
momentum in the $AdS$ space is given by the global coordinates in
terms of $E$ and $J$.  In terms of dual CFT, these are actually the
energy and $R$ charge of a state of field theory on $S^3\times R$,
where $S^3$ has unit radius. One can then identify 
$E=\Delta$ as the conformal dimension of an operator in $R^4$. 
The precise identification is given by:
\bea
2 p^- &=& -p_{+} = \Delta - J, \cr
& \cr
2 p^+ &=& -p_{-} = {{\Delta + J}\over R^2}. 
\eea      
In terms of Yang-Mills parameter the lightcone hamiltonian can be
written and one can find out the contribution of each oscillator:
\bea
(\Delta - J)_n = w_n = \sqrt{1+{4\pi g N n^2\over J^2}},
\eea 
with $gN\over J^2$ remaining fixed in the $gN\rightarrow \infty$
limit.

It looks quite promising at the moment to study string theory in
pp-wave background with RR and NS-NS flux being turned on, in order 
to understand string theory-gauge theory duality by using 
AdS/CFT like correspondence. String theory in this background 
is easy to handle due to the presence of natural lightcone gauge
and in many cases it can be quantized exactly. 
Study of $D$-branes: classical solutions, 
open string spectrum and supersymmetry in these backgrounds with
both constant and nonconstant flux has also been a subject
of intense discussion in the recent past.

\sect{\bf Plan of the Thesis}

We briefly outline the content of the subsequent chapters below.

In {\it chapter 2}, we construct new $D$-brane bound states using
charged macroscopic type IIB string solutions. Starting from
$D=9$ charged macroscopic string solutions, we obtain solutions in
$D=10$, which are interpreted as $D$-brane bound states 
carrying ($F$, $D0$, $D2$) charges as well
as nonzero momenta. The mass-charge relationship of these bound states 
are explicitly shown to 
satisfy the non-threshold bound of $1/2$ BPS objects.

In {\it chapter 3}, we present classical solutions of $Dp$ as well as 
$Dp-Dp'$ branes in pp-wave spacetime, with constant NS-NS and RR flux.
The supersymmetric properties of the solutions are analyzed by 
solving the type II killing spinor equations explicitly. We also
discuss the worldsheet construction of the $p-p'$ branes.
We then generalize the above 
solutions to the cases where the pp-wave background is supported by
nonconstant flux as well. 

In {\it chapter 4}, we analyze various open and mixed sector tree-level
amplitudes of $N=2$ strings in a space-time with (2,2) signature, 
in the presence of constant $B$ field. We have been able to reproduce
the expected `topological' nature of string amplitudes in the
open sector. In the mixed sector, we compute 
a 3-point function and show that the corresponding field
theory is written in terms of a generalized *-product.

In {\it chapter 5}, we present the overview and conclusion.

\chapter{$D$-BRANE BOUND STATES}
\markright{Chapter 2. $D$-BRANE BOUND STATES}

\sect{Introduction}
Bound states of $D$-branes \cite{witten,mli,mrd,green,russo-tsey,
costa,araf,jabb,kamal} have been an interesting area of investigation due to 
their applications in understanding the non- perturbative aspects 
of various string and gauge theories \cite{malda-russo,hashi,yoz}. 
In particular, the supergravity configurations of  supersymmetric $D$-branes
and their bound states were used extensively in testing various 
conjectures which involve knowledge of string theory or gauge
theory beyond their perturbative regime. They are also of importance
in understanding the physics of black holes from a microscopic
point of view. In view of their widespread
applications, it is often very useful to generalize D-brane bound
state constructions. As a step in this direction, in \cite{kamal1}, 
we have constructed generalized $D0 - D2$ bound state using
charged macroscopic strings \cite{sen92,kumar}. These strings carry, 
in general,
many parameters associated with the charges and currents. As a consequence,
the $D$-brane bound states have many nontrivial charges. We argue
that these  configurations can also in general carry 
$F$-string charges and momenta along their worldvolume directions. 
We then further extend our result by presenting the generalized 
$(Dp-D(p+2))$ bound states. In this context we give an explicit
example for the $D1-D3$ case.  
A method for obtaining the bound state of various D-branes is
described explicitly in \cite{rcmyers}. This is done  by smearing the brane 
of type IIA (IIB) theory along certain transverse direction
and mixing it with  a longitudinal direction through a 
coordinate transformation to construct `tilted' brane in $D=10$. Finally,
an application of $T$-duality on the configuration leads to a bound state
solution in the dualized IIB (IIA) theory. As stated earlier,  
in this paper we concentrate mainly on $D0-D2$ bound state
of the IIA theory. They are constructed by starting with a 
$D$-string solution in the IIB theory in $D=10$, which is also 
delocalized along one of the transverse directions. These 
delocalized solutions can also be obtained by 
decompactifying a fundamental string solution in $D=9$ to $D=10$.
Further, one applies an $S$- duality transformation in $D=10$ to 
construct $SL(2, Z)$ multiplets.
One, however, knows the existence of more 
general string solutions in heterotic as well as type II strings in 
all dimensions $4 \leq D \leq 9$\cite{sen92,kumar}, 
which carry vectorial charges and currents. We make use of such 
charged macroscopic string solutions
to generate  a generalized smeared (delocalized) $D$-string in $D=10$. 
We then apply the procedure for obtaining the $D$-brane bound states,
as described above, on such delocalized configurations. 
For the special case when vectorial charges are set to zero, our delocalized  
solutions reduce to the $D=10$ smeared string solutions of  
\cite{rcmyers}. We therefore have a generalization of the $D0 - D2$
bound state by starting with the charge macroscopic string 
solutions in $D=9$. 

We also show that by smearing some of the transverse directions of our
`generalized' $D0-D2$ bound state discussed above, and applying 
$T$-duality along these additional directions, we can construct 
new $Dp-D(p+2)$ bound states as well. We work out the 
case, $p=1$, explicitly. 
In fact even more general bound states can be constructed 
by using $D<9$ charged macroscopic string solutions. We however 
mainly restrict ourselves to the $D=9$ solutions. 
We also explicitly verify the (non-threshold) BPS condition in 
all the examples discussed here.

\sect{Review of Charged Macroscopic String Solutions}
Study of string theory in early $90$'s focussed mainly on the
construction of classical solutions in string theory. In order to
study the nonperturbative aspects of string theory, one needs to
include the solitonic states present in the theory in addition to
the standard Fock space states. In particular, some underlying 
symmetries of string theory become manifest only after including 
these solitonic states. This also gives new insight into the study
of black hole evaporation in quantum gravity. In \cite{dabh1,dabh2},
a specific example of such solution was constructed, which represents
fields around a macroscopic heterotic strings\cite{wit}. It was found that
various properties of these strings resemble with those of the 
string solutions found earlier, such as the force between two
parallel strings vanish and so on. The classical solutions of the low energy
effective field theory equations of motion were constructed which
represent multi-string solutions. These strings carry
both electric and magnetic charges. The study
of charged macroscopic strings are useful in the context of black hole
physics \cite{jmalda} and strong/weak duality applications. In the
context of dualities, they played an important role in $SL(2,Z)$
duality in ten dimension and a string-string duality between $K3$
compactification of IIA and $T^4$ compactification of heterotic string
theory. The support for this duality involved the construction of
certain BPS states carrying (1-form) gauge field charges. Such states
were obtained  by using charged macroscopic string solution of
heterotic strings with 1/2 supersymmetry and then by mapping them to 
type II theory. 

In case of heterotic string theory, besides the string coordinate
$X^{\mu}$ there are other degrees of freedom as well. In particular,
it contains 16 internal degrees of freedom $Y^I$, which are
responsible for giving the gauge field in the spectrum of the theory.
Thus if one considers a string source, where the worldsheet momenta
conjugate to the coordinate $Y^I$ is nonvanishing, it would
represent a macroscopic string carrying finite amount of charge per
unit length. Charged macroscopic string solutions
are generated from the neutral string solution by applying a solution
generating technique\cite{sen92} and are in general parametrized
by a group $(O(d-1, 1;d-1, 1))$ arising out of one time and
$d-1$ spatial isometries. These parameters also appear in the 
charged macroscopic string solutions. 
In particular, the solution described in \cite{sen92,kumar} 
is parametrized by two nontrivial parameters of the transformation 
namely $\alpha$ and $\beta$. The general solution for arbitrary values
of $\alpha$ and $\beta$ for $D < 10$ is given in \cite{kumar}.
Explicit supersymmetry property of these solutions are given 
for $ \beta=0$, $\alpha\ne0$ and $\alpha = -\beta$ in \cite{kumar}. 
For algebraic simplifications, in this paper, we will deal 
with these choices of $\alpha$ and $\beta$. Decompactifications, 
$SL(2, Z)$ transformations and $T$- duality operations for general 
$\alpha$ and $\beta$ are possible to write down, but algebraically 
more complicated.  

The bosonic backgrounds associated with charged
macroscopic strings in spacetime dimension $D$ have been obtained from
similar solutions in heterotic strings\cite{sen92} by turning of the 
sixteen gauge fields associated with the right moving bosonic sector. 
The above is possible because the right moving sector is proportional 
to the NS-NS sector of type II theory in ten dimensions. The solution
is given by:

\bea 
ds^2 &=& r^{D-4} \Delta^{-1} [ -(r^{D-4}+C) dt^2 + C (\cosh\alpha -
\cosh\beta) dt dx^{D-1} \cr
& \cr
&+& (r^{D-4} + C \cosh\alpha \cosh\beta) (dx^{D-1})^2]\cr
& \cr
&+& (dr^2 + r^2 d\Omega_{D-3}^2),     
\label{e13}
\eea
\be \label{e14}
B_{ (D-1)t} = {C\over 2\Delta} (\cosh\alpha + \cosh\beta)
\{ r^{D-4} + {1\over 2} C (1 + \cosh\alpha \cosh \beta) \} , 
\ee
\be \label {e15}
e^{-\Phi} = {\Delta^{1/2} \over r^{D-4}} , 
\ee
\bea \label{e16}
A^{\ab}_{t} &=& -{n^{\ab} \over 2 \sqrt 2 \Delta} C \sinh \alpha
\{ r^{D-4} \cosh\beta + {1\over 2} C (\cosh\alpha + \cosh \beta) \}
\nonumber \\
&& \qquad \qquad \hbox{for} \qquad 1 \le a \le (10-D) \, , \nonumber \\
&=& -{p^{(a-10+D)} \over 2 \sqrt 2 \Delta} C \sinh \beta
\{ r^{D-4} \cosh\alpha + {1\over 2} C (\cosh\alpha + \cosh \beta) \}
\nonumber \\
&& \qquad \qquad \hbox{for} \qquad (10-D)+1 \le a \le (20 - 2D) \, , 
\nonumber \\
\eea
\bea \label{e17}
A^{\ab}_{D-1} &=& - {n^{\ab} \over 2\sqrt 2 \Delta} C \sinh \alpha 
\{r^{D-4}
+ {1\over 2} C \cosh \beta (\cosh\alpha + \cosh \beta) \} \nonumber \\
&& \qquad \qquad \hbox{for} \qquad
1\le a \le (10-D) \, , \nonumber \\
&=&  {p^{(a-10+D)} \over 2\sqrt 2 \Delta} C \sinh \beta \{ r^{D-4}
+ {1\over 2} C \cosh \alpha (\cosh\alpha + \cosh \beta) \} \nonumber \\
&& \qquad  \qquad \hbox{for} \qquad
(10-D)+1 \le a \le (20-2D) \, , \nonumber \\
\eea
\be \label{e18}
M_D = I_{20-2D} + \pmatrix{ P nn^T & Q np^T \cr Q pn^T & P pp^T \cr} \, ,
\ee
where,
\be \label{e19}
\Delta = r^{2(D-4)} + Cr^{D-4} ( 1 + \cosh\alpha \cosh\beta) + {C^2 \over 4}
(\cosh\alpha + \cosh\beta)^2 \, ,
\ee
\be \label{e19a}
P = {C^2 \over 2\Delta} \sinh^2 \alpha \sinh^2 \beta \, ,
\ee
\be \label{e19b}
Q = - C \Delta^{-1} \sinh\alpha \sinh\beta \{ r^{D-4} + {1\over 2} C
(1 + \cosh\alpha \cosh \beta) \} \, .
\ee
with $n^{(a)}$, $p^{(a)}$ being the components of $(10-D)$-dimensional 
unit vectors. 
$A_{\mu}$'s in eqns. (\ref{e16}), (\ref{e17})
are the gauge fields appearing due to the Kaluza-Klein (KK)
reductions of the ten dimensional metric and the 2-form antisymmetric
tensor coming from the NS-NS sector, $B_{\mu \nu}$ is 
the NS-NS 2-form field and $\Phi$ is the dilaton in the D-dimensional
space-time. In the above configuration, $C$ is 
a constant related to tension of the string. The matrix $M_D$ parameterizes
the moduli fields. The exact form of this parameterization 
depends on the form of the $O(10-D, 10-D)$ metric used. 
The above solution has been written for a diagonal
metric of the form: 
\be
 L_D = \pmatrix{ -I_{10-D} & \cr & I_{10-D}}. \label{eld}
\ee
One sometimes also uses an off-diagonal metric convention
(as in eqn.(\ref{decomp}) below:
\be
L = \pmatrix{ & I_{10-D} \cr I_{10-D} & }. \label{el}
\ee
These two conventions are however related by:
\be
 L_D = \hat{P} L \hat{P}^T, \>\>\> M_D = \hat{P} M \hat{P}^T, \label{eleld}
\ee
where 
\be
   \hat{P} = {1\over \sqrt{2}}\pmatrix{-I_{10-D} & I_{10-D} \cr 
I_{10-D} & I_{10-D}}. \label{hatp}
\ee 

The gauge fields in two conventions are related as:
\be
  \pmatrix{A^1_{\mu} \cr A^2_{\mu}} = 
\hat{P} \pmatrix{\hat{A}^1_{\mu} \cr \hat{A}^2_{\mu}}, 
            \label{aredef}
\ee
with $A^{1,2}_{\mu}$'s in the above equation
being $(10-D)$-dimensional columns consisting 
of the gauge fields $A_{\mu}$'s defined in (\ref{e16}-\ref{e17}), 
and coming from the left and the right-moving sectors of string theory.  

We now discuss the decompactification of the 
D-dimensional solutions (\ref{e13})-(\ref{e19b}) to $D=10$. 
As stated earlier, we now restrict ourselves to specific values of
the parameters: $\alpha = -\beta$, $\alpha = \beta $ and 
$\beta =0, \alpha \neq 0$ for algebraic simplifications. 
These special cases, in particular the latest
possibility, encompasses all the nontrivialities of our construction. 
We also restrict ourselves to $D=9$ for constructing $D0-D2$ bound state. 
First we start with the seed solution (with $\beta =0$,
$\alpha$ arbitrary) given as:
\bea
ds^{2} &=& {1\over \cosh^2{\alpha\over 2} e^{-E} -\sinh^2{\alpha\over
2}} (-dt^2 +(dx^{D-1})^2) \cr
& \cr
&+& {\sinh^2{\alpha\over 2} (e^{-E}-1)\over
(\cosh^2{\alpha\over 2} e^{-E} -\sinh^2{\alpha\over
2})^2} (dt+dx^{D-1})^2 + \sum_{i=1}^{D-2} dx^i dx^i , \cr
& \cr
B_{(D-1)t} &=& {\cosh^2{\alpha\over 2}(e^{-E}-1)\over \cosh^2{\alpha\over
2} e^{-E} -\sinh^2{\alpha\over2}} , \cr
& \cr
A^{(1)}_{D-1} &=& A^{(1)}_t = - {1\over {2\sqrt{2}}}\times
{\sinh\alpha (e^{-E}-1)\over
\cosh^2{\alpha\over 2} e^{-E} -\sinh^2{\alpha\over
2}} , \cr 
& \cr
\Phi &=& -\ln(\cosh^2{\alpha\over 2}e^{-E} -\sinh^2{\alpha\over 2}) , 
\label{betaeq0}
\eea
with $e^{-E}$ being the Green function in the 5-dimensional 
transverse space:
\be
e^{-E} = (1 + {C\over r^5}) ,  \label{green}
\ee
and constant $C$ determines the string tension. 

Now, to construct delocalized solutions in $D=10$, we decompactify the 
above solution back to ten dimensions. 
The decompactification exercise is done following a set of notations 
given in \cite{senijmp}. When restricted to the NS-NS sector of 
type II theories, they can be written as:
\bea\label{decomp}
&& \hat {G}_{a b}  = G^{\ten}_{[a+(D-1), b+(D-1)]}, 
\quad  \hat B_{a b}  =
B^{\ten}_{[a+(D-1), b+(D-1)]}, 
\nonumber \\
&& \hat{A}^{(a)}_{\bar{\mu}}  = {1\over 2}\hat G^{ab} 
G^{\ten}_{[b+(D-1),\bar{\mu}]}, \nonumber \\
&&  \hat{A}^{(a+(10-D))}_{\bar{\mu}} = {1\over 2}
B^{\ten}_{[a+(D-1), \bar{\mu}]} - \hat B_{ab} A^{(b)}_{\bar{\mu}}, 
\nonumber \\
&& G_{\bar{\mu}\bar{\nu}} = G^{\ten}_{\bar{\mu}\bar{\nu}} 
- G^{\ten}_{[(a+(D-1)), \bar{\mu}]} 
G^{\ten}_{[(b+(D-1)), \bar{\nu}]} \hat
G^{ab}, \nonumber \\
&& B_{\bar{\mu}\bar{\nu}} = 
B^{\ten}_{\bar{\mu}\bar{\nu}} - 4\hat B_{ab} A^{(a)}_{\bar{\mu}}
A^{(b)}_{\bar{\nu}} - 
2 (A^{(a)}_{\bar{\mu}} A^{(a+(10-D))}_{\bar{\nu}} - A^{(a)}_{\bar{\nu}} 
A^{(a+(10-D))}_{\bar{\mu}}),
\nonumber \\
&& \Phi = \Phi^{\ten} - {1\over 2} \ln\det \hat G, \quad 
\quad 1\le a, b \le 10-D, \quad
0\le {\bar{\mu}}, \bar{\nu} \le (D-1).
\eea

We now start with the nine-dimensional ($D=9$) solution in 
(\ref{betaeq0}) and 
following the Kaluza-Klein (KK) compactification mechanism summarized above,
write down the solution directly in ten dimensions\cite{kumar}. 
For $\beta = 0$, only nonzero background fields are then given by
\bea \label{10dg}
ds^2 &=&  {1\over {\cosh^2 {\alpha\over 2} e^{-E} - \sinh^2 {\alpha\over 2}}}
         ( -dt^2 + (dx^8)^2 ) \cr
      &+& {\sinh^2{\alpha\over 2}(e^{-E} - 1)\over 
        {\cosh^2 {\alpha\over 2}e^{-E} - \sinh^2 {\alpha\over 2}}}
        (dt + dx^{8})^2  \cr 
& \cr
&+& {{\sinh \alpha (e^{-E} - 1)}\over 
                  {\cosh^2 {\alpha\over 2}e^{-E} - \sinh^2 {\alpha\over 2}}}
     dx^9 (dt + dx^8) +  \sum_{i=1}^{7} dx^i dx^i + (dx^9)^2,
\eea
\bea \label{10db}
B_{8 t}  &=& {{\cosh^2 {\alpha\over 2} (e^{-E} - 1)}
              \over {\cosh^2 {\alpha\over 2}e^{-E} - 
\sinh^2 {\alpha\over 2}}},\cr
B_{9 t} &=& - {\sinh\alpha\over 2}{{(e^{-E} - 1)}\over 
{\cosh^2 {\alpha\over 2}e^{-E} - \sinh^2 {\alpha\over 2}}} = B_{98}.
\eea
The dilaton in ten dimensions remains same as the one in 
(\ref{betaeq0}):
\be \label{10phi} 
\Phi^{(10)} = - \ln ({\cosh^2 {\alpha\over 2} e^{-E}
- \sinh^2 {\alpha\over 2}}).
\ee

For $\alpha= -\beta$, on the other hand, we have $D=9$ solutions given
by a metric:
\be
ds^2 = -{1\over {1 + {{C \cosh^2\alpha}\over r^5}}} {dt}^2 +
        {1\over {1 + {C\over r^5}}} {(dx^8)}^2 + \sum_{i=1}^{7}
        dx^i dx^i.  \label{al=beg}
\ee
The only non-zero component of the antisymmetric tensor is of the
form
\be
B_{t 8} = - {C \cosh\alpha\over 2}\left[{1\over {(r^5 + C)}} 
+ {1\over {(r^5 + C \cosh^2\alpha)}}\right].
          \label{al=beb}
\ee
We also have a nontrivial modulus parameterizing the $O(1, 1)$
matrix $M_D$ in eqn.(\ref{e18}):
\be
   \hat{G}_{99} \equiv \hat{g} = {{1+ {C \cosh^2\alpha\over r^5}}\over
              {1 + {C\over r^5}}}. 
           \label{hatg}
\ee
The two gauge fields appearing in equations (\ref{decomp})  
for $D=9$ are of the form:
\be
\hat{A}^1_t = {{C \sinh\alpha \cosh\alpha}\over { 2 
(r^5 + C \cosh^2\alpha)}}, \>\>\> \hat{A}^1_8 = 0, 
\ee
\be
\hat{A}^2_t = 0, \>\>\>
\hat{A}^2_8 = {-{C \sinh\alpha}\over { 2 
(r^5 + C)}}. 
\ee

The decompactified solution for $\alpha = -\beta$ case in $D=10$ 
is given by:

\begin{eqnarray}
ds^2 &=& - {(1 - {C \over r^5}{\rm sinh}^2 {\alpha})
     \over {1 + {C \over r^5}}}(dt)^2   
    + {1 + {C \over r^5}{\rm cosh}^2{\alpha}\over {1 + {C \over
      r^5}}}({dx^9})^2 \cr
& \cr  
&+& {{2 C \over r^5}{\rm cosh}{\alpha}{\rm sinh}{\alpha}
       \over{1 + {C \over r^5}}}{dx^9}dt + {1 \over{1 + {C \over
         r^5}}}({dx^8})^2 + \sum_{i=1}^7 (dx^i)^2,
\label{fstring}
\end{eqnarray}
antisymmetric NS-NS $B_{\mu \nu}$:
\begin{equation}
B_{98} = - C {{{\rm sinh}\alpha}\over{r^5 + C}}, \>\>\>
B_{t8} = - C {{{\rm cosh}\alpha}\over{r^5 + C}}, 
\label{b_munu}
\end{equation} 
and by the dilaton:
\begin{equation}
\Phi^{(10)}= - {\rm ln} (1 + {C \over r^5}). 
\label{iiphi}
\end{equation} 
For $\alpha = \beta$, the $D = 9$ background metric and 
antisymmetric tensors are identical to the one in (\ref{al=beg}). 
The modulus field is now given by, 
\be
\hat{g} = {{1+ {C \over r^5}}\over
              {1 + {C \cosh^2\alpha\over r^5}}}.
           \label{hatg-dual}
\ee
Finally the components of the gauge fields are now:
\be
\hat{A}^1_t = 0,\>\>
\hat{A}^1_8 = {{C \sinh\alpha}\over { 2 (r^5 + C)}},
\ee
\be
\hat{A}^2_t = {-{C \sinh\alpha \cosh\alpha}\over { 2 
(r^5 + C cosh^2\alpha)}}, \>\>\>\hat{A}^2_8 = 0.
\ee
The $10$ dimensional metric along with the antisymmetric tensor
fields and dilaton are given by:
\bea
ds^2 &=& - {1\over {1 + {C\over r^5} \cosh^2\alpha}}dt^2 + {1 +
  {C\over r^5} \over {1 + {C\over r^5}\cosh^2\alpha}} {(dx^9)}^2
+ {{1 + {C\over r^5}\sinh^2\alpha} \over{1 + {C\over
      r^5}\cosh^2\alpha}} {(dx^8)}^2 \cr
\cr
&+& 2 {{{C\over r^5}\sinh\alpha}\over{1 + {C\over r^5}\cosh^2\alpha}}
(dx^8)(dx^9) + \sum^7_{i = 1}{(dx^i)}^2,\cr
& \cr
B_{9t} &=& - {{C\over r^5}\over {1 + {C\over r^5}\cosh^2\alpha}} \sinh
\alpha\cosh\alpha,\>\>\>B_{8t} = {{C\over r^5} \over{1 + {C\over r^5}
\cosh^2\alpha}}\cosh\alpha\cr 
& \cr
\phi^{(10)} &=& -\ln (1 + {C\over r^5}\cosh^2\alpha). 
\label{aleqbe}
\eea   
Before applying $SL(2, Z)$ transformation to the solutions in eqns.
(\ref{10dg})-(\ref{10phi}), (\ref{fstring})-(\ref{iiphi}) and
(\ref{aleqbe}), we now describe the general construction of 
$Dp - D(p+2)$ bound states starting from $D$- strings of type IIB
\cite{rcmyers}.

\sect{Construction of $D$-brane bound states}

In this section we first review the construction of $D0 - D2$ 
non-threshold bound states in type IIA string theories from 
the (neutral)  $D$-strings in type IIB. Our starting point
in this case is the $D$-string solution, smeared (delocalized) along
a transverse direction $x \equiv x^9$. This solution is given as:
\begin{eqnarray}
ds^2 &=& \sqrt{H} \left( {{-dt^2 + dy^2}\over H} +  dx^2
+ \sum_{i=1}^7 (dx^i)^2 \right), \cr
& \cr
A^{(2)} &=& \pm \left( {{1\over H} - 1 }\right) dt \wedge dy, \cr
& \cr
e^{\Phi_b^{(10)}} &=& H.
\label{d1neutral}
\end{eqnarray}
\begin{equation}
H = 1 + {C\over r^5}.
\label{eh}
\end{equation}
is the solution of the Greens' function equation:
\begin{equation}
\Delta H = C A_{N-1} \pi_{i=1}^N \delta(x^i),
\label{greenfunction}
\end{equation}
where $N$ indicates the transverse directions which are not smeared
and $A_N$ is the
area of $S^N$ orthogonal to the brane. 
To compare with the solutions in section-2, one identifies $H = e^{-E}$.  
Then a rotation is performed between $x$ and $y$ directions:
\begin{eqnarray}
\pmatrix{x\cr y} = \pmatrix{ \cos\phi & -\sin\phi \cr
                         \sin\phi & \cos\phi} 
                   \pmatrix{ \tilde{x}\cr \tilde{y}}
\label{rotation}
\end{eqnarray}
to mix the longitudinal and transverse coordinates of the above
solution. 
The solution in eqn.(\ref{d1neutral}) then transforms to:
\begin{eqnarray}
ds^2 &=& \sqrt{H}\Big[ {-dt^2\over H} + ({\cos^2\phi\over H} 
+ \sin^2\phi) d\tilde{y}^2 +({\sin^2\phi\over H} 
+ \cos^2\phi) d\tilde{x}^2 \cr
& \cr
&+& 2 \cos\phi \sin\phi ({1\over H} - 1)d\tilde{y}d\tilde{x} 
+ \sum_{i=2}^8 (d {x}^i)^2 \Big],  \cr
& \cr
A^{(2)} &=& \pm \left( {{1\over H} - 1 }\right) dt 
\wedge (\cos\phi d\tilde{y} + \sin\phi{d\tilde{x}} ), \cr
& \cr
e^{\Phi_b^{(10)}} &=& H.
\label{d1rot}
\end{eqnarray}
Finally, a  T-duality transformation \cite{chull,rcmyers} on 
coordinate $\tilde{x}$ gives 
the following classical solution in the type IIA theory:
\begin{eqnarray}
ds^2 &=& \sqrt{H} \left( {-{dt^2\over H}}+
{{d{\tilde x}^2 + d{\tilde y}^2}\over {1+ (H-1)\cos\phi}} +  
\sum_{i=2}^8 (dx^i)^2 \right), \cr
& \cr
A^{(1)} &=& \pm \left( {1\over H} - 1 \right)\sin\phi dt, \cr
& \cr
A^{(3)} &=& \pm { (H-1)\cos \phi \over{1 + (H-1)\cos^2 \phi}}
dt\wedge d\tilde{x} \wedge d\tilde{y}, \cr
& \cr
B^{(a)} &=& { (H-1)\cos\phi \sin\phi \over{1 + (H-1)\cos^2 \phi}}
d\tilde{x} \wedge d\tilde{y}, \cr
& \cr
e^{\Phi_a^{(10)}} &=& { H^{3\over 2} \over{1 + (H-1)\cos^2 \phi}}.
\label{d0-d2-2a}
\end{eqnarray}
Solution in eqn. (\ref{d0-d2-2a}) can be interpreted as a $D0 - D2$ bound
state. The $D0$ and $D2$ charge densities carried by the above
bound state solution are given as:
\begin{eqnarray}
Q_0 =  5 C \sin\phi A_6, \cr
Q_2 =  5 C \cos\phi A_6, \cr
\label{02charge}
\end{eqnarray}
where $A_6$ is the  area of $S^6$ orthogonal to the brane.
The ADM mass \cite{lu,rcmyers} is defined by the expression:
\begin{equation}
m = \int \sum_{i=1}^{9-p} n^i
\left[\sum_{j=1}^{9-p} (\partial_j h_{i j} - \partial_i h_{j j}  )
 - \sum_{a=1}^p \partial_i h_{aa} \right] r^{8 - p} d\Omega,
\label{adm-mass}
\end{equation}
where $n^i$ is a radial unit vector in the transverse space and 
$h_{\mu \nu}$ is the deformation of the Einstein-frame metric
from flat space in the asymptotic region. 
We also should mention here that in order to write mass and charges, we
have tuned the gravitational constant to a suitable value.
The ADM mass density in the present context is found to be:
\begin{equation}
m_{0, 2} = 5 C A_6.
\label{02mass}
\end{equation}
Mass and charge densities given in eqns. (\ref{02charge}) and (\ref{02mass})  
above satisfy the BPS condition:
\begin{equation}
(m_{0,2})^2 = (Q_0^2 + Q_2^2).
\label{mass-charge}
\end{equation}

To generalize the results to other D-brane bound states, we 
smear one more transverse direction, $x^i \sim x^7$ in eqn. 
(\ref{d0-d2-2a}). Finally applying $T$-duality along 
this direction, one is able to construct a $D1-D3$ bound state. 
In the dualized (IIB) theory, one can write down  
all the field  components trivially. One finds an exact matching 
with the $D1-D3$ solution of \cite{rcmyers}. In particular, 
matching of the $4$-form field can be seen by  using the 
identity: 
\begin{eqnarray}
{A^{(4)}} &\equiv & {(H - 1) \cos{\phi}\over {1 + (H - 1)\cos^2{\phi}}} 
- {1 \over2}{(H - 1)\over H} {\sin{\phi}(H - 1) \cos{\phi}\sin{\phi}
\over{1 + (H - 1)\cos^2{\phi}}}\cr
& \cr
&=& {(H - 1) \cos{\phi}\over 2 H}
\left[{1 + {H \over{1 + (H - 1)\cos^2{\phi}}}}\right].
\label{d1-d3-myers}
\end{eqnarray} 
We give the final $D1 - D3$ solution \cite{rcmyers}
for completeness as well as for
later use:
\begin{eqnarray}
ds^2 &=& \sqrt{H}\Big\{{- d t^2 +(d y^2)^2 \over H} + 
{ d \tilde y^2 + d \tilde x^2 \over 1 + (H-1) \cos^2 \phi} + d r^2\cr
& \cr
&+& r^2(d \theta^2 + \sin^2 \theta(d \phi_1^2 +
\sin^2 \phi_1 (d \phi_2^2 + \sin^2 \phi_2( d \phi_3^2 
+\sin^2 \phi_3 d \phi_4^2))))
\vphantom{1\over H}\Big\}\cr
& \cr
A^{(4)} &=& \mp{\cos \phi\over2}{H-1 \over H}
\left(1+  {H \over
1 + (H-1) \cos^2 \phi}\right)\times d t \wedge d \tilde y \wedge
d y^2 \wedge d \tilde x \cr
& \cr
&\pm& 4 C \cos\phi\,\sin^4\theta\sin^3\phi_1\sin^2\phi_2
\cos\phi_3 d\phi\wedge d\phi_1\wedge d\phi_2\wedge d\phi_4, \cr
& \cr
A^{(2)} &=&\pm {H-1 \over H}\, \sin \phi\, d t \wedge d y^2,\cr
& \cr
B^{(b)} &=& { (H -1) \cos \phi \sin \phi 
\over  1 + (H-1) \cos^2 \phi} d \tilde x \wedge d \tilde y, \cr
& \cr
e^{\Phi_b^{(10)}} &=&\, {H \over  1 + (H-1) \cos^2 \phi},
\label{d1-d3-myers2}
\end{eqnarray}
where $H = 1 + {{C \over r^4}}$.

One can now repeat this process to generate all the bound states
of $Dp-D(p+2)$ type in a similar manner. 
We now apply the general procedure described above 
to the charged macroscopic string solutions of section-2.2.

\sect{Generalized $(Dp - D(p+2))$ bound state}

In this section we construct non-threshold bound states which are the 
generalization of the $D0-D2$ bound state presented in the previous
section.  Here we discuss the $\alpha =  -\beta$ and $\alpha = \beta$
solutions of section-2 and postpone the discussion of 
$\beta = 0, \alpha \neq 0$ solutions to section-2.5.  

\subsect{$\alpha = - \beta$ Solution and Generalization of $D0-D2$ Bound
  States}

The delocalized elementary string solution is given in 
eqns.(\ref{fstring})-(\ref{iiphi}). A delocalized $D$-string 
in $D=10$ can be generated from this
solution by an application of $S$-duality transformation \cite{jhschwarz}
which transforms an elementary string into a $D$-string. 
The metric, antisymmetric 2-form
$(B_{\mu \nu})$ and the dilaton for the delocalized 
$D$-string solution are given by:
\begin{eqnarray}
ds^2 &=& - {{1 - {C\over r^5}{\rm sinh^2}{\alpha}}\over{\sqrt{1 +
      {C\over r^5}}}}{(dt)}^2 + 
{{1 + {C\over r^5}{\rm cosh^2}{\alpha}}\over{\sqrt{1 +{C\over
        r^5}}}}{(dx^9)}^2 \cr
& \cr
&+& {2 {C\over r^5} {\rm sinh}{\alpha}{\rm cosh}{\alpha}\over{\sqrt{ 1
      + {C\over r^5}}}} {dx^9}{dt} + {1 \over{\sqrt{1 + {C\over
    r^5}}}}{(dx^8)}^2 \cr
&\cr
&+& {\sqrt{ 1 + {C\over r^5}}}\sum_{i=1}^7
{(dx^i)}^2,
\label{iibm}
\end{eqnarray}
\begin{equation}
B^{(2)}_{9 8} = B_{9 8},~B^{(2)}_{t 8} = B_{t 8}, 
~e^{{\Phi}_b^{(10)}} = {{ 1 + {C\over r^5}}},
\end{equation}
with $B_{9 8}$ and $B_{t 8}$ as given in eqn. (\ref{b_munu}) and 
the superscript on $B$ denotes the R - R nature of the 2-form
antisymmetric tensors. The next step of our construction is to apply
rotation in $(x^9- x^8)$- plane by an angle $\phi$ which gives the following
configuration:
\begin{eqnarray}
ds^2 &=& -{{1 - {C\over r^5}{\rm sinh^2}{\alpha}\over{\sqrt{1 + {C\over
          r^5}}}}{dt^2} + {1 + {C\over r^5}{\rm cosh^2}{\alpha}{\rm
      cos^2}{\phi}\over{\sqrt{1 + {C\over r^5}}}}}{(d \tilde x^9)}^2 
\cr
& \cr
&+& {1 + {C\over r^5}{\rm cosh^2}{\alpha}{\rm
      sin^2}{\phi}\over{\sqrt{1 + {C\over r^5}}}}{(d \tilde x^8)}^2 
+ {2 {C\over r ^5}{\rm cosh}{\alpha}{\rm sinh}{\alpha}{\rm
    cos}{\phi}\over{\sqrt{1 + {C\over r^5}}}}{d \tilde x^9}{dt}\cr
& \cr
&-& {2 {C\over r ^5}{\rm cosh}{\alpha}{\rm sinh}{\alpha}{\rm
    sin}{\phi}\over{\sqrt{1 + {C\over r^5}}}}{d \tilde x^8}{dt} -
{{2{C\over r^5}{\rm cosh^2}{\alpha}{\rm sin}{\phi}{\rm
    cos}{\phi}}\over{\sqrt{1 +{C\over r^5}}}}{(d \tilde x^9)}{d
\tilde x^8}\cr
& \cr
&+& {\sqrt{1 +{C\over r^5}}}\sum_{i=1}^7 {(dx^i)}^2,
\end{eqnarray}
\begin{eqnarray}
B^{(2)}_{\tilde 8 t} &=& {C \over {C + r^5}}{\rm cosh}{\alpha}{\rm
  cos}{\phi},\cr
& \cr
B^{(2)}_{\tilde 9 t} &=& {C \over {C + r^5}}{\rm cosh}{\alpha}{\rm
  sin}{\phi},\cr
& \cr
B^{(2)}_{\tilde 9 \tilde 8} &=& {-C \over {C + r^5}}{\rm
sinh}{\alpha}.
\end{eqnarray}
Finally we apply T-duality along 
the $\tilde x^9$-direction. By following the prescription as given in
\cite{rcmyers,chull}, we end up with the structure for the
metric, NS- NS $B_{\mu \nu}$, as well as 1- form and 3- form fields 
of type IIA theory:
\begin{eqnarray}
d{\cal{S}}^2 &=& -{1 + {C\over r^5}(1 - {\rm cosh^2}{\alpha}\sin^2{\phi}) 
\over{\sqrt{1 + {C\over r^5}}(1 + {C \over r^5}\cosh^2{\alpha}
\cos^2{\phi})}}{dt^2} \cr
& \cr
&+&{{\sqrt{1 + {C\over r^5}}}\over{1 + {C\over r^5}}
 {\rm cosh^2}{\alpha}{\rm cos^2}{\phi}}{(d \tilde x^9)}^2 
+{{1 + {C\over r^5}{\rm cosh^2}{\alpha}}\over{\sqrt{1 +{C\over
      r^5}}}(1 + {C\over r^5}{\rm cosh^2}{\alpha}{\rm cos^2}
{\phi})}{(d \tilde x^8)}^2 \cr
& \cr
&-& {{C \over r^5}{\rm sinh}{\alpha}{\rm cosh}{\alpha}{\rm
  sin}{\phi}\over{\sqrt{{1 + {C\over r^5}}}
(1 + {C\over r^5}{\rm cosh^2}{\alpha}
{\rm cos^2}{\phi})}} dt{d \tilde x^8}
+ {\sqrt{1 + {C \over r^5}}}\sum_{i=1}^7 {(dx^i)}^2,
\label{d0-d2-metric}
\end{eqnarray}
\begin{eqnarray}
e^{ \Phi_b^{(10)}} &=& {{(1 +{C\over r^5})^{3/2}}\over{1 + {C\over
r^5}{\rm
    cosh^2}{\alpha}{\rm cos^2}{\phi}}},\cr
& \cr
{\cal{A}}_t &=& {C \over{C + r^5}}{\rm cosh}{\alpha}{\rm
  sin}{\phi},~~~~{\cal{A}}_{\tilde x^8} = {-C\over {C + r^5}}
{\rm sinh}{\alpha}\cr
& \cr
{\cal{B}}_{\tilde 9 t} &=& {{{-C \over r^5}{\rm sinh}{\alpha}{\rm
  cosh}{\alpha}{\rm cos}{\phi}}\over{1 + {C \over r^5}{\rm
    cosh^2}{\alpha}{\rm cos^2}{\phi}}},~~~~ 
{\cal{B}}_{\tilde 9 \tilde 8}  =  {{{C \over r^5}{\rm sin}{\phi}{\rm
  cos}{\phi}{\rm cosh^2}{\alpha}}\over{1 + {C \over r^5}{\rm
    cosh^2}{\alpha}{\rm cos^2}{\phi}}},\cr
& \cr 
{\cal{A}}_{\tilde 9 t \tilde 8} &=& - {C \over r^5}{\rm cosh}{\alpha}{\rm
  cos}{\phi}\over{1 + {C \over r^5}{\rm cosh^2}{\alpha}
 {\rm cos^2}{\phi}}.
\label{iiamat}
\end{eqnarray}
This solution is a generalization of the 
$D0-D2$ bound state, where in addition we have turned on 
NS-NS $2$-form as well. To show that this indeed represents a
non-threshold $1/2$ BPS state, in the next sub-section, we 
explicitly verify the mass-charge relation of these bound states.
\subsubsection{Mass-Charge Relationship}

To show the BPS nature of our solution, we perform a dimensional 
reduction of our solution along $\tilde{x}^8$ and 
$\tilde{x}^9$ directions. By dimensional reduction, one avoids any 
ambiguity that may arise due to the presence of 
purely spatial components of the $p$-form fields in 
equation (\ref{d0-d2-metric}), (\ref{iiamat}) above. 
As a result, all the nonzero charges
in our case arise from the temporal part of $1$-form field components 
only, in this eight dimensional theory \cite{roy}. They are
$A^1_t \sim {\cal{A}}_t$, $A^2_t \sim {\cal{B}}_{\tilde 9 t}$, 
$A^3_t \sim {\cal{A}}_{\tilde 9 t
  \tilde 8}$ from the components of the $p$-form 
fields, as well as off-diagonal component 
${g_{t \tilde 8}}/{g_{\tilde 8\tilde 8}}$ of the metric.
The charges associated with these field strengths and metric
components can be read from the solutions 
(\ref{d0-d2-metric}), (\ref{iiamat}). They are :
\begin{eqnarray}
Q_1 &=& 5 C {\rm cosh}{\alpha}~{\sin}{\phi},\cr
& \cr
Q_2 &=& - 5 C {\rm sinh}{\alpha}~{\cosh}{\alpha}~{\rm cos}{\phi}, \cr
& \cr
Q_3 &=&  - 5 C {\rm cosh}{\alpha}~\cos \phi, \cr
& \cr
P &=& 5 C {\rm sinh}{\alpha}~{\cosh}{\alpha}~{\rm sin}{\phi},
\label{02charges}
\end{eqnarray} 
where $Q_i$'s are the charges corresponding to 
the field strengths of $A^i_t$ that we just defined 
and $P$ can be interpreted as the momentum along
$\tilde{x}^8$ direction in the ten-dimensional theory.
The mass-density of our bound state can be computed
using (\ref{adm-mass}) and is given by
\begin{equation}
m_{(0, 2)} = 5 C {\rm cosh^2}{\alpha}.
\label{mtwo}
\end{equation} 
Comparing (\ref{02charges}) and (\ref{mtwo}), we get the standard 
BPS condition as:
\begin{equation}
(m_{0, 2})^2 = {Q^2}_1 + {Q^2}_2 + {Q^2}_3 + P^2.
\end{equation}
This, in turn, implies the supersymmetric nature of the bound state.

\subsubsect{Further Generalization to $Dp-D(p+2)$}

We now obtain a generalization of the $D1-D3$ bound state solution 
presented in eqn. (\ref{d1-d3-myers2}) by applying T-duality along 
$x^7$ direction on the generalized $D0-D2$ solution
(\ref{d0-d2-metric}), (\ref{iiamat}) presented earlier in this section. 
The final result is given by:
\noindent
\begin{eqnarray}
d{\cal{S}}^2 &=& -{1 + {C\over r^4}(1 - \cosh^2{\alpha}\sin^2{\phi}) 
\over{\sqrt{1 + {C\over r^4}}(1 + {C\over r^4}\cosh^2{\alpha}
\cos^2{\phi})}}{dt^2} + {{\sqrt{1 + {C\over r^4}}}\over{1 + {C\over r^4}}
 {\rm cosh^2}{\alpha}{\rm cos^2}{\phi}}{(d \tilde x^9)}^2 \cr
& \cr
&+&{{1 + {C\over r^4}{\rm cosh^2}{\alpha}}\over{\sqrt{1 +{C\over
      r^4}}}(1 + {C\over r^4}{\rm cosh^2}{\alpha}{\rm cos^2}
{\phi})}{(d \tilde x^8)}^2 + {1\over\sqrt{1 + {C\over
    r^4}}}{(dx^7)}^2\cr
& \cr 
&-& {{C \over r^4}{\rm sinh}
{\alpha}{\rm cosh}{\alpha}{\rm sin}{\phi}\over{\sqrt{{1 + {C\over r^4}}}
(1 + {C\over r^4}{\rm cosh^2}{\alpha} {\rm cos^2}{\phi})}} 
dt{d \tilde x^8} + {\sqrt{1 + {C \over r^4}}}\sum_{i=1}^6 {(dx^i)}^2,\cr
& \cr
{\cal{A}}^{(2)}_{7 t} &=& {C\sin \phi \cosh \alpha\over(r^4 +
C)},~~~~~ {\cal{A}}^{(2)}_{7 \tilde 8} =
- {C\sinh\alpha\over(r^4 + C)}, \cr
& \cr
{\cal{A}}^{(4)}_{\tilde 9 t \tilde 8 7 } 
&=& - {({C\over r^4} \cosh \alpha \cos \phi) \over { 2( 1 + {C\over r^4})}}
\left[1 + {{1 + {C\over r^4} \over{{1 + {C\over r^4}}
\cosh^2\alpha\cos^2\phi}}}\right],\cr
& \cr
{\cal{B}}_{\tilde 9 t} &=& {{{-C \over r^4}{\rm sinh}{\alpha}{\rm
cosh}{\alpha}{\rm cos}{\phi}}\over{1 + {C \over r^4}{\rm
cosh^2}{\alpha}{\rm cos^2}{\phi}}},~~~~ 
{\cal{B}}_{\tilde 9 \tilde 8}  =  {{{C \over r^4}{\rm sin}{\phi}{\rm
cos}{\phi}{\rm cosh^2}{\alpha}}\over{1 + {C \over r^4}{\rm
cosh^2}{\alpha}{\rm cos^2}{\phi}}},\cr
& \cr
e^{{\Phi}_{b}^{(10)}} &=& {1 + {C\over r^4}\over{1 + {C\over
r^4}\cosh^2\alpha\cos^2\phi}}.
\end{eqnarray}
Once again, for the special case $\alpha = 0$, our solution
reduces to the one in \cite{rcmyers}. We have therefore presented 
a generalization of the $D1-D3$ bound state to include new 
charges and momenta. The BPS nature of the new solution can be 
again examined by looking at the leading behavior of the 
gauge fields when the above solution is reduced along all its
isometry directions: $x^7, \tilde{x}^8, \tilde{x}^9$. They are:
\begin{eqnarray}
Q_1 &=& 4 C \cosh \alpha \sin \phi, \cr
& \cr 
Q_2 &=& - 4 C \cosh \alpha \cos \phi, \cr
& \cr
Q_3 &=& - 4 C \sinh \alpha \cosh \alpha \cos \phi, \cr
& \cr
P &=& - 4 C \sinh \alpha \cosh \alpha \sin \phi,
\end{eqnarray}
where $Q_1$, $Q_2$, $Q_3$ and $P$ are the charge 
associated with  ${\cal{A}}^{(2)}_{7 t}$, ${\cal{A}}^{(4)}_{\tilde 9 t 
\tilde 8 7 }$, ${\cal{B}}_{\tilde 9 t}$ and $G_{t \tilde 8}$, respectively.

In order to compute the ADM mass-density, we find:
\begin{eqnarray}
h_{77} &=& {C\over r^4}\left({-3\over4} + {1\over 4}\cosh^2 \alpha \cos^2
  \phi\right),\cr
& \cr
h_{\tilde 8 \tilde 8} &=& {C\over
  r^4}\left(\cosh^2\alpha - {3\over 4} \cosh^2 \alpha \cos^2 \phi
  -{3\over 4}\right),\cr
& \cr
h_{\tilde 9 \tilde 9} &=& {C \over r^4}\left({1\over 4} - {3 \over
    4}\cosh^2\alpha\cos^2 \phi\right),\cr
& \cr
h_{i j} &=& {C\over r^4}\left({1 \over 4} + {1\over 4}\cosh^2 
\alpha \cos^2 \phi\right){\delta}_{i j},
\end{eqnarray} 
with $h_{i j}$'s being the deformations of the Einstein metric
above flat background. One then gets, using (\ref{adm-mass}), 
the mass-density of the $D1-D3$ system as:
\begin{eqnarray}
m_{1,3} = 4 C \cosh^2 \alpha.
\end{eqnarray}
We therefore once again have:
\begin{equation}
m_{1,3}^2 = Q_{1}^2 + Q_{2}^2 + Q_{3}^2 + P^2,
\end{equation}
showing the BPS nature of the new bound state. Further generalization 
to higher $Dp-D(p+2)$ bound states can similarly be worked out.
We therefore skip the details. 

\subsect{$\alpha = \beta$ Solution and Generalized 
$D0-D2$ Bound State}

The delocalized $D$-string solution, in this case can be constructed 
by using the $S$-duality transformation on the generalized 
$F$-string solution given in eqn. (\ref{aleqbe}). The classical
solution for the delocalized $D$-string is given by:
\bea
ds^2 &=& - {1\over \sqrt{({1 + {C\over
        r^5}\cosh^2\alpha})}} {(dt)}^2 + 
        {{1 + {C\over r^5}}\over{\sqrt{({1 + {C\over
        r^5}\cosh^2\alpha})}}} {(dx^9)}^2 \cr
& \cr   
&+& {{1 + {C\over r^5} \sinh^2\alpha} \over {\sqrt{({1 + {C\over r^5}
\cosh^2\alpha})}}} {(dx^8)}^2 + 2 {{{C\over r^5}\sinh\alpha}\over
{\sqrt{(1 + {C\over r^5}\cosh^2\alpha)}}} (dx^8)(dx^9) \cr
& \cr
&+& \sqrt{(1 + {C\over r^5}\cosh^2\alpha)} \sum^7_{i = 1}{(dx^i)}^2,\cr
& \cr
e^{\Phi^{(10)}} &=& 1 + {C\over r^5}\cosh^2\alpha,\cr
& \cr
B^{(2)}_{9t} &=& - {{C\over r^5} {\sinh \alpha \cosh \alpha}
\over{1 + {C\over r^5}\cosh^2\alpha}},\>\>\> 
B^{(2)}_{t8} = - {{C\over r^5} \cosh \alpha
\over{1 + {C\over r^5}\cosh^2\alpha}}. 
\eea 
Now, as already explained in the previous case, we apply a rotation
in the $(x^9 - x^8)$ plane by an angle $\phi$, which gives the
following configuration: 
\bea
ds^2 &=& - {1\over \sqrt{({1 + {C\over r^5}\cosh^2\alpha})}} {(dt)}^2
+ {{1 + {C\over r^5}(\cos\phi + \sinh \alpha \sin\phi)^2}
\over {\sqrt{({1 + {C\over r^5}\cosh^2\alpha})}}} {(d\tilde x^9)}^2 \cr
& \cr
&+& {2{C\over r^5}(\cos\phi + \sin\phi \sinh\alpha)(\sinh\alpha \cos
  \phi- \sin\phi)\over {\sqrt {1 + 
      {C\over r^5}\cosh^2\alpha}}}(d\tilde x^8) (d\tilde x^9)\cr 
& \cr
&+&{{1 + {C\over r^5}(\sin\phi - \sinh \alpha \cos\phi)^2}
\over {\sqrt{({1 + {C\over r^5}\cosh^2\alpha})}}} {(d\tilde x^8)}^2
+ \sqrt{1 + {C\over r^5}\cosh^2\alpha} \sum^7_{i = 1}{(dx^i)}^2,\cr
& \cr
e^{\Phi^{(10)}} &=& 1 + {C\over r^5}\cosh^2\alpha,\cr
& \cr
B^{(2)}_{\tilde{9} t} &=& - {{C\over r^5} \cosh \alpha \over {1 + {C\over
      r^5}\cosh^2 \alpha}}(\cos \phi \sinh \alpha - \sin \phi)\cr
& \cr
B^{(2)}_{\tilde{8} t} &=&  {{C\over r^5} \cosh \alpha \over {1 + {C\over
      r^5}\cosh^2 \alpha}}(\sin \phi \sinh \alpha + \cos \phi)
\eea
Next, we apply $T$-duality along $\tilde x^9$ direction as the final
step of our construction. The metric, dilaton and various field
strengths of the resulting configuration is given by:
\bea
d{\cal{S}}^2 &=&  - {1\over \sqrt{({1 + {C\over r^5}\cosh^2\alpha})}} 
{(dt)}^2 + {\sqrt{1 +{C\over r^5}\cosh^2\alpha}
\over{1 + {C\over r^5}(\cos\phi +
    \sinh\alpha \sin\phi)^2}} {(d\tilde x^9)}^2 \cr
\cr
&+&  {\sqrt{1 + {C\over r^5}\cosh^2\alpha}\over{1 +
    {C\over r^5}(\cos\phi + \sinh\alpha\sin\phi)^2}}{(d\tilde x^8)}^2 
+ \sqrt{1 + {C\over r^5}\cosh^2\alpha} \sum^7_{i = 1}{(dx^i)}^2 \cr
& \cr
e^{\Phi^{(10)}} &=& {(1 + {C\over r^5})^{3\over 2}\over {1 + {C\over
      r^5}(\cos\phi + \sinh\alpha \sin\phi)^2}}\cr
& \cr
{\cal{A}}_{t} &=& -{{C\over r^5}\cosh\alpha \over {1 + {C\over
      r^5}\cosh^2\alpha}}(\cos\phi \sinh\alpha - \sin \phi)\cr
& \cr
{\cal{A}}_{\tilde 9 \tilde 8 t} &=&{{C\over r^5}\cosh\alpha \over {1 + {C\over
      r^5}\cosh^2\alpha}}(\sin\phi \sinh\alpha + \cos \phi)\cr
& \cr
{\cal{B}}_{\tilde 9\tilde 8} &=&-{{{C\over r^5}(\cos\phi +
    \sin\phi\sinh\alpha)(\sinh\alpha\cos\phi - \sin\phi)} 
\over {1 + {C\over
      r^5}(\cos\phi + \sinh\alpha \sin\phi)^2}}  
\label{conf}
\eea 
This solution describes the generalized $D0-D2$ bound state which
contains additional NS-NS fields in addition to the 1-form and
3-form RR fields.   
To show that the above solution really represents the non-threshold
bound of $D0-D2$ branes, which preserve $1/2$ supersymmetry, in the 
next sub-section we will verify the mass-charge relationship
explicitly. 
\subsubsect{Mass-Charge Relationship}

To show the BPS nature of the solution, described in the earlier
sub-section, we will apply dimensional reduction along the two
directions $x^{\tilde 8}$ and $x^{\tilde 9}$. As a result all
non zero charges comes from the temporal component of $1$-form field
components only in the eight dimensional theory. They are
$A^1_t \sim {\cal{A}}_t$, $A^2_t \sim {\cal{A}}_{\tilde 9\tilde 8 t}$ 
from the components of the $p$-form fields. The charges associated
with these field strengths can be read of from the corresponding
expressions given in eqn. (\ref{conf}). They are:
\bea
Q_{1} &=& - 5 C\cosh \alpha(\cos \phi \sinh\alpha - \sin \phi),\cr
& \cr
Q_{2} &=&  5C\cosh \alpha (\cos\phi + \sin\phi\sinh\alpha) 
\label{02charges1}\eea    
where $Q_i$'s are the charges corresponding to the field strengths 
of $A^i_t$ $(i=1,2)$ that we have defined. The mass-density of our bound 
state can be computed by using (\ref{adm-mass}) and is given by:
\begin{equation}
m_{(0, 2)} = 5 C {\rm cosh^2}{\alpha}.
\label{mtwo1}
\end{equation} 
Comparing (\ref{02charges1}) and (\ref{mtwo1}), we get the $1/2$
BPS condition as:
\begin{equation}
(m_{0, 2})^2 = Q^2_1 + Q^2_2.
\end{equation}
This, in turn, implies the supersymmetric nature of the bound state.
As described earlier, one can also find out all other $Dp-D(p+2)$ bound 
states by applying $T$-duality along the transverse directions
of the $(D0-D2)$ solution described in eqn. (\ref{conf}). 
\sect{$\beta = 0, \alpha \neq 0$ Solutions and Generalization 
to $(p, q)$- Strings}

In this section, we discuss even more nontrivial examples, using 
$\beta =0, \alpha \neq 0$ cases, discussed in 
section-2.2. We now also perform further generalization by applying 
$T$-duality on delocalized $(p, q)$-strings rather than 
$(0, 1)$ or D-strings. 

\subsection{Construction of SL(2, Z) Multiplets}
The delocalized elementary string solution in ten-dimensions
for $\beta = 0, \alpha \neq 0$ situation  is given 
in eqns.(\ref{10dg})-(\ref{10phi}). This solution represents a string
along $x^8$, which is delocalized along $x^9$. One can use it to 
construct generalized bound states in the manner described in the 
last section. We, however, make further generalization by 
using the $SL(2,Z)$ symmetry of type IIB string theories in $D=10$. 
We can easily construct 
a $(p,q)$ multiplet of delocalized string from (\ref{10dg})-(\ref{10phi}). 
The procedure of constructing such configuration is discussed 
in \cite{jhschwarz}. Instead of giving the detail, we write down the 
final form of the configuration.

The metric is given by:
\begin{eqnarray}
ds^2 &=& {A\over \Delta}\Big(- [1 - {\rm sinh}^2 {\alpha\over 2} (e^{-E}
-1)]
dt^2 + [1 + {\rm sinh}^2 {\alpha\over 2} (e^{-E} -1)] (dx^8)^2\nonumber\\
&+& \Delta (dx^9)^2 
+  {\rm sinh}\alpha (e^{-E} -1)dx^9 dt +
{2} {\rm sinh}^2{\alpha\over 2} (e^{-E} -1) dt dx^8
\nonumber\\
&+&  {\rm sinh}\alpha (e^{-E} -1) dx^8 dx^9 +
\Delta \sum_{i=1}^7 dx^i dx^i \Big),
\label{pqdeloca}
\end{eqnarray}
where we have defined 
\begin{equation} 
\Delta = {\rm cosh}^2{\alpha\over 2}e^{-E} - {\rm sinh}^2{\alpha\over 2},
~~{\rm and}~~ A = {{\sqrt{p^2 + q^2 \Delta}}\over{{\sqrt{p^2 + q^2}}}}.
\end{equation}
Furthermore, the NS-NS and R-R two forms are given by
\begin{eqnarray}
B^{(1)}_{8t} = {p\over{\sqrt{p^2 + q^2}}} B_{8t},
~~B^{(1)}_{9t} = {p\over{\sqrt{p^2 + q^2}}} B_{9t} ~= ~B^{(1)}_{98}, \cr
& \cr
B^{(2)}_{8t} = {q\over{\sqrt{p^2 + q^2}}} B_{8t},
~~B^{(2)}_{9t} = {q\over{\sqrt{p^2 + q^2}}} B_{9t} ~= ~B^{(2)}_{98}.
\label{pqdelocb}
\end{eqnarray}
Here $B_{8t}, B_{9t}, B_{98}$ are given by (\ref{10db}). The superscripts 
$1$ and $2$ indicate that these two forms are NS-NS and R-R
in nature respectively. 
Also, after $SL(2,Z)$ transformation, we  have the dilaton and axion as:
\begin{equation} 
\Phi_b^{(10)} = - 2 ~{\rm ln}\Big({(p^2 + q^2) \Delta ^{1\over 2}
\over{p^2 + q^2 \Delta}}\Big), ~~\chi = {pq(\Delta -1)\over{p^2 + q^2 
\Delta}}.
\label{pqdlocc}
\end{equation}
\subsection{Construction of Bound States}

Now, we perform, as before, a rotation on the above $(p,q)$ string solution
in the
$x^9-x^8$ plane. This is again given by
\begin{eqnarray}
dx^9 &=& {\rm cos} \phi ~d{\tilde x}^9 - {\rm sin}\phi ~d{\tilde x^8},
\nonumber\\
dx^8 &=& {\rm cos}\phi ~d{\tilde x}^8 + {\rm sin}\phi ~d{\tilde x}^9
\label{rot}.
\end{eqnarray}
Under this rotation, (\ref{pqdeloca}) - (\ref{pqdelocb}) become:
\begin{eqnarray}
ds^2 &=& {A\over \Delta}\Big( -(1 - \delta_\alpha )dt^2 
 +  [  {\rm sin}^2 \phi (1 + \delta_\alpha)
  + \gamma_\alpha {\rm sin}\phi {\rm cos}\phi + \Delta
   {\rm cos}^2\phi] (d\tilde x^9)^2 \nonumber\\
 &+&  [  {\rm cos}^2 \phi (1 + \delta_\alpha)
  - \gamma_\alpha {\rm sin}\phi {\rm cos}\phi + \Delta 
   {\rm sin}^2\phi] (d\tilde x^8)^2\nonumber\\
&+& [ \gamma_\alpha {\rm cos} 2\phi 
   + 2 (\delta_\alpha - \Delta +1) {\rm cos}\phi {\rm sin}\phi ] 
    d\tilde x^9 d\tilde x^8 
+ [ 2\delta_\alpha {\rm sin}\phi + \gamma_\alpha
 {\rm cos}\phi ] d\tilde x^9 dt\nonumber\\ 
&+& [ 2\delta_\alpha {\rm cos}\phi - \gamma_\alpha
 {\rm sin}\phi ] d\tilde x^8 dt + \Delta \sum_{i=1}^7 (dx^i)^2\Big),
\label{pqrota}
\end{eqnarray}
and 
\begin{eqnarray}
B^{(1)}_{\tilde 8 t} &=& {p \over{\Delta \sqrt{p^2 + q^2}}}[
{\rm cos}\phi (e^{-E}-1 + \delta_\alpha ) + {{\rm sin}\phi
\gamma_\alpha\over 2}],\cr
& \cr
B^{(1)}_{\tilde 9 t} &=& {p \over{\Delta \sqrt{p^2 + q^2}}}[
{\rm sin}\phi (e^{-E}-1 + \delta_\alpha ) - {{\rm cos}\phi
\gamma_\alpha\over 2}],\cr
& \cr
B^{(1)}_{\tilde 9 \tilde 8} &=& -{p \gamma_\alpha\over{2\Delta
\sqrt{p^2 + q^2}}}, \cr
& \cr
B^{(2)}_{\tilde 8 t} &=& {q \over{\Delta \sqrt{p^2 + q^2}}}[
{\rm cos}\phi (e^{-E}-1 + \delta_\alpha ) + {{\rm sin}\phi
\gamma_\alpha\over 2}],\cr
& \cr
B^{(2)}_{\tilde 9 t} &=& {q \over{\Delta \sqrt{p^2 + q^2}}}[
{\rm sin}\phi (e^{-E}-1 + \delta_\alpha ) - {{\rm cos}\phi
\gamma_\alpha\over 2}],\cr
& \cr
B^{(2)}_{\tilde 9 \tilde 8} &=& -{q \gamma_\alpha\over{2\Delta
\sqrt{p^2 + q^2}}}.
\label{pqrotb}
\end{eqnarray}
Here, we have defined 
\begin{equation}
\delta_\alpha = {\rm sinh}^2{\alpha\over 2}(e^{-E} -1),
~~\gamma_\alpha = {\rm sinh}\alpha (e^{-E} -1),
\nonumber
\end{equation}
Furthermore, the dilaton and axion remain same as (\ref{pqdlocc}). 
Our final step of the construction is to apply T-duality along $\tilde
x^9$ direction. The resulting solution would then correspond to 
a bound state in type IIA theory. The T-duality 
map between IIA and IIB theories is given explicitly in \cite{hull,
rcmyers}. Applying this map, we end up with resulting configuration:
\begin{eqnarray}
d{\cal{S}}^2 &=& {1\over{ \Delta \sigma{\sqrt{p^2 + q^2}} {\sqrt{p^2 + q^2
   \Delta}}}}[p^2\{{\rm sin}\phi (e^{-E}  + \delta_\alpha)\nonumber\\
&&   ({\rm sin}\phi (e^{-E} -2 + \delta_\alpha) - {\rm cos}\phi
   \gamma_\alpha ) - \Delta {\rm cos}^2\phi (1 -
\delta_\alpha)\}\nonumber\\
  && - q^2\Delta\{ ({\rm sin}\phi + {\gamma_\alpha\over 2} {\rm
   cos}\phi)^2 
   + \Delta {\rm cos^2}\phi (1 - \delta_\alpha)\}] (dt)^2 \nonumber\\
&&+{1\over{ 2\Delta \sigma{\sqrt{p^2 + q^2}} {\sqrt{p^2 + q^2
  \Delta}}}}[\Delta\{-\gamma_\alpha^2 q^2 + 4 \delta_\alpha (
  p^2 + \Delta q^2)\}{\rm cos}\phi \nonumber\\
&&- 2 \gamma_\alpha
  \{ \delta_\alpha p^2 + e^{-E} p^2 + \Delta q^2\}{\rm sin}\phi]   
  dt d{\tilde x}^8\nonumber\\
&&+ {\Delta {\sqrt{p^2 + q^2}}\over{\sigma {\sqrt{p^2 + q^2 
  \Delta}}}}d\tilde x^9 d\tilde x^9\nonumber\\
&&- {p \over {\sigma{\sqrt{p^2 + q^2 \Delta}}}}[
  2 {\rm sin}\phi (e^{-E} -1 + \delta_\alpha) - \gamma_\alpha {\rm
  cos}\phi]d{\tilde x}^9 dt\nonumber\\
&&+ {1\over{\Delta \sigma {\sqrt{p^2 + q^2 \Delta}}{\sqrt{p^2 + q^2}}}}
  [ (\Delta + \delta_\alpha \Delta - {\gamma_\alpha^2\over 4})
  (p^2 + q^2 \Delta) \nonumber\\
&&+ p^2 \{{-\gamma_\alpha\over 2} {\rm cos}\phi
  + (e^{-E} -1 + \delta_\alpha) {\rm sin}\phi\}^2]
  d\tilde x^8 d\tilde x^8\nonumber\\
&& + {p \gamma_\alpha\over{\sigma{\sqrt{p^2 + q^2 \Delta}}}}d\tilde x^8
d\tilde x^9\nonumber\\
&& + {{\sqrt{p^2 + q^2 \Delta}}\over{\sqrt{p^2 + q^2}}}
\sum_{i=1}^{7}dx^i dx^i.
\label{iiametric}
\end{eqnarray}
The IIA dilaton is
\begin{equation}
e^{{\Phi}_a^{(10)}} = \Big({{{p^2 + q^2\Delta}}\over{{p^2 +
q^2}}}\Big)^{3\over 2}{1\over \sigma}.
\end{equation}
The two form components are given by
\begin{eqnarray}
{\cal{ B}}_{t \tilde 8} &=& -{{p\over {2\Delta\sigma\sqrt{p^2 +
  q^2}}}}
  [-\gamma_\alpha (e^{-E} + \delta_\alpha){\rm sin}\phi 
  +2 \Delta (-e^{-E} +1 - 
  \delta_\alpha){\rm cos}\phi],\nonumber\\
{\cal {B}}_{t \tilde 9} &=& {1\over \sigma}[\delta_\alpha {\rm
  sin}\phi +
  {\gamma_\alpha\over 2}{\rm cos}\phi],\nonumber\\
{\cal {B}}_{\tilde 8 \tilde 9} &=& {1\over \sigma}
[{\gamma_\alpha\over 2}{\rm cos} 2\phi + (\delta_\alpha- \Delta +1)
{\rm cos}\phi {\rm sin}\phi].
\label{iiab}
\end{eqnarray}
The nonzero components of 1-form and 3-form fields in IIA theory 
are given by
\begin{eqnarray}
{\cal {A}}_{t} &=& {q{\sqrt{p^2 + q^2}}\over{p^2 + q^2 \Delta}}
  [{\rm sin}\phi (e^{-E} -1 + \delta_\alpha ) - \gamma_\alpha{{\rm
  cos}\phi\over 2}],\nonumber\\
{\cal {A}}_{\tilde 8} &=& -{q \gamma_\alpha {\sqrt{p^2 + q^2}}\over
  {2(p^2
  + \Delta q^2)}} \nonumber\\
{\cal{A}}_{\tilde 9} &=& -{p q (\Delta -1)\over {p^2 +
  q^2\Delta}}\nonumber\\
{\cal{A}}_{t \tilde 8 \tilde 9} &=& {{q\over {2\sigma\Delta\sqrt{p^2
  +
  q^2}}}} [-\gamma_\alpha (e^{-E} + \delta_\alpha){\rm
  sin}\phi +2 \Delta (-e^{-E} +1 - 
  \delta_\alpha){\rm cos}\phi],
\label{iiaa}
\end{eqnarray}
where 
\begin{equation}
\sigma = {\rm sin}^2\phi ( 1 + \delta_\alpha) +
         \gamma_\alpha {\rm sin}\phi {\rm cos}\phi + 
         \Delta {\rm cos}^2\phi,
\end{equation}
and 
\begin{equation}
e^{-E} = 1 + {C \over r^5}. 
\end{equation}
We, once again, have multiple nonzero components of the 
NS-NS $2$-form as well as R-R $1$-form and $3$-form fields, in addition to 
having nonzero momenta coming from the off-diagonal components of the 
metric. We have explicitly checked that the above configuration 
reduces to known D-brane bound states in appropriate limits.
\subsection {Mass-Charge Relationship}

As in section-(2.4), we now verify that mass and charge associated 
with the above solution satisfy the expected mass-charge relation of
$1/2$ (non-threshold) BPS bound states. As in section-(2.4),we 
once again reduce the solution along both its spatial isometry 
directions. In the resulting theory in $D=8$, nonzero charges
are associated with gauge fields from the 
IIA field reductions. We have the non zero gauge fields which are of 
electric type, given as:
$A^1_t = {\cal{A}}_{t}$, 
$A^2_t = {\cal{A}}_{t\tilde 8 \tilde 9}$,
$A^3_t = -{\cal{B}}_{t \tilde 8}$, 
$A^4_t = {\cal{B}}_{t \tilde 9}$,
$A^5_t = g_{t \tilde 8}/g_{\tilde 8 \tilde 8}$, 
$A^6_t = g_{t \tilde 9}/g_{\tilde 9 \tilde 9}$ \cite{roy}. 
The charges can be read off from 
the leading order behavior of the above expressions, and are as
follows:
\begin{eqnarray}
Q_{1} &=& {{5 C q\over{\sqrt{p^2 + q^2}}}\times{{\rm cosh}{\alpha \over2}}
 ({\rm sin}{\phi}{\rm cosh}{\alpha\over2} 
- {\rm cos}{\phi}{\rm sinh}{\alpha\over2})},\cr
\cr
Q_{2} &=& -{{5 C q\over{\sqrt{p^2 + q^2}}}\times{{\rm cosh}{\alpha
\over2}}
({\rm cos}{\phi}{\rm cosh}{\alpha \over2} 
+ {\rm sin}{\phi}{\rm sinh}{\alpha \over2})},\cr
\cr
Q_{3} &=& - {{5 C p\over{\sqrt{p^2 + q^2}}}\times{{\rm cosh}{\alpha \over2}}
({\rm cos}{\phi}{\rm cosh}{
\alpha \over2} + {\rm sin}{\phi}{\rm sinh}{\alpha \over2})},\cr
\cr
Q_{4} &=& {5 C \times{{\rm sinh}{\alpha \over2}} ({\rm sin}{\phi}
{\rm sinh}{\alpha \over2} +
{\rm cos}{\phi}{\rm cosh}{\alpha \over2})},\cr
\cr
P_{1} &=& {5 C \times{{\rm sinh}{\alpha \over2}}({\rm cos}{\phi}{\rm
    sinh}{\alpha\over2} -
{\rm sin}{\phi}{\rm cosh}{\alpha \over2})},\cr
\cr
P_{2} &=& {{5 C p\over{\sqrt{p^2 + q^2}}}\times{{\rm cosh}{\alpha \over2}}
({\rm sin}{\phi}{\rm cosh}{\alpha \over2} - {\rm cos}{\phi}{\rm
  sinh}{\alpha \over2})},
\label{charges}
\end{eqnarray}
where $Q_i$ corresponds to the charge associated with
$A^i_t$, $(i=1,..,4)$ and the $P_i$, $(i=1,2)$ correspond to
the charges with respect to the gauge fields 
$A^5_t$ and $A^6_t$. They also correspond to the 
momenta along $\tilde{x}^8$ and $\tilde{x}^9$ in the original 
uncompactified theory. 
The ADM mass density of the bound state, that we have constructed 
is 
\begin{equation}
m^{(p, q)}_{(0,2)} = 5 C {\rm cosh}{\alpha}.
\label{massform}
\end{equation}
Therefore we have 
\begin{equation}
\left(m^{(p, q)}_{(0,2)}\right)^2 = {Q_{1}}^2 + {Q_{2}}^2 + 
{Q_{3}}^2 + {Q_{4}}^2 + {P_{1}}^2 + {P_{2}}^2.
\label{masscharge} 
\end{equation}
This relation implies that the bound state constructed above do satisfy
the BPS bound for the system.
As in section-2.4, one can also give a generalization 
of $Dp-D(p+2)$ bound states in a similar manner, as discussed above, 
by smearing 
directions $x^7$ etc. in the solution (\ref{iiametric} - \ref{iiaa}) 
and then by applying T-duality along these directions. We, however,
skip these details.   

It is interesting to note that we have been able to generate all six
gauge charges in $D=8$, using our procedure. As is known, these gauge
charges form a $(3, 2)$ representation of $SL(3, Z) \times SL(2, Z)$ 
$U$-duality symmetry in $D=8$. 
We can therefore rewrite the RHS of equation (\ref{masscharge})
in a $U$-duality invariant form by exciting general moduli as in 
\cite{jhschwarz,roy}. This is not surprising, as solution generating 
technique used by us is known to be equivalent to the one using 
$U$-duality transformations\cite{townsend,papado,mcvetic}. 
In particular, in examples where we 
consider particle-like states in $D=8$ for writing down the 
mass-charge relations, the relevant $U$-duality group is 
$SL(3, Z)\times SL(2, Z)$. In type IIB examples, this $SL(2, Z)$ in 
$D=8$ originates from the group of constant coordinate transformations
along the two internal directions, whereas $SL(3, Z)$ is a
combination of the $D=10$ S-duality group together with the
$T$-duality along the $x^8$ and $x^9$. $(3, 2)$ multiplet of states
mentioned above are then generated by applying these transformations
to a seed solution in $D=8$, 
originating from the (delocalized) F-string solution 
in $D=10$. All our transformations can also be mapped to 
appropriate ones lying inside the $U$-duality group. 

\sect{Summary}

We have explicitly constructed the nontrivial bound states
of $D$-branes starting with charged macroscopic strings. In particular, we 
were able to construct configurations in ten dimensions which carry 
$(F, D0, D2)$ charges as well as non-zero momenta. We also found that
these bound states are supersymmetric. We have also checked that all our 
solutions reduce to known bound states in appropriate limits.  We 
further generalized our results to $(Dp - D(p+2))$ bound states in 
IIA/B theories.

The bound states of $D$-branes, when compactified to lower dimensions, 
often allow us to understand various properties of black hole including
Bekenstein-Hawking entropy in a microscopic way 
\cite{ashokesen,wald,callan-malda}. It would be interesting to
investigate as to what new insight we gain from our configurations 
along this direction. A first step may then be to reduce the solutions that 
we presented in sections-(2.4) and (2.5) to $D=8$ and $D=7$ 
respectively to interpret them as  charged  extremal black holes in
lower dimensions. Various excitations of the higher dimensional bound 
states along the compactified direction may then correspond to the 
required degrees of freedom responsible for the entropy associated 
with the black hole. 
It would also, perhaps, be interesting to understand the conformal field
theory descriptions of these bound states. They are typically described by
boundary states of the underlying open string theory. Construction of
these boundary states often turned out to be very useful in the past,
see for example \cite{divecchia,lerda}.

\chapter{$D$-BRANES IN PP-WAVE BACKGROUND}
\pagenumbering{arabic}
\markright{Chapter 3. $D$-BRANES IN PP-WAVE BACKGROUND}
\vspace{-1cm}
\sect{Introduction}
PP-waves \cite{penrose,tsyt1,guven,metsaev,
blau,tsyt2,malda,mukhi,GO,tseytlin,sonn,mohsen,chu,lee,Bala,Taka}
are known to be an interesting class of supersymmetric solutions of
type IIB supergravity, with wide applications to gauge theory.
It is known that pp-wave spacetime yields exact classical backgrounds
for string theory since all curvature invariants and therefore 
all $\alpha'$ corrections vanish\cite{klim,steif}. 
Hence pp-wave spacetimes correspond to exact conformal field theory 
backgrounds. These backgrounds are shown to be exactly solvable in 
light cone gauge\cite{met0,met1,met2}. They are also  known to be 
maximally supersymmetric solutions of supergravities in 
various dimensions\cite{meessen,pope,hull}, and are known to  
give rise to solvable string theories from worldsheet 
point of view as well. Many of these backgrounds are obtained 
from $AdS_p\times S^q$ type geometries 
in {\it Penrose} limit and are also maximally supersymmetric
\cite{penrose,blau}. The particular case of $AdS_5\times S^5$ is 
of special interest, with applications to $N=4$, $D=4$ gauge theories 
in the limit of large conformal dimensions and R-charges\cite{malda}.
$D$-branes known as nonperturbative objects in
string theory also survive in this limits and also have an appropriate 
representation in such gauge theories, in terms of operators
corresponding to `giant gravitons' and `defects'\cite{lee,Bala}.  
Keeping in view of the importance of $D$-branes in understanding
the nonperturbative as well as the duality aspects of string theory,
it is desirable to study them in general backgrounds with flux
being turned on. Motivated by these facts, in \cite{bkp,ks},
we continue the search for 
explicit $D$-brane supergravity solutions in string theories 
in pp-wave background. Our study has been concentrated mainly on
the NS-NS and R-R pp-waves arising out of $AdS_3\times S^3$ 
geometry \cite{tseytlin}. Explicit supergravity solution 
of $D$-branes, along with open string spectrum, has been  
studied in \cite{dabh,akumar,sken,alishah,bain,singh,pal,michi,
OPS,NP,HNP,NPS}.
Branes oriented at relative $SU(2)$ angle in pp-wave background 
along with the $\kappa$-symmetry
and supersymmetry has also been discussed in literature\cite{rashmi}.
We give the supergravity realization of $D$-branes in the pp-wave
background of $AdS_3\times S^3 \times R^4$ with constant and
nonconstant three form flux.
We also find it interesting to note that, unlike the case of 
$AdS_5\times S^5$ PP-wave, in our case we are able to obtain several 
`localized' $D$-brane solutions. $Dp$-branes in R-R PP-wave background
are obtained by applying a set of $S$ and $T$-duality transformations
on the brane solutions in NS-NS background. Such background PP-wave 
configurations were already discussed in the context of $D7$-branes in 
\cite{akumar}. The worldsheet construction of these branes  
then follows from the results in \cite{akumar}, and will also be 
discussed below. We, however, point out that there is a crucial 
difference between the solutions presented in this paper and 
those in \cite{akumar}. In the present case, only light-cone 
directions of the pp-wave are along the branes, 
the remaining four are always transverse to them. On the other hand, 
for the solutions in \cite{akumar} the directions transverse to the
branes are flat. One of the important result of our construction is
that the $Dp$-brane ($p=1,...,5$) solutions, are fully localized, 
whereas the $Dp-Dp^{\prime}$ are localized in the common transverse space.
In addition to the above solutions, we also present the $D$-brane 
solutions in the other pp-waves such as the ones in little string
theories etc.     

PP-wave backgrounds with nonconstant flux have also been studied in
\cite{maoz,russo,hikida,kim,bonelli}. The corresponding worldsheet theory 
is described by nonlinear sigma model. These theories provide 
examples of interacting theories in light cone gauge. 
The sigma model with R-R five form flux is supersymmetric and 
one can have linearly realized `supernumerary' supersymmetries in 
these backgrounds\cite{cvetic}. The corresponding sigma model 
with 3-form NS-NS 
and RR fluxes is non-supersymmetric\cite{russo} unless there exists
some target space isometry and corresponding Killing vector fields,
which ensure the worldsheet supersymmetry\cite{kim}. Also the
supernumerary supersymmetries are absent in this case due to the structure 
of gamma matrices. The bosonic part is same in both cases and the
two models share many properties, e.g. both are exact string theory
backgrounds and integrability structure is same. We
generalize our construction of $D$-branes with nonconstant flux as
well. 
\sect{Review of $D$-brane solutions in pp-wave background}

The pp-wave arising out of the Penrose limit of $AdS_5 \times S^5$
has the following metric and surviving constant field strength:
\bea
ds^2 &=& -4dx^+ dx^- - \mu^2 x^2_i (dx^+)^{2} + \sum^8_{i=1}(dx^i)^{2}\cr
& \cr
F_{+1234} &=& F_{+3456} = Constant \times \mu,
\eea  
where $x^{\pm} = {1\over 2}(x^0 \pm x^9)$ are the lightcone coordinates.
The metric part of the solution has isometry group 
$SO(1,1)\times SO(8)$ which is further broken to $SO(1,1)\times
SO(4)\times SO(4)$ due to the presence of null RR five form field.
This space is nonisotropy and homogeneous. On the other hand,
the pp-wave arising out of $AdS_3 \times S^3\times R^4$ kind 
of geometry has the following structure which is supported by constant
RR 3-form flux:
\bea
ds^2 &=& 2dx^+ dx^ - - \mu^2 x^2_i (dx^+)^{2} + \sum^4_{i=1}(dx^i)^{2} 
+ \sum^8_{a=5}(dx^a)^{2}\cr
& \cr
F_{+12} &=& F_{+34} = 2 \mu.
\eea

$D$-branes, the extended objects in string theory, are important
in understanding various duality conjectures in string and gauge
theories. In AdS/CFT duality, they correspond to nonperturbative objects
in gauge theory side or defects on which a lower dimensional conformal
field theory lives. To study the field theory dynamics in the presence
of such objects using duality, it is often useful to have access to 
the supergravity solutions for $D$-branes in AdS space. Unfortunately
not enough is known about these objects in curved space. But it is
easy to treat them in pp-wave background, which is closer to the 
flat space. For instance, while the CFT construction of
$D$-brane is rather difficult in AdS
space\cite{bachas,malda-moore,ponsot}, it is much easier to treat 
this problem in pp-wave background\cite{billo,alis-jab}. So the hope 
is that even the supergravity solution will be more tractable 
in pp-wave background. Performing this exercise might tell us how
to tackle the same in AdS space.
There are two ways to construct the explicit supergravity
solutions of these objects in pp-wave background. First one is by 
{\it direct search}: Writing down the appropriate ansatz for
$D$-branes in pp-wave background and then solve low energy
supergravity equations of motion and the Bianchi identities etc..
The second approach is by taking `Penrose' limit of the near horizon
geometry of intersecting $D$-brane solutions. The problem with
the second approach is to find out the `localized' $D$-brane solutions
in pp-wave background. Very few examples of these in AdS and
pp-wave spacetime are
known\cite{faya1,faya2,lupo,boe,papa,akumar,alishah}. For example:
The $D$-brane solutions found out in \cite{akumar} represents 
localized $D5$ and $NS5$ branes in pp-wave background. These solutions 
were obtained by applying `Penrose' limit to the near horizon geometry
of intersecting brane solutions of the type $NS1-NS1'-NS5-NS5'$
discussed in\cite{papa}. The pp-wave
background arising out of $AdS_5\times S^5$ type of geometry supported
by nonzero five form RR-field strength are nonisotropy and
homogeneous. Due to the nonisotropy nature of the space,
$D$-branes with different orientations preserve different amount of
supersymmetry. The homogeneity of space implies that $D$-branes can
pass through any point in space at a given moment of time.

Using $D$-brane approach, recently various embeddings of $D$-branes 
in pp-wave background is explored. There are three kinds of
branes in Hpp-wave background: {\it longitudinal $D$-branes}, for which
the pp-wave propagates along the brane, {\it transversal $D$-branes},
for which one of the direction is transverse to the pp-wave, but the
time like direction is always along the brane and {\it instantonic
$D$-branes}, for which both the directions along which the pp-wave
propagates and the time like direction are transverse to the brane. 
In the maximal pp-wave background, there are two families of $D$-brane
solutions exist. The first family is the {\it delocalized}, 
supersymmetric $D$-brane solutions and the second family is the 
{\it localized}, nonsupersymmetric solutions. These solutions were
verified in \cite{bain}. The peculiarity of the nonsupersymmetric
solutions is that the gravity becomes repulsive close to the core of
the $D$-branes. If we embed a $Dp$-brane in this background in such a 
way that the pp-wave propagates along its worldsheet, the worldvolume
coordinates splits into three sets: the lightcone coordinates $(x^+,
x^-)$, $m$ coordinates along the first $SO(4)$ subspace $(x^1,...x^4)$
and $n$ coordinates along second $SO(4)$ subspace $(x^5,...x^8)$. For a 
$Dp$ brane $(m + n = p-1)$, and this embedding is called in the
literature as $(+, -, m, n)$. We summarize the supersymmetry of
various $D$-branes in these kind of embeddings\cite{sken,bain} in
table (3.2). 
\begin{table}[h]
\begin{center}
\begin{tabular}{|c|c|c|}
\hline
brane& embedding & SUSY\\
\hline 
D1& (+, -, 0, 0)& 1/4\\
\hline
D3 & (+, -, 2, 0)& 1/4\\

   & (+, -, 1, 1) & - \\
\hline
D5 & (+, -, 3, 1)& 1/4\\

   & (+, -, 2, 2) & - \\
\hline                 
D7 & (+, -, 4, 2)& 1/4\\

   & (+, -, 3, 3) & - \\
\hline
\end{tabular}
\caption{Amount of supersymmetry preserved by various $D$-brane
  supergravity solutions.}
\end{center}
\end{table} 

In the next section we will discuss the $Dp$ as well as $Dp-Dp'$
branes in the pp-wave background arising out of $AdS_3\times S^3$
type of geometry and supported by NS-NS and RR three form flux.
As it will be clear that all the $Dp$-branes, for $(p=1,..,5)$ are
fully localized where as the $Dp-Dp^{\prime}$ are localized along 
common transverse directions. We will also discuss the supersymmetry 
properties of these solutions and the worldsheet construction for the 
$p-p^{\prime}$ system.      
\sect{$D$-Branes in PP-wave Background with Constant NS-NS
  Flux}
\subsect{Supergravity Solutions}

We start by writing down the $D$-string solutions in the pp-wave 
background originating from the Penrose limit of NS-NS  
$AdS_3\times S^3\times R^4$ \cite{tseytlin}. 
The supergravity solution of a system of $N$ $D$-strings in such a 
background is given by:
\begin{eqnarray}
ds^2&=&f^{-{1\over2}}_1(2 dx^+dx^- - \mu^2{\sum_{i=1}^{4}}x_i^2(dx^+)^2)
+ f^{1\over2}_1{\sum_{a=1}^{8}}(dx^a)^2  \cr
& \cr
e^{2\phi}&=& f_1,~~~~H_{+12} = H_{+34}= 2\mu, \cr
& \cr
F_{+-a}&=&\partial_a f^{-1}_1,~~~~f_1 = 1 + {N g_s l^6_s\over r^6},
\label{d1-ads3}
\end{eqnarray}
with $f_1$ satisfying the Green function equation in the
8-dimensional transverse space.
We have explicitly verified that the solution presented in
(\ref{d1-ads3}) satisfy the type IIB field equations
(see e.g.\cite{duff,oz}). One notices that in this case,
constant NS-NS 3-forms along the PP-wave direction are
required precisely to cancel the $\mu$-dependent part of $R_{++}$
equation of motion. 

Starting from the $D$-string solution in eqn. (\ref{d1-ads3}), one 
can write down all the $Dp$-brane solutions ($p=1,..,5$)
in NS-NS PP-wave background by applying successive T-dualities 
along $x^5, ..,x^8$. As is known, this procedure also involves
smearing of the brane along these directions. For example, 
a $D3$-brane solution has a form: 
\begin{eqnarray}
ds^2&=&f^{-{1\over2}}_3(2 dx^+dx^- -\mu^2{\sum_{i=1}^{4}}x_i^2(dx^+)^2+
(dx_5)^2+(dx_6)^2)\cr
& \cr
&+& f^{1\over2}_3{\sum_{a=1..4,7,8}}(dx_a)^2  \cr
& \cr
H_{+12}&=&H_{+34}= 2\mu, \cr
& \cr
F_{+ - 5 6 a}&=& \partial_a {f_3}^{-1},~~~~e^{2\phi} = 1,
\label{d3-bkgrd}
\end{eqnarray}
with $f_3$ being the harmonic function in the transverse space of the
$D3$-brane.

Now we present the supergravity solution of intersecting 
($Dp-Dp^{\prime}$)-brane system in PP-wave background. These solutions are
described as `branes lying within branes'. In particular for $D1 - D5$ 
case, the solution is given by:
\begin{eqnarray}
ds^2&=&(f_1 f_5)^{-{1\over 2}}(2 dx^+dx^-
-\mu^2{\sum_{i=1}^{4}}x_i^2 (dx^+)^2) + \Big({f_1 \over f_5}\Big)^{1\over
  2}{\sum_{a=5}^{8}}(dx^a)^2 \cr
& \cr
&+&(f_1 f_5)^{1\over 2}{\sum_{i=1}^{4}}(dx_i)^2, \cr
& \cr
e^{2\phi}&=&f_1\over f_5, \cr
& \cr
H_{+ 1 2}&=&H_{+ 3 4} = 2\mu,\cr
& \cr
F_{+ - i}&=&\partial_{i}{f_1}^{-1},~~~~F_{m n p} = \epsilon_{m n p l}
\partial_{l}f_5,
\label{d1-d5}
\end{eqnarray}  
with $f_1$ and $f_5$ satisfying the Green function equations for $D1$ and 
$D5$-branes respectively. 
Eqn. (\ref{d1-d5}) provides one of the
main result of our paper and shows that, as in the flat space,
intersecting brane solutions are possible in the PP-wave background
as well. We have once again checked that the solution
presented above do satisfy type IIB field equations of motion.

One can now apply T-duality transformations to generate more intersecting 
brane solutions starting from the one given in eqn. (\ref{d1-d5})
\cite{myers,kamal2}. Note that the directions $x^5,..,x^8$ are
transverse to the $D$-string in eqn. (\ref{d1-d5}), 
whereas they lie along the longitudinal directions of $D5$. 
As a result, one can easily obtain solutions of the type 
$D2-D4$ as well as $D3-D3^{\prime}$ in this pp-wave
background. These solutions will give a  
PP-wave generalization of the intersecting solutions given in 
\cite{tseyt6}. We however skip the details of this analysis.

\subsect{Supersymmetry Analysis}

In this section we will present the supersymmetry of the solutions
described earlier in the previous subsection. 
The supersymmetry variation of dilatino and 
gravitino fields of type IIB supergravity in ten dimension, 
in string frame, is given by\cite{schwarz,fawad}:
\begin{eqnarray}
\delta \lambda_{\pm} &=& {1\over2}(\Gamma^{\mu}\partial_{\mu}\phi \mp
{1\over 12} \Gamma^{\mu \nu \rho}H_{\mu \nu \rho})\epsilon_{\pm} + {1\over
  2}e^{\phi}(\pm \Gamma^{M}F^{(1)}_{M} + {1\over 12} \Gamma^{\mu \nu
  \rho}F^{(3)}_{\mu \nu \rho})\epsilon_{\mp},
\label{dilatino}
\end{eqnarray}
\begin{eqnarray}
\delta {\Psi^{\pm}_{\mu}} &=& \Big[\partial_{\mu} + {1\over 4}(w_{\mu
  \hat a \hat b} \mp {1\over 2} H_{\mu \hat{a}
  \hat{b}})\Gamma^{\hat{a}\hat{b}}\Big]\epsilon_{\pm} \cr
& \cr
&+& {1\over 8}e^{\phi}\Big[\mp \Gamma^{\mu}F^{(1)}_{\mu} - {1\over 3!}
\Gamma^{\mu \nu \rho}F^{(3)}_{\mu \nu \rho} \mp {1\over 2.5!}
\Gamma^{\mu \nu \rho \alpha \beta}F^{(5)}_{\mu \nu \rho \alpha
  \beta}\Big]\Gamma_{\mu}\epsilon_{\mp},
\label{gravitino}
\end{eqnarray}
where we have used $(\mu, \nu ,\rho)$ to describe the ten
dimensional space-time indices, and hat's represent the corresponding
tangent space indices. $\Gamma^{\mu}$'s are the ten dimensional 
Dirac matrices.
Solving the above two equations for the solution describing a D-string
as given in eqn. (\ref{d1-ads3}), we get several conditions. First, the   
dilatino variation gives:
\begin{eqnarray}
\Gamma^{\hat{a}}\epsilon_{\pm} -
\Gamma^{\hat{+}\hat{-}\hat{a}}\epsilon_{\mp}=0,
\label{dilt-1}
\end{eqnarray}
\begin{eqnarray}
(\Gamma^{\hat{+}\hat{1}\hat{2}}+\Gamma^{\hat{+}\hat{3}\hat{4}})
\epsilon_{\mp}=0.
\label{dilt-2} 
\end{eqnarray}
In fact, both the conditions (\ref{dilt-1}) and (\ref{dilt-2}), are required
for satisfying dilatino variation condition.
Gravitino variation gives the following conditions on the spinors:
\begin{eqnarray}
\delta \psi_+^{\pm} &\equiv &\partial_{+}\epsilon_{\pm}\mp{\mu\over2}
f^{-{1\over2}}_1(\Gamma^{\hat{1}\hat{2}}+\Gamma^
{\hat{3}\hat{4}})\Gamma^{\hat{-}}\epsilon_{\pm}=0,\>\>\>
\delta \psi_-^{\pm} \equiv \partial_{-}\epsilon_{\pm}=0,\cr
& \cr 
\delta \psi_a^{\pm}&\equiv &\partial_{a}\epsilon_{\pm}
= - {1\over 8} {f_{1,a}\over f_1} \epsilon_{\pm},\>\>\>
\delta \psi_i^{\pm} \equiv \partial_{i}\epsilon_{\pm}=
- {1\over 8} {f_{1,i}\over f_1} \epsilon_{\pm}.
\label{1_i}
\end{eqnarray}
In writing the above set of equations, we have also 
imposed a necessary condition:
\begin{eqnarray}
\Gamma^{\hat +}\epsilon_{\pm} = 0,
\end{eqnarray} 
in addition to (\ref{dilt-1}). Further, by using
\begin{eqnarray}
(1 - \Gamma^{\hat 1\hat 2\hat 3\hat 4})\epsilon_{\pm} = 0,
\end{eqnarray}
all the supersymmetry conditions are solved by spinors:
$\epsilon_{\pm} = exp(-{1\over 8} ln f_1)\epsilon^0_{\pm}$, with
$\epsilon^0_{\pm}$ being a constant spinor.
$D$-string solution in eqn. (\ref{d1-ads3}) therefore preserves $1/8$ 
supersymmetry. All other $Dp$-branes ($p=1,..,5$), obtained by 
applying T-dualities as discussed above, will also preserve same 
amount of supersymmetry. 

Next, we will analyze the supersymmetry properties of the intersecting
branes. We will concentrate on the $(D1-D5)$-case explicitly.
The dilatino variation gives the following conditions on the spinors:
\begin{eqnarray}
\Gamma^{\hat i}\epsilon_{\pm} - \Gamma^{\hat +\hat -\hat
  i}\epsilon_{\mp} = 0,
\label{d1-d5-dila1}
\end{eqnarray}
\begin{eqnarray}
\Gamma^{\hat i}\epsilon_{\pm} + {1\over 3!}\epsilon_{\hat i\hat j\hat
  k\hat l}\Gamma^{\hat j\hat k\hat l}\epsilon_{\mp} = 0,
\label{d1-d5-dila2}
\end{eqnarray}
\begin{eqnarray}
(\Gamma^{\hat +\hat 1\hat 2} + \Gamma^{\hat + \hat 3 \hat
  4})\epsilon_{\mp} = 0.
\label{d1-d5-dila3}
\end{eqnarray}
One needs to impose all the three conditions, specified above for the
dilatino variation to vanish.
On the other hand, the gravitino variation gives:
\begin{eqnarray}
\delta\psi_+^{\pm}&\equiv &\partial_+ \epsilon_{\pm} \mp {\mu\over
  2}(f_1 f_5)^{-{1\over 2}}(\Gamma^{\hat 1\hat 2} + \Gamma^{\hat 3 \hat
  4})\Gamma^{\hat -}\epsilon_{\pm} = 0,\>\>\> 
\delta\psi_-^{\pm} \equiv \partial_- \epsilon_{\pm} = 0,\cr
& \cr
\delta\psi_i^{\pm}&\equiv &\partial_i \epsilon_{\pm} = 
 - {1\over 8} \left[{f_{1,a}\over f_1} \epsilon_{\pm} +
{f_{5,a}\over f_5}\right] \epsilon_{\pm} ,\>\>\>
\delta\psi_a^{\pm} \equiv \partial_a \epsilon_{\pm} = 0.
\end{eqnarray}
In writing down the above gravitino variations we have once again
made use of the projection $\Gamma^{\hat +}\epsilon_{\pm} = 0$.
The above set of equations can be solved by imposing: 
\begin{eqnarray}
(1 - \Gamma^{\hat 1\hat 2\hat 3\hat 4})\epsilon_{\pm} = 0,
\end{eqnarray} 
in addition to (\ref{d1-d5-dila1}) and the solution is given as:
$\epsilon_{\pm} = exp(-{1\over 8} ln (f_1f_5))\epsilon^0_{\pm}$.
One therefore has $1/8$
supersymmetry for the $(D1-D5)$ solution presented in eqn. (\ref{d1-d5}).

\sect{\bf $ p-p^{\prime}$ Branes in PP-Wave Background with Constant
  RR-Flux}
\subsect{Supergravity Solutions}

In this section, we will present the $Dp$ as well as
$(Dp-Dp^{\prime})$-branes in R-R PP-wave of $AdS_3\times
S^3\times R^4$. 
These backgrounds can be  obtained from the solutions given in 
the last section by applying
$S$ and $T$-duality transformations in several steps.
For example, from the $D3$-brane solution in NS-NS PP-wave background
(\ref{d3-bkgrd}), one gets a $D3$-brane in RR PP-wave
background under $S$-duality transformation. 
Now applying $T$-duality along the directions ($x^5, x^6$), 
we can generate a $D$-string solution. 
On the other hand, by applying $T$-duality along two transverse 
directions, $(x^7, x^8)$, of the $D3$-brane, 
one gets a $D5$-brane lying along $(x^+, x^-, x^5,..,x^8)$ directions.
Supergravity solution of a system of  
$N$ $D$-strings is then given explicitly by:
\begin{eqnarray}
ds^2&=&f^{-{1\over2}}_1(2 dx^+dx^- - \mu^2{\sum_{i=1}^{4}}x_i^2(dx^+)^2)
+ f^{1\over2}_1{\sum_{a=1}^{8}}(dx^a)^2  \cr
& \cr
e^{2\phi}&=& f_1,~~~~F_{+1256} = F_{+3456}= 2\mu, \cr
& \cr
F_{+-a}&=&\partial_a f^{-1}_1,~~~~f_1 = 1 + {N g_s l^6_s\over r^6},
\label{d1rr-ads3}
\end{eqnarray}
with $f_1$ satisfying the Green function in $8$-dimensional transverse
space. One notices that the solution has a constant $5$- form field
strength.

The supergravity solution of $D5$-brane is given by:
\begin{eqnarray}
ds^2&=&f^{-{1\over2}}_5(2 dx^+dx^- - \mu^2{\sum_{i=1}^{4}}x_i^2(dx^+)^2
 + {\sum_{a=5}^{8}}(dx^a)^2) + 
f^{1\over2}_5{\sum_{i=1}^{4}}(dx^i)^2  \cr
& \cr
e^{2\phi}&=& f^{-1}_5,~~~~F_{+1256} = F_{+3456} = F_{+1278} =
F_{+3478} = 2\mu, \cr
& \cr
F_{mnp}&=&\epsilon_{m n p q}\partial_q f_5,
~~~~f_5 = 1 + {N g_s l^2_s\over r^2},
\label{d5rr-ads3}
\end{eqnarray}
with $f_5$ satisfying the Green function in the transverse directions
$(x^1,...,x^4)$.
Now we will present the $(Dp-Dp^{\prime})$-brane solutions in RR
PP-wave background. In particular, to write down the supergravity
solution of a $(D1-D5)$ system, we made an ansatz which combines 
the $D$-string of eqn. (\ref{d1rr-ads3}) and $D5$-brane given 
in eqn. (\ref{d5rr-ads3}). The final configuration is as follows:
\begin{eqnarray}
ds^2&=&(f_1 f_5)^{-{1\over 2}}(2 dx^+dx^-
-\mu^2{\sum_{i=1}^{4}}x_i^2 (dx^+)^2) + \Big({f_1 \over f_5}\Big)^{1\over
  2}{\sum_{a=5}^{8}}(dx^a)^2 \cr
& \cr
&+&(f_1 f_5)^{1\over 2}{\sum_{i=1}^{4}}(dx_i)^2, \cr
& \cr
e^{2\phi}&=&f_1\over f_5, \cr
& \cr
F_{+ 1 2 5 6}&=&F_{+ 3 4 5 6} = F_{+ 1 2 7 8} = F_{+ 3 4 7 8} = 2\mu,\cr
& \cr
F_{+ - i}&=&\partial_{i}{f_1}^{-1},~~~~F_{m n p} = \epsilon_{m n p l}
\partial_{l}f_5,
\label{d1-d5-rr}
\end{eqnarray}  
with $f_1$ and $f_5$ being the Green function in the common transverse
space. One can check that the solution presented above do satisfy the
type IIB field equations. Once again, more $p-p^{\prime}$ branes
can be obtained from the $D1-D5$ solution in (\ref{d1-d5-rr}) by
applying $T$-dualities. 

One may also attempt to find a $(D1-D5)$ solution by 
taking a decoupling limit, followed by the Penrose scaling, 
of the solution presented 
in\cite{papa} in a similar way as the $D5$-brane solution in\cite{kumar}.
The starting solution along which one would take the Penrose limit
is as follows:
\begin{eqnarray}
ds^2&=&{1\over{({H^{\prime}_1}H^{\prime}_5)^{1/2}}}[{r^2\over {R^2_1}}(-dt^2
+ dx^2)] + \Big({H^{\prime}_1\over H^{\prime}_5}\Big)^{1/2}{R^2_1\over 
  r^2} dr^2\cr 
& \cr
&+&\Big({H^{\prime}_1\over H^{\prime}_5}\Big)^{1/2}(d\psi^2 + \sin^2\psi
d\Omega^2_2) + (H^{\prime}_1 H^{\prime}_5)^{1/2} (dy^2 + y^2
d\Omega^2_3),
\end{eqnarray}
where 
\begin{equation}
H_1 = 1 + {R_1^2\over x^2},~~~H_5 = 1 + {R_5^2\over x^2},~~~
H^{\prime}_1 = 1 + {{R^{\prime}}_1^2\over y^2},
~~~H^{\prime}_5 = 1 + {{R^{\prime}}_5^2\over y^2}.
\end{equation}  
One however notices that different terms in the metric above
come with different powers of $H^{\prime}_1$,
leading to difficulty in choosing a `null geodesic' to define
an appropriate Penrose limit and find brane solutions. 

\subsect{Supersymmetry Analysis}
Now, we will present the supersymmetry of the $Dp$ as well as
$Dp-Dp^{\prime}$ branes in R-R pp-wave background given in 
section-(2.2). First we will discuss the
supersymmetry of the $D$-string in
eqn. (\ref{d1rr-ads3}). The dilatino variation (\ref{dilatino}),
gives:
\begin{eqnarray}
\Gamma^{\hat a}\epsilon_{\pm} - \Gamma^{\hat +\hat -\hat a} 
\epsilon_{\mp} = 0.
\label{d1rr-dila}
\end{eqnarray}  
Gravitino variation gives the following conditions on the spinors:
\begin{eqnarray}
\delta \psi_+^{\pm} &\equiv &\partial_{+}\epsilon_{\pm}\mp{\mu\over8}
f^{-{1\over2}}_1\left((\Gamma^{\hat +\hat 1\hat 2\hat{5}\hat{6}}+\Gamma^
{\hat +\hat 3\hat 4\hat{5}\hat{6}}) +(\Gamma^{\hat +\hat 1\hat
  2\hat{7}\hat{8}}+\Gamma^{\hat +\hat 3\hat 4\hat{7}\hat{8}})\right)
\Gamma^{\hat{-}}\epsilon_{\pm}=0,\cr
& \cr
\delta \psi_-^{\pm} &\equiv& \partial_{-}\epsilon_{\pm}=0,\cr
& \cr 
\delta \psi_a^{\pm}&\equiv& \partial_{a}\epsilon_{\pm}
= - {1\over 8} {f_{1,i}\over f_1} \epsilon_{\pm},\>\>\>
\delta \psi_i^{\pm} \equiv \partial_{i}\epsilon_{\pm}
= - {1\over 8} {f_{1,a}\over f_1} \epsilon_{\pm},
\label{d1rr-susy}
\end{eqnarray}
where we have once again imposed a
necessary condition: $\Gamma^{\hat +}\epsilon_{\pm} = 0$, in addition
to (\ref{d1rr-dila}). Further, by using the condition:
\begin{eqnarray}
(1 - \Gamma^{\hat 1\hat 2\hat 3\hat 4}) = 0,
\end{eqnarray}
all the supersymmetry conditions are satisfied, thus 
preserving $1/8$ unbroken supersymmetry.

Next, we present the supersymmetry property of the 
$D1-D5$ solution written in eqn. (\ref{d1-d5-rr}).
The dilatino variation (\ref{dilatino}) gives the following 
conditions on spinors:
\begin{eqnarray}
\Gamma^{\hat i}\epsilon_{\pm} - \Gamma^{\hat +\hat -\hat
  i}\epsilon_{\mp} = 0,
\label{d1-d5rr-dila1}
\end{eqnarray}
\begin{eqnarray}
\Gamma^{\hat i}\epsilon_{\pm} + {1\over 3!}\epsilon_{\hat i\hat j\hat
  k\hat l}\Gamma^{\hat j\hat k\hat l}\epsilon_{\mp} = 0.
\label{d1-d5rr-dila2}
\end{eqnarray}  
On the other hand, the gravitino variation (\ref{gravitino}), gives the
following conditions:
\begin{eqnarray}
\delta\psi_+^{\pm}&\equiv &\partial_+ \epsilon_{\pm} \mp {\mu\over
  8}(f_1 f_5)^{-{1\over 2}}\left((\Gamma^{\hat +\hat 1\hat 2\hat 5\hat 6}
+ \Gamma^{\hat +\hat3 \hat 4 \hat 5\hat 6}) 
+(\Gamma^{\hat +\hat 1\hat 2\hat 7\hat 8}
+\Gamma^{\hat +\hat3 \hat 4 \hat 7\hat 8})\right) 
\Gamma^{\hat -}\epsilon_{\pm} = 0,\cr
& \cr 
\delta\psi_-^{\pm} &\equiv& \partial_- \epsilon_{\pm} = 0,\>\>\>
\delta\psi_a^{\pm} \equiv \partial_a \epsilon_{\pm} = 0,\cr
& \cr
\delta\psi_i^{\pm}&\equiv& \partial_i \epsilon_{\pm} = 
- {1\over 8} \left[{f_{1,i}\over f_1} + {f_{5,i}\over f_5}\right] 
\epsilon_{\pm},
\end{eqnarray}
where we have once again used a necessary condition: $\Gamma^{\hat
  +}\epsilon_{\pm} = 0$ along with the ones in 
(\ref{d1-d5rr-dila1}) and (\ref{d1-d5rr-dila2}).
The above set of equations can be solved 
by imposing further:
\begin{eqnarray}
(1 - \Gamma^{\hat 1\hat 2\hat 3\hat 4})\epsilon_{\pm} = 0. 
\end{eqnarray}
One therefore has $1/8$ supersymmetry for the $D1-D5$ system described
in eqn. (\ref{d1-d5-rr}) as well.

\sect{Worldsheet Construction of $p-p^{\prime}$ Branes}

\subsect{NS-NS PP-wave}

In this section, we will discuss the $(D1-D5)$-brane system, 
constructed earlier in the paper, from 
the point of view of first quantized string theory in Green-Schwarz
formalism, in light-cone gauge. In the present case, in flat
directions $x_{\alpha} (\alpha = 5,...,8)$, we have the Dirichlet boundary
condition at one end and Neumann boundary condition at 
the other end of the open string.  
Along $x_i$ $(i = 1,...,4)$ directions, one has the usual 
Dirichlet boundary condition. The relevant classical action to be
studied in our case (after imposing the light-cone gauge conditions
on fermions and bosons \cite{tseytlin} )
is as follows \cite{michi}:
\begin{eqnarray}
L=L_b + L_f,
\end{eqnarray}
where
\begin{eqnarray}
L_b&=&\partial_+ u\partial_- v - m^2 x^2_i + \partial_+ x_i\partial_-
x_i + \partial_+ x_{\alpha}\partial_- x_{\alpha}\cr
& \cr
&+& \mu \sum_{(i,j) = (1, 2), (3, 4)} x^i (\partial_+ u \partial_- x^j
- \partial_- u \partial_+ x^j ), 
\end{eqnarray}
\begin{eqnarray}
L_f = i S_R (\partial_+ - m M )S_R 
+ i S_L (\partial_-  + m M )S_L,
\end{eqnarray}
with
\begin{eqnarray}
m\equiv \alpha^{\prime}p^{u}\mu = 2\alpha^{\prime}p_v \mu, 
\end{eqnarray}
\begin{eqnarray}
M = -{1\over 2}(\gamma^{1 2} + \gamma^{3 4}).
\end{eqnarray}
Eight component real spinors $(S_L, S_R)$ have been obtained from 
16-component Majorana-Weyl spinors in the left and the right sector
after solving the light-cone gauge conditions.

The equations of motion and boundary conditions for bosons 
$x^i, x^{\alpha}$ in our case are as follows:
\begin{eqnarray}
\partial_+\partial_- x_{i_1} + m^2x_{i_1} - 
m \epsilon^{i_1 j_1} (\partial_- x^{j_1} - \partial_+ x^{j_1})
= 0,~~~~~~~
\partial_+\partial_-x_{\alpha} = 0,
\end{eqnarray}  
\begin{eqnarray}
\partial_{\sigma} x^{\alpha}{\Big|_{\sigma =0}} = \partial_{\tau}
x^{\alpha}{\Big|_{\sigma =\pi}} = 0,\;\;\;
x^i {\Big|_{\sigma =0,\pi}}=  constant. 
\end{eqnarray}
The solutions to the bosonic equations of motion, with the boundary
conditions specified above, 
is given by (defining $X^{\hat{1}} = {1\over \sqrt 2} (x^1 + i x^2)$ and  
$X^{\hat{2}} = {1\over \sqrt 2} (x^3 + i x^4)$) \cite{michi} : 
\begin{eqnarray}
X^{\alpha}(\sigma,\tau) = i \sum_{r\in (z + {1\over 2})}
{1\over r}\alpha_r^{\alpha} e^{-i r \tau}\cos r\sigma,\;\; 
(\alpha = 5,..,8),
\label{long-boson}
\end{eqnarray} 
\begin{eqnarray}
X^{\hat{i}}(\sigma,\tau) = e^{-2 i m\sigma}\left[x_0 +(x_1 e^{2 i
  m\sigma}-x_0){\sigma\over \pi} + i \sum_{n\ne 0}{1\over n}\alpha_n^i
e^{-i n \tau}\sin n\sigma\right].
\label{trans-boson}
\end{eqnarray}

To consider the equations of motion of fermions, we note that 
the matrix $M$ evidently breaks the $SO(8)$ symmetry further, 
and thereby splits the fermions in the $8\rightarrow 4 + 4$ way:
$S_L\rightarrow(\tilde S_L,\hat S_L), S_R\rightarrow(\tilde S_R,\hat
S_R)$:
\begin{eqnarray}
\gamma^{1234} \pmatrix{\tilde{S}_{L,R} \cr \hat{S}_{L,R}} = 
\pmatrix{- \tilde{S}_{L,R} \cr \hat{S}_{L,R}}. \label{ga1234}
\end{eqnarray}
In this connection, one also introduces $4\times 4$ matrices 
$\Lambda$ and $\Sigma$:
\begin{eqnarray}
\gamma^{12}\pmatrix{\tilde{S}_{L,R} \cr \hat{S}_{L,R}} = 
-\pmatrix{\Lambda \tilde{S}_{L,R} \cr\Sigma \hat{S}_{L,R}}, \label{gamma12}
\end{eqnarray}
with $\Lambda^2 = \Sigma ^2 = -1$. 
$\Lambda$ and $\Sigma$ in the above equation are $4\times 4$ 
antisymmetric matrices with eigenvalues $\pm i$. Using these 
notations, one has:
\begin{eqnarray}
M \pmatrix{\tilde{S}_{L,R} \cr \hat{S}_{L,R}} 
= \pmatrix{\Lambda \tilde{S}_{L,R} \cr  0}.
\end{eqnarray}
The equations of motion written in terms of $(\tilde{S}_L,{\hat{S}}_L)$ and 
$(\tilde{S}_R,{\hat{S}}_R)$ are then of the form: 
\begin{equation}
\partial_{+}(e^{2 m\tau} \tilde{S}_{R}) =0, \>\>\>\>
\partial_{-}(e^{- 2m\tau }\tilde{S}_{L}) =0,
\label{long-ferm}
\end{equation}
\begin{equation}
\partial_{+}\hat{S}_{R} =0, \>\>\>\>
\partial_{-}\hat{S}_{L} =0.
\label{trans-ferm}
\end{equation}
Now we will write down the boundary conditions for the fermions
in the mixed sector. As the equations of motion and the 
boundary condition for the components $\hat{S}_{L,R}$ are identical to the
ones in flat space, we only concentrate on finding explicit solution for 
$\tilde{S}_{L,R}$ below.
Following\cite{tseytlin,akumar,michi}, one can write down
the boundary conditions for the fermions as:
\begin{eqnarray}
\tilde{S_L}\big|_{\sigma=0} = - \tilde{S_R}\big|_{\sigma=0},\>\>\>\>
\end{eqnarray}
\begin{eqnarray}
\tilde{S_L}\big|_{\sigma=\pi} = \tilde{S_R}\big|_{\sigma=\pi},\>\>\>\>
\end{eqnarray}
\begin{eqnarray}
\hat{S_L}\big|_{\sigma=0,\pi} = \hat{S_R}\big|_{\sigma=0,\pi}.
\end{eqnarray}
The solution for $\tilde{S}_{L,R}$ equations of motion
(\ref{long-ferm}), with the above boundary condition,
can be read from \cite{michi}, and has the following form:
\begin{eqnarray}
\tilde{S}_L = - e^{-2 m\sigma\Lambda}\sum_{r\in (z + {1\over 2})}s_r
e^{-i r(\tau + \sigma)}, 
\end{eqnarray}
\begin{eqnarray}
\tilde{S}_R =  e^{-2 m\sigma\Lambda}\sum_{r\in (z + {1\over 2})}s_r
e^{-i r(\tau - \sigma)}. 
\end{eqnarray}

The canonical quantization conditions as well as the worldsheet 
hamiltonian for the $D1-D5$ system discussed above can also be
written in a straightforward manner following the procedure in 
\cite{michi}. We skip these details. 

\subsect{R-R PP-wave}
Now, we present the worldsheet analysis of the $(D1-D5)$-system
discussed earlier in section-(2.2). This can be done by 
realizing that the PP-wave background for these solutions 
is given by a $T$-dual configuration of the ones presented
in \cite{tseytlin,akumar}. More explicitly, in the worldsheet
action in the present case:
\begin{eqnarray}
L=L_B + L_F,
\end{eqnarray}
where
\begin{eqnarray}
L_B = \partial_+ u\partial_- v - m^2 x^2_i + \partial_+ x_i\partial_-
x_i + \partial_+ x_{\alpha}\partial_- x_{\alpha}, 
\end{eqnarray}
\begin{eqnarray}
L_F = i {S}_R \partial_+ {S}_R 
+ iS_L \partial_+ S_L - 2im {S}_L 
M {S}_R,
\end{eqnarray}
with
\begin{eqnarray}
m\equiv \alpha^{\prime}p^{u}\mu = 2\alpha^{\prime}p_v \mu, 
\end{eqnarray}
\begin{eqnarray}
M = -{1\over 2}(\gamma^{1 2} + \gamma^{3 4}) \gamma^{5 6},\>\>\>
\label{Matrix}
\end{eqnarray}
the terms involving fermions are easily seen to be related to the
ones in \cite{tseytlin,akumar} through $T$-dualities along 
$x^5$ and $x^6$. This in fact leads to the relation: 
\begin{eqnarray}
{S^{\prime}}_R = \gamma^{5 6}S_R,
\label{t-fermion}
\end{eqnarray}
and reproduces the original action in \cite{tseytlin}. The mode 
expansion for fermions as well as canonical quantization conditions
can therefore be also written down in a straightforward manner. 
We end this section by pointing out that, since the $D$-brane solutions
found in this paper are preserving less than $1/2$ supersymmetry,
some of the restrictions on the brane directions, imposed using 
zero mode considerations\cite{dabh} do not directly apply above. 

\sect{Branes in `Little String Theory' Background}

\subsect{Supergravity Backgrounds} 
In this section, we discuss the branes in the Penrose limit of 
`little string theory'(LST). PP-waves of non-local theories have been
discussed recently in the literature\cite{rangamani,sakai,kumar1}.
Among them, `little string theory' arises on the world volume of $NS5$-brane
when a decoupling limit, $g_s\rightarrow 0$ with fixed
$\alpha^{\prime}$, is taken\cite{berkooz,seiberg}. To construct our
solution, we start with the NS5-brane solution given by the metric and
dilaton:
\begin{eqnarray}
ds^2 &=& -dt^2 + dy^{2}_5 + H(r)(dr^2 + r^2d\Omega^{2}_3),\cr
& \cr
e^{2\Phi} &=& g^{2}_s H(r),
\end{eqnarray}  
with $H(r) = 1 + {Nl^{2}_s \over r^2}$. 
The near horizon limit of the above solution is 
the linear dilaton geometry, which in the string frame is given by,
\begin{eqnarray}
ds^2= Nl^{2}_s\big(-d\tilde{t}^2 + \cos^2\theta d\psi^{2} + d\theta^{2}
+ \sin^2\theta d\phi^{2} +{dr^2 \over r^2}\big) + dy^{2}_5,
\label{little-1}
\end{eqnarray}
with $t = \sqrt{N}l_s \tilde{t}$. PP-wave background of LST is then
found by applying Penrose limit to (\ref{little-1}). We however
consider the case after applying S-duality on eqn. (\ref{little-1}). 
Applying S-duality transformation and then taking Penrose limit
(as described in\cite{rangamani}), the background solution is given by: 
\begin{eqnarray}
ds^2&=& -4dx^{+}dx^{-} - \mu^{2}\vec{z}^{2}{dx^+}^2 + (d\vec{z})^2 +
dx^2 + dy^{2}_5 \cr
& \cr
e^{2\phi}&=&Const, \cr
& \cr
F_{+ 1 2}&=&C \mu,
\end{eqnarray}  
where $F_{+ 1 2}$ is the 3-form field strength in the $\vec{z}$-plane.
Now we proceed to analyze the existence and stability
of branes in this background. The supergravity solution for a system of 
D5-branes in this background is given by:
\begin{eqnarray}
ds^2&=&f^{-{1\over2}}\Big(-4 dx^+dx^- -\mu^2\vec{z}^2(dx^+)^2+
{\sum_{i =1}^{2}}(dz_i)^2+{\sum_{p = 1}^{2}}(dy_p)^2\Big)\cr
& \cr
&+& f^{1\over2}{\sum_{a=1}^{4}}(dx_a)^2,  \cr
& \cr
e^{2\Phi}&=&f^{-1},~~~~~F_{+ 1 2} = 2\mu, \cr
& \cr
F_{m n p}&=&\epsilon_{m n p r}\partial_{r}f,~~~ f = 1 + {Ng_s l^2_s\over 
  r^2}.
\label{litt-d5}
\end{eqnarray}
One notices that the background has only one constant 3-form field
strength $(F_{+ 1 2})$. We have once again verified that the solution 
presented above satisfies type IIB field equations.

One can then write down (by applying S-duality on (\ref{litt-d5}))
the NS5-brane in a PP-wave background of the
`little string theory'\cite{rangamani} as:
\begin{eqnarray}
ds^2&=&-4 dx^+dx^- -\mu^2\vec{z}^2(dx^+)^2+
{\sum_{i =1}^{2}}(dz_i)^2+{\sum_{p = 1}^{2}}(dy_p)^2 
+f{\sum_{a=1}^{4}}(dx_a)^2 \cr
& \cr
e^{2\Phi}&=&f,~~~~~H_{+ 1 2} = 2\mu, \cr
& \cr
H_{m n p}&=&\epsilon_{m n p r}\partial_{r}f,
\end{eqnarray}
with $H$'s being the NS- sector 3-form field strengths. 
\subsect{Supersymmetry Analysis}
We will now consider the dilatino and gravitino
variation of the solution presented in eqn. (\ref{litt-d5})
to study the supersymmetry properties.
The dilatino variation gives equations:
\begin{eqnarray}
\Gamma^{\hat a}\epsilon_{\pm} + {1\over 3!}\epsilon_{\hat a \hat b
  \hat c \hat d}\Gamma^{\hat b \hat c \hat d}\epsilon_{\mp} = 0,
\label{lstdila1}
\end{eqnarray}
\begin{eqnarray}
\Gamma^{\hat + \hat1 \hat2}\epsilon_{\mp} =0.
\label{lstdila2}
\end{eqnarray}
The gravitino variation leads to the equations:
\begin{eqnarray}
\delta\Psi^{\pm}_+ \equiv \partial_{+}\epsilon_{\pm} + 
{\mu^2 z_{\hat i} \over 2}\Gamma^{\hat + \hat i}\epsilon_{\pm} 
+ {1\over 16}\mu^2 \vec{z}^2 {f,\hat a\over f^{3\over 2}}
\Gamma^{\hat +\hat a}\epsilon_{\pm} - 
{\mu \over 4}\Gamma^{\hat +\hat1\hat2}\Gamma^{-}\epsilon_{\mp}=0,
\end{eqnarray}
\begin{eqnarray}
\delta\Psi^{\pm}_{-}\equiv \partial_{-}\epsilon_{\pm}=0,
\end{eqnarray} 
\begin{eqnarray}
\delta\Psi^{\pm}_{i}\equiv\partial_{i}\epsilon_{\pm} - {\mu\over
  4}\Gamma^{\hat{+}\hat1\hat2}\delta_{i\hat{i}}\Gamma^{\hat{i}}
\epsilon_{\mp}=0.
\end{eqnarray}
\begin{eqnarray}
\delta\Psi^{\pm}_{p}\equiv\partial_{p}\epsilon_{\pm} - 
{\mu\over 4}\Gamma^{\hat + \hat 1\hat 2}
\delta_{p\hat{p}}\Gamma^{\hat{p}}\epsilon_{\mp}=0.
\end{eqnarray}
\begin{eqnarray}
\delta\Psi^{\pm}_{a}\equiv\partial_{a}\epsilon_{\pm}  + 
{1\over 8} {f,a\over f}\epsilon_{\pm} - {\mu\over
4}f^{1\over 2}\Gamma^{\hat{+}\hat1\hat2}\delta_{a\hat{a}}\Gamma^{\hat{a}}
\epsilon_{\mp}=0,
\end{eqnarray}
In writing these set of equations we have used the
condition (\ref{lstdila1}). Imposing the condition $\Gamma^{\hat
  +}\epsilon =0$, we further reduce them to:
\begin{eqnarray}
\partial_{+}\epsilon_{\pm} - {\mu \over 4}\Gamma^{\hat
  +\hat1\hat2}\Gamma^{\hat-}\epsilon_{\mp} = 0,
\label{d5-soln}
\end{eqnarray}
\begin{eqnarray}
\partial_{-}\epsilon_{\pm}=0,~~~~\partial_{i}\epsilon_{\pm} = 0,~~~~~
\partial_{p}\epsilon_{\pm} = 0,~~~\partial_{a}\epsilon_{\pm} 
= - {1\over 8} {f,a\over f}\epsilon_{\pm}. \label{last-eqn.}
\end{eqnarray}
Since eqns. (\ref{d5-soln}) and (\ref{last-eqn.})
are integrable ones, hence in this case we get 
$1/4$ supersymmetry. It will also be nice to give a worldsheet
construction for such $D$-branes. 

\sect {$D$-Branes in PP-wave Spacetime with Nonconstant NS-NS Flux}

\subsect{Supergravity Solutions}

In this section we present classical solutions of $Dp$ as well as
$Dp-Dp^\prime$ -branes with nonconstant $NS-NS$ three form flux transverse 
to brane worldvolume.
We start by writing down the supergravity solutions of $D$-string in
pp-wave background  with nonconstant NS-NS 
three form flux. The metric, dilaton and the field strengths are given by:   
\begin{eqnarray}
ds^2&=&f^{-{\half}}_1\left(2 dx^+dx^- + K(x_i)(dx^+)^2\right) \cr
& \cr
&+& f^{\half}_1{\sum^{8}_{m=1}}(dx^m)^2,\>\>\>
(i = 1,...,4),\cr
& \cr
H & = & \p_1 b_2(x_i)\>dx^+\wedge dx^1\wedge dx^2 + \p_3 b_4(x_i)\>dx^+\wedge
dx^3\wedge dx^4,\cr 
& \cr
e^{2\phi}&=& f_1, ~~~~F_{+-n}=\p_n f^{-1}_1,
\label{d1}
\end{eqnarray}
\noindent
with $b(x_i)$ and $K(x_i)$ satisfying the equations $\Delta b
(x_i) = 0$ and 
$ \Delta K (x_i) = -(\p_i b_j)^2$ respectively and 
$f_1 = 1 + {Q_1\over r^{6}}$ 
is the harmonic function in the transverse
space. We have checked that the above solution satisfies type IIB field 
equations. For constant three form flux this
solution reduces to that of ref \cite{bkp}. All other
$Dp$-brane ($p=2,...,5$) solutions can be found out by applying
$T$-duality along $x^5,...,x^8$ directions. For example:     
the classical solution for a system of $D3$-brane in such a background
is given by:
\begin{eqnarray}
ds^2&=&f^{-{\half}}_3\left(2 dx^+dx^- + K(x_i)(dx^+)^2 + (dx^{7})^2 
+ (dx^{8})^2\right) \cr
&\cr
&+& f^{\half}_3{\sum^{6}_{m=1}}(dx^m)^2,\>\>\> (i = 1,...,4),\cr
& \cr
H & = & \p_1 b_2(x_i)\>dx^+\wedge dx^1\wedge dx^2 + \p_3 b_4(x_i)\> dx^+\wedge
dx^3\wedge dx^4, \cr
& \cr
e^{2\phi}&=& 1, ~~~~F_{+-78n}=\p_n f^{-1}_3,
\label{d3}
\end{eqnarray}
\noindent
with $b(x_i)$ and $K(x_i)$ satisfying the equations $\Delta b
(x_i) = 0$ and 
$ \Delta K (x_i) = -(\p_i b_j)^2$ respectively and 
$f_3 = 1 + {Q_3\over r^{4}}$ is  
the harmonic function satisfying the Green function equation 
in the transverse space. 

Now we present classical solution of $D1-D5$ system as an example of
$p-p^{\prime}$ bound state in these background. The supergravity 
solution for a such system is given by:  
\begin{eqnarray}
ds^2&=&(f_1 f_5)^{-{1\over 2}}(2 dx^+dx^-
+ K(x_i) (dx^+)^2) + \Big({f_1 \over f_5}\Big)^{1\over
  2}{\sum_{m=5}^{8}}(dx^m)^2 \cr
& \cr
&+&(f_1 f_5)^{1\over 2}{\sum_{i=1}^{4}}(dx^i)^2, \cr
& \cr
e^{2\phi} &=& {f_1\over f_5} ,\cr
& \cr
H & = & \p_1 b_2(x_i)\>dx^+\wedge dx^1\wedge dx^2 + \p_3 b_4(x_i)\> dx^+\wedge
dx^3\wedge dx^4, \cr
& \cr
F_{+ - i}&=&\partial_{i}{f_1}^{-1},~~~~F_{i j k} = \epsilon_{i j k l}
\partial_{l}f_5,
\label{d1-d5nc}
\end{eqnarray} 
with $b(y_j)$ and $K(x_i)$ satisfying the equations $\Delta b
(x_i) = 0$ and 
$ \Delta K (x_i) = -(\p_i b_j)^2$ respectively and 
$f_1 = 1 + {Q_1\over
r^{2}}$ and  $f_5 = 1 + {Q_5\over
r^{2}}$ are the harmonic functions of $D1$ and $D5$-branes in common
transverse space. One can check that the above ansatz do satisfy type
IIB field equations.

\subsect{Supersymmetry Analysis}

In this section we present the supersymmetry of the solutions
described earlier in section (2).
Solving the above two equations for D-string solution as given 
in (\ref{d1}), we get several conditions on the spinors. 
First, the dilatino variation given in eqn. (\ref{dilatino})gives:
\begin{eqnarray}
\frac{f_{1,\hat{n}}}{f_1}
\Gamma^{\hat{n}}\epsilon_{\pm}
\mp f^{\frac{1}{4}}_1(\p_{\hat{i}} b_{\hat{j}})\Gamma^{\hat{+}\hat{i}\hat{j}} 
\epsilon_{\pm} -
\frac{f_{1,\hat{n}}}{f_1}
\Gamma^{\hat{+}\hat{-}\hat{n}}\epsilon_{\mp} = 0,
\label{dilt-1nc}
\end{eqnarray}
Gravitino variation (\ref{gravitino}) gives the following 
conditions on the spinors:
\begin{eqnarray}
\label{g+}
\delta \psi_+^{\pm} &\equiv &\left(\partial_{+}+{1\over 4}
{\p_{\hat n} (K f^{-{1\over 4}}_{1})} \Gamma^{\hat +\hat n}   
\mp{1\over4}(\p_{\hat i} b_{\hat j})(\Gamma^{\hat{i}\hat{j}})\right)
\epsilon_{\pm} \cr
& \cr
&-& {1\over 8} \Gamma^{\hat +\hat -
\hat n}{K f_{1,\hat n} \over f^{5/4}_1}\Gamma^{\hat
+}\epsilon_{\mp}=0 \\
\label{g-}
\delta \psi^{\pm}_{-} &\equiv& \partial_{-}\epsilon_{\pm}=0 \\
\label{gn}
\delta \psi_n^{\pm}&\equiv& \left(\partial_{n}
 - {1\over 8} {f_{1,n}\over f_1}\right) \epsilon_{\pm}, ~~~~
(n=5,...,8)\\
\label{gi}
\delta \psi_i^{\pm} &\equiv& \left(\partial_{i}\mp\frac{\delta_{i \hat i}}{4}f^\half_1(\p_{\hat i}b_{\hat j})
\Gamma^{\hat{+}\hat{j}} - {1\over 8} {f_{1,i}\over f_1}\right) 
\epsilon_{\pm},~~~~(i=1,...,4).
\end{eqnarray}
In writing the above equations we have used the brane supersymmetry
condition:
\begin{eqnarray}
\Gamma^{\hat +\hat -}\epsilon_{\pm} = \epsilon_{\mp}.
\label{susyb}
\end{eqnarray}
Taking derivative of the eqn. (\ref{gi}) with respect to $\p_{\hat k}$
and subtracting the derivative of $\p_{\hat k}$ equation with respect to
$\p_{\hat i}$, we get
\be
(\p_{\hat k}\p_{\hat i} b_{\hat j})\Gamma^{\hat{+}\hat{j}}\epsilon_{\pm}=0,
\ee
which can be satisfied for nonconstant $\p_{\hat i}b_{\hat j}$ only if $\Gamma^{\hat 
  +}\epsilon_{\pm} = 0$.
Using $\Gamma^{\hat 
  +}\epsilon_{\pm} = 0$. and brane supersymmetry condition
(\ref{susyb}), the dilatino variation (\ref{dilt-1nc}) is satisfied. 
Using  $\Gamma^{\hat 
  +}\epsilon_{\pm} = 0$, the supersymmetry conditions (\ref{gn}) 
and (\ref{gi}) are solved by spinors: $\epsilon_{\pm} = exp(-{1\over
  8} ln f_1)\epsilon^0_{\pm}$, with $\epsilon^0_{\pm}$ being a function of 
 $x^+$ only. Since $\epsilon^0_{\pm}$ is independent of $x^i$ and
 $(\p_ib_j)$ is a function of $x^i$ only, the gravitino variation
 gives the following conditions to have nontrivial solutions:
\be
(\p_{\hat i}b_{\hat j})(\Gamma^{\hat{i}\hat{j}})\epsilon^0_{\pm}=0
\label{gamma}
\ee
and
\be
\p_+\epsilon^0_\pm = 0.
\ee
The condition $\Gamma^{\hat +}\epsilon_{\pm} = 0$ breaks sixteen 
supersymmetries. The number of remaining supersymmetries depend 
on the existence of constant 
$\epsilon^0_\pm $ solutions of the equation (\ref{gamma}). For the particular
case when $H_{+12} = H_{+34}$, the equation (\ref{gamma}) gives the condition:
\be
(1 - \Gamma^{\hat 1\hat 2\hat 3\hat 4})\epsilon^0_{\pm} = 0.
\label{hat}
\ee 
Therefore in this case, the $D$-string solution
(\ref{d1}), preserves  $1/8$ supersymmetry. 
Similarly, one can show that the $D3$-brane solution 
(\ref{d3}) also preserves $1/8$ supersymmetry. 

Next, we will analyze the supersymmetry properties of  $(D1-D5)$
system that is described in eqn. (\ref{d1-d5nc}) of the previous section.

The dilatino variation gives the following conditions on the spinors:
\begin{eqnarray}
\delta\lambda_{\pm}&\equiv&
{f_{1,\hat i}\over f_1}
\left(\Gamma^{\hat i}\epsilon_{\pm} - \Gamma^{\hat +\hat -\hat
i}\epsilon_{\mp}\right)\mp 
(f_1 f_5)^{\frac{1}{4}}(\p_{\hat i}b_{\hat j})\Gamma^{\hat
  {+}\hat{i}\hat{j}}\epsilon_\pm \cr
& \cr
&-&{f_{5,\hat i}\over f_5}
\left(\Gamma^{\hat i}\epsilon_{\pm} + {1\over 3!}\epsilon_{\hat i \hat 
    j \hat k \hat l}\Gamma^{\hat j\hat k\hat l}\epsilon_{\mp}\right)=0
\label{d1-d5-dila}
\end{eqnarray}

On the other hand, the gravitino variation gives:
\begin{eqnarray}
\delta\psi_+^{\pm}&\equiv &\partial_+ \epsilon_{\pm}+ 
{1\over 4}
{\p_{\hat n} (K (f_1 f_5)^{-{1\over 4}})} \Gamma^{\hat +\hat n}\mp {1\over
  4}
(\p_{\hat i}b_{\hat j})\Gamma^{\hat{i}\hat{j}}\epsilon_\pm \cr
&\cr
&-& {1\over 8}(f_1 f_5)^{-\frac{1}{4}}\left( \Gamma^{\hat +\hat - \hat n}{K
    f_{1,\hat n} \over f_1} + \Gamma^{\hat m_1...\hat m_3}\epsilon_{\hat
    {m_1},...,\hat{m_3},\hat n} {K f_{5,\hat n} \over f_5}\right)\Gamma^{\hat+}\epsilon_{\mp}=0 \\
\label{d1-d5-g+}
\delta\psi_-^{\pm} &\equiv& \partial_- \epsilon_{\pm} = 0,~~~~~
\delta\psi_m^{\pm} \equiv \partial_m \epsilon_{\pm} = 0,\\
\label{d1-d5-g-}
\delta\psi_i^{\pm}&\equiv &\partial_i \epsilon_{\pm} 
 - {\delta_{i\hat i}\over 4}(f_1 f_5)^{\frac{1}{2}} (\p_{\hat i}b_{\hat j})\Gamma^{\hat{+}\hat{j}}+ \frac{1}{8}\left[{f_{1,i}\over f_1} +
{f_{5,i}\over f_5}\right] \epsilon_{\pm} = 0.\>\>\>
\label{d1-d5-gi}
\end{eqnarray}
In writing down the above gravitino variations we have once again
made use of the brane conditions:
\be
\Gamma^{\hat i}\epsilon_{\pm} - \Gamma^{\hat +\hat -\hat
i}\epsilon_{\mp} = 0,
\label{d1-d5-susy}
\ee
and 
\be
\Gamma^{\hat i}\epsilon_{\pm} + {1\over 3!}\epsilon_{\hat i \hat 
    j \hat k \hat l}
\Gamma^{\hat j\hat k\hat l}\epsilon_{\mp} = 0.
\label{d1-d5-susy1}
\ee

Taking derivative of the eqn. (\ref{d1-d5-gi}) with respect to $\p_{\hat k}$
and subtracting the derivative of $\p_{\hat k}$ equation with respect to
$\p_{\hat i}$, we get
\be
(\p_{\hat k}\p_{\hat i} b_{\hat j})\Gamma^{\hat{+}\hat{j}}\epsilon_{\pm}=0,
\ee
which can be satisfied for nonconstant $\p_{\hat i}b_{\hat j}$ only if $\Gamma^{\hat 
  +}\epsilon_{\pm} = 0$.

Using $\Gamma^{\hat 
  +}\epsilon_{\pm} = 0$ and brane supersymmetry conditions
(\ref{d1-d5-susy}) and(\ref{d1-d5-susy1}) , the dilatino condition 
(\ref{d1-d5-dila}) is satisfied. Using  $\Gamma^{\hat 
 +}\epsilon_{\pm} = 0$, the supersymmetry condition
(\ref{d1-d5-gi}) is solved by spinors: $\epsilon_{\pm} = exp(-{1\over 8} ln
 (f_1f_5))\epsilon^0_{\pm}$, with $\epsilon^0_{\pm}$ being a function of 
 $x^+$ only. Since $\epsilon^0_{\pm}$ is independent of $x^i$ and
 $(\p_ib_j)$ is a function of $x^i$ only, the gravitino variation
 gives the following conditions to have nontrivial solutions:
\be
(\p_{\hat i}b_{\hat j})(\Gamma^{\hat{i}\hat{j}})\epsilon^0_{\pm}=0
\label{d1-d5-susy2}
\ee
and
\be
\p_+\epsilon^0_\pm = 0.
\ee

Once again, the number of supersymmetries depend on the existence of
solutions of equation (\ref{d1-d5-susy2}). For the particular case when 
$H_{+12} = H_{+34}$, the $D1-D5$ bound state solution (\ref{d1-d5nc})
also preserves $1/8$ supersymmetry.

\sect{Summary}

In this chapter, we have presented several supersymmetric 
$Dp$ and $Dp-Dp^{\prime}$-brane configurations in 
pp-wave background with constant and nonconstant three form flux.
and analyzed their supersymmetry properties. 
We have presented $(D1-D5)$-brane construction from the
point of view of massive Green-Schwarz formalism in the light cone 
gauge in NS-NS and R-R pp-wave of $AdS_3\times S^3\times R^4$ with 
constant three form flux. It will
be interesting to study the gauge theory duals of the branes
presented in this paper by using operators such as `defects'. 
One could possibly also look at the black hole physics using the 
$(D1-D5)$ system presented here in an attempt to understand their properties.
All the solutions presented here with the constant NS-NS and RR flux
are shown to preserve $1/8$ supersymmetry, whereas the $D$-branes
in the background of little string theory preserve $1/4$
supersymmetry. The supernumerary
supersymmetry is absent for the background with nonconstant flux.
$D$-brane solutions with nonconstant $RR$ flux can be found out 
by applying $S$-duality transformation on the solutions presented
here, which will be generalization of those given in \cite{kamal2}. 
The $D$-brane solutions with nonconstant flux have 
the interpretation of $D$-branes in nonsupersymmetric sigma model 
of \cite{russo}. It is also desirable to analyze them from the worldvolume 
point of view following the procedure of \cite{hikida}.

\chapter{NONCOMMUTATIVE $N=2$ STRINGS}
\pagenumbering{arabic}
\markright{Chapter 4. NONCOMMUTATIVE $N=2$ STRINGS}
\vspace{-1cm}
\sect{Geometry of $N=2$ Strings: A brief review}
It is well known fact that properties of string theory are directly
attributable to the properties of underlying 2d conformal field
theory. More structure on the worldsheet therefore gives rise to
more interesting string theory. For example: Bosonic strings have no 
worldsheet supersymmetry $(N=0)$. Introduction of one local
supersymmetry on the worldsheet $(N=1)$, gives rise to the far more
interesting superstring theory that lives in $(9,1)$ spacetime
dimensions. Then it is natural to extend the idea of introducing more
local supersymmetry on the worldsheet and look for string theories in
diverse dimensions. It turns out that $N=2$ superconformal
supersymmetry on the worldsheet gives rise to an interesting string theory,
known as {\it $N=2$ strings} \cite{bruce,ademo}. These are the
kind of string theories which provide a consistent quantum
theory of self-dual gravity in four dimensions. It was realized by
Ooguri and Vafa \cite{OV,OV1} that $N=2$ string lives in 2-complex 
spacetime dimensions and have only one physical particle, 
a massless scalar in its spectrum.
Then the next question to ask is what is the signature of the space-time ?
Unfortunately the signature can't be $(3,1)$ because it breaks
$N=2$ worldsheet supersymmetry, which requires the underlying manifold 
to be a complex K\"{a}hler manifold. The correct structure is
(2,2) which is consistent with keeping the complex structure.
This theory lives intrinsically in a K\"{a}hler two-complex dimensional
space-time, with Lorentz group $U(1,1)= SL(2,R)\bigotimes U(1)$. They 
also have the interesting property that all $n$-point amplitudes $(n\ge4)$
vanish, thereby showing the topological or integrable nature of the
theory. But the 3-point function does not vanish and therefore 
the theory has a nontrivial S matrix. The massless scalar of the
theory was argued as the K\"{a}hler potential of space-time. It
satisfies a quadratic equation of motion, the Plebanski equation
\cite{pleb}, which is the condition that the spacetime has 
a self-dual Riemann tensor. Thus the closed $N=2$ string can be 
thought of as a theory of self-dual gravity on (2,2) dimensional 
spacetime. 

For flat space-time with (2,2) signature metric, $ds^2 = dx^1 d{\bar
  x}^1 - dx^2 d{\bar x}^2$, the action for the $N=2$ nonlinear sigma
model on the worldsheet is $S_0 = \int d^2 z d^2 \theta d^2\bar \theta
K_0(X, \bar X)$, where $X^i (i= 1,2)$ is an $N=2$ chiral superfield,
\bea
X^i (Z, \bar Z; \theta, \bar \theta) = x^i(Z, \bar Z) + 
\psi^i_R (Z, \bar Z) \theta + \psi^i_L (Z, \bar Z) \bar \theta 
+ F^i(Z,\bar Z)\theta \bar \theta
\eea
and $K_0$ is a K\"{a}hler potential for the flat space-time, 
$K_0 = X^1 \bar X^{1} - X^2 \bar X^{2}$. 
Now to find out the meaning of
the massless modes, one deforms the action while preserving the $N=2$
supersymmetry on the worldsheet. That can only be done by changing the
K\"{a}hler potential $K_0\rightarrow K = K_0 + \phi$. This
deformation of the  K\"{a}hler potential corresponds to the massless
scalar. The spectrum of the closed string consists of a single 
massless scalar. Metric degrees of
freedom is also present here and encoded in terms of massless scalar
field, by the formula $g_{i \bar j} = \partial_i \bar \partial_j K$.   
\subsect{Vertex Operators and Interactions}

The superspace-vertex operator for
emitting a closed-string scalar of momentum $k$ is\footnote{Here the
`dots' indicate contractions over the two complex coordinates.}
\bea
V_c = {\kappa \over \pi} e^{i(k.\bar X + \bar k.X )},
\eea
where $X$ is an $N=2$ chiral superfield.  Thus, the vertex at
$\theta= \bar \theta =0$ is simply the momentum insertion
\bea
V_c^0 |_{\theta=\bar \theta = 0} = {\kappa\over \pi}
e^{i(k.\bar x + \bar k.x)},
\label{ver0}
\eea
while the vertex integrated over the fermionic coordinates is
\bea
V_c^{int} = {\kappa \over \pi} 
( i k.~ \partial \bar x -i \bar k.~\partial x 
- k. ~\bar \psi_R \bar k.~ \psi_R )
( i k.~\bar \partial \bar x -i \bar k.~\bar \partial x - k. ~\bar \psi_L
~\bar k.~\psi_L ) e^{i(k.\bar x + \bar k.x )}
\label{vint}
\eea

To obtain the three-point function, one needs to fix three bosonic
coordinates and two $\theta$'s (and $\bar \theta$'s).
Thus one calculates the correlation function of
one $V_c^{int}$ and two $V_c^0$'s, at fixed positions.  The result
is:
\bea
& & A_{ccc} = \langle V_c|_{\theta=0}(0) V_c^{\rm int}(1)
V_c(\infty)|_{\theta=0}\rangle
\nonumber\\
& & = \kappa c^2_{12},
\eea
where 
\be
c_{12} \equiv (k_1.\bar k_2 - \bar k_1 .k_2) 
\ee
is the invariant product of momenta in 2-complex dimensions.
Note that $c_{ij}$ is antisymmetric with respect to its two
indices, and is additive in the sense that $c_{i, j} + c_{i, k} =
c_{i, j+k}$. Using momentum conservation, one sees that $A_{ccc}$ is 
completely antisymmetric. The three point amplitude
of the 2-complex dimensional closed string therefore implies a truly
nontrivial S-matrix element, and not just some unphysical vertex.   

The four-point function can be calculated similarly, and the result
is given by:
\bea 
A_{cccc} &=& \int_0^1\langle V_c|_{\theta=0}(0) V^{\rm
  int}_c(x) V^{\rm int}_c(1)V_c|_{\theta=0}(\infty)\rangle \cr
& \cr
&=& {\kappa^2\over \pi} F^2 {{\Gamma(1-s/2)\Gamma(1-t/2)\Gamma(1-u/2)} 
  \over {\Gamma(s/2)\Gamma(t/2)\Gamma(u/2)}}.
\label{cl-f}
\eea
Ooguri and Vafa noted however that, in the scattering of massless
particle in $(2,2)$ dimensions, there is an identity:
\be
F \equiv 1 - {c_{12}c_{34}\over su} - {c_{23}c_{41}\over tu} = 0.
\ee
Therefore the four point function (\ref{cl-f}) vanishes onshell. Ooguri
and Vafa proposed that the vanishing of amplitude comes from some kind of
topological nature of the theory. They have conjectured that all
higher order amplitudes should also vanish.

One sees that the local three-point function and vanishing of the
four point function can be deduced from the following action:
\be
{\cal L}_c = \int d^4 x \big({1\over 2} \partial^{i} \phi \bar
\partial_{i} \phi
+ {2\kappa \over 3}\phi \partial  \bar\partial \phi \wedge \partial 
\bar \partial \phi \big), 
\ee 
The equation 
of motion  derived from this action is the Plebanski equation,
the equation for self-dual gravity written in terms of K\"ahler
potential in (2,2) space. So the
$N=2$ closed string describes self-dual gravity. The next question to ask
of course is that whether the open string and heterotic strings
describe self dual Yang-Mills (SDYM) ? Soon it was
realized for open $N=2$ strings by N. Marcus\cite{mar}. As in the 
closed string case, the spectrum of the open string also consists 
of a massless scalar. However, as usual we would like to append the 
chan paton factor to the string amplitudes. The superspace vertex 
operator to emit these scalars is equal to that of the closed string 
case and is given by:
\be
V_o = g e^{i(k.\bar X + \bar k.X)}.
\ee    
Again after integrating out the fermionic coordinates one obtains the
integrated vertex operator as:
\be
V^{int}_o = \int d^2 \theta V_0 = {g\over 2} (ik.~\partial_{\sigma}\bar 
x - i\bar k.~\partial_{\sigma}x - 4k.~\bar{\psi}~\bar{k}.~\psi)
e^{i(k.\bar x + \bar k.x)}.
\ee 
The three-point open-string amplitude is found by calculating the
correlation function of one $V^{int}_o$ and two $V_o$ at fixed
positions. The amplitude is given by:
\be
A_{ooo} = \langle V_o|_{\theta=0}(0) V_o^{\rm int}(1)
V_o(\infty)|_{\theta=0}\rangle = gc_{12}f^{abc}.
\label{3-fu}
\ee
The amplitude is totally antisymmetric with respect to the three
scalars without the insertion of the group theory factor $f^{abc}$, 
so one needs to insert the group theory factor in order to
prevent the vanishing of the amplitude. At this stage, $f^{abc}$ is an
unspecified totally antisymmetric tensor. 
Similarly, the four point amplitude depends upon the cyclic
ordering of the vertices on the boundary, which we choose to place
at $0$, $x$, $1$ and $\infty$, with $x$ integrated between $0$ and
$1$. Two of the vertices are of $V^{int}_o$ and two are $V_o$. The
four point amplitude is given by:
\bea
A_{oooo} &\equiv& \int_0^1\langle V_0|_{\theta=0}(0) V^{\rm
  int}_o(x) V^{\rm int}_o(1)V_o|_{\theta=0}(\infty)\rangle \cr
& \cr
&=& {g^2 \over 4} F{\Gamma(1-2s)\Gamma(1-2t)\over\Gamma(2u)}, 
\eea  
and as in the case of the closed string this amplitude also vanishes 
due to the $F$ factor. Because of the vanishing of this amplitude, the
usual unitarity constraint on the gauge group factors does not apply
here. However, one can still find a constraint on them by 
demanding the consistency between the string theory four-point function
and the same calculated form the effective field theory. The three
point amplitude of eqn \ref{3-fu} can be obtained from the Lagrangian:
\be
{\cal{L}}_{3o} = \int d^4 x\Big( {1\over 2}\partial^i\varphi^a\bar
\partial_{\bar i}\varphi^a - i {g\over 3}f^{abc}\varphi^a\partial^i
\varphi^b\bar\partial_{\bar i}\varphi^c \Big)
\label{ac-tp}
\ee
The field theory four-point function can be calculated by sewing or by 
gluing together two three point amplitudes and by adding some local
or contact four point vertex $V_{4o}$. This gives:
\be
A_{ooooFT} = - g^2 \Big\{ Af^{abx}f^{xcd} + B f^{bcx}f^{xda} +
  A f^{cax} f^{xbd} \Big\} - V_{4o},
\ee  
where 
\bea
A = {c_{12}c_{34}\over s}\>\>\>B = {c_{23}c_{41}\over t}\>\>\>
C = {c_{31}c_{24}\over u}.
\eea
The kinematic relation $F=0$ and its permutation imply that 
\bea
B = u - A \>\>\> and \>\>\> C = t + A.
\eea
Now the four point amplitude looks like
\bea
A_{ooooFT} &=& - g^2 \Big\{ A \Big(f^{abx}f^{xcd} -  f^{acx}f^{xbd} -
  f^{bcx}f^{xda}\Big) \cr
& \cr
&+& u f^{bcx}f^{xda} + t f^{cax}f^{xbd}\Big\} - V_{4o}.
\label{4-fi}
\eea
Since $V_{4o}$ must be a local vertex, the factor multiplying $A$ must 
vanish. This factor is the Jacobi identity. Thus the $f^{abc}$'s are 
the structure constants of the semi simple group times a product of
an arbitrary number of $U(1)$ factors. 
Thus the resulting quartic interaction determined by eqn. \ref{4-fi}
is now:
\be
{\cal L}_{4o}  = \int d^4 x \Big( - {g^2 \over 6} f^{adx} f
^{xbc}\partial^i \varphi ^a\varphi ^b\bar \partial _{\bar
  i}\varphi^c\varphi^d\Big).
\label{ac-fp}
\ee 

The computation of the three and four point function for the mixed
sector has been carried out in\cite{mar}. We, however, present 
the results here.
The amplitude of two open and one closed stings is given by:
\bea
A_{ooc} &\equiv& \int_{-\infty}^\infty db \langle V_o|_{\theta=0}(b)
V^{\rm int}_c(z=x+iy)
V_o|_{\theta=0}(\tau\rightarrow\infty)\rangle \cr
& \cr
&=& {\kappa \over \pi} \delta^{ab} c^2_{12} \int_{-\infty}^\infty db
{y\over {(x-b)^2 + y^2}}\cr
&\cr
&=&\kappa \delta^{ab}c^2_{12}
\label{mix-3}
\eea
This vertex is same as the self interaction of the gravitational
vertex $A_{ccc}$, showing the universality in the couplings of various 
fields to gravity.
Now one proceeds to find out the four point amplitude with one closed
and three open strings. The simplest gauge fixing here would be to fix
the three open string vertices to be $0$, $1$ and $\infty$, and to fix
the $\theta$'s of two vertices at  $0$ and $\infty$. One needs then to 
integrate the closed vertex over the UHP. As the integrand is symmetric 
under the exchange of $z$ and $\bar z$, the integral in the UHP can be 
transformed by the integral over the full complex plane and the
result is given by:
\bea
A_{oooc}&=&\int\int_{\rm UHP} dz d{\bar z}
\langle V_o|_{\theta=0}(0)V^{\rm int}_o(1)V_o|_{\theta=0}(\infty) 
V^{\rm int}_c(z,{\bar z})\rangle \cr
&=& {i\over 2} \kappa g f^{a b c} F {{\Gamma(s)\Gamma(t)\Gamma(u)}\over
{\Gamma(-s)\Gamma(-t)\Gamma(-u)}} (c_{12} t + c_{23} s),
\eea
which again vanishes because of the $F$ factor. Since the open string
carries a group index, the amplitude with one open string and
arbitrary closed ones should vanish. The amplitude of two open and two
closed strings should also vanish again because of the presence of
$F$ factor. In the field theory action which reproduces the above
three and four point amplitudes, there are mixed quartic terms in the
action: 
\be
{\cal L}_{oooc} = \int \Big( -{4\over 3} i g \kappa f^{abc} \partial
\bar\partial \phi \wedge \varphi^a \partial \varphi^b \bar \partial
\varphi^c\Big).
\ee
This finishes a brief
review of string theory with $N=2$ superconformal symmetry on the
worldsheet. In the next section, we will analyze the 
$N=2$ string theory in the presence of constant antisymmetric 
tensor background.

\sect{$N=2$ Strings in Constant NS-NS Background}

\noindent
In this section,  we study $N=2$ strings in constant 
NS-NS antisymmetric tensor ($B$) background\cite{kamal3},
in view of interesting developments  in noncommutative string 
theory\cite{DOUG,SW,SOLITONS,OMNCSOL}. In this regard, 
we have also been motivated by the fact that 
noncommutative $N=2$ strings are expected to have interesting
implications in possible generalizations of 
M(atrix) and F theories\cite{ketov} to include noncommutativity. 
It is known that antisymmetric tensor backgrounds can be incorporated 
in $N=2$ superspace formalism using chiral and twisted-chiral 
superfields\cite{HULL}. In this manner, one has an $N=2$ worldsheet 
supersymmetry without having a K\"{a}hler metric \cite{HULL}. 
One now obtains a noncommutative complex manifold as the target space geometry.

There are two main highlights 
of the $N=2$ noncommutative field theory obtained here. 
The first is a nontrivial constraint, satisfied by the noncommutative
theory, originating from the requirement 
of the absence of poles in the 4-point amplitude in the 
open sector of the $N=2$ strings for nonzero $B$. 
The second is in the mixed sector. Here the noncommutative field
theory involves a generalized *-product\cite{GAROUSI,MEHEN}, and
$B$ explicitly appears with the open string
metric in two (left/right) linear combinations for contracting 
the target space indices of the open-string fields. We finally 
analyze a 4-point mixed sector amplitude in the extreme noncommutative limit. 
Nontriviality in the computation of the tree-level string amplitudes 
in the mixed sector, stems from the fact that due to
the absence of $z\rightarrow{\bar z}$-symmetry in the presence of
$B$, one can not use the  generalized Koba-Nielson integrals of 
\cite{OV,mar,KAW} which are relevant when the domain of integration is
the full complex plane.

One can consider $N=2$ string action (in presence of
$B$), written in an $N=1$ superspace notation in \cite{HULL}:
\begin{equation}
\label{eq:1}
S=\int d^2x \int d\theta_L d\theta_R\biggl[g_{IJ}D^\alpha X^I
D_\alpha X^J
+B_{IJ}D^\alpha X^I(\gamma^5D)_\alpha X^J\biggr],
\end{equation}
where the superspace field $X^I$ ($I\equiv1, 2,
{\bar 1}, {\bar 2}$) represents $X, Y, {\bar X}, {\bar Y}$, and  
$\alpha\equiv L,R$. The closed string metric and the 
antisymmetric background fields are denoted by 
$g_{IJ}$ and $B_{IJ}$ respectively. We would like to point out
that the action specified above is not $N=2$ supersymmetric without
adding boundary terms. The equations motion for the bosons and 
fermions can be derived from the above action. However, for 
deriving the boundary conditions for fermions, one needs to
add $B$-dependent boundary terms in the above action to restore 
supersymmetry. The field equations, boundary conditions and canonical 
commutation relations, including those for fermions derived in $N=2$ case 
turn out to be similar to the ones written in 
\cite{CH,SW}. Since the vacuum energy of the bosonic and 
fermionic oscillators remains same as for $B=0$\cite{SW}, 
the spectrum of the theory once again consists of a scalar 
($\varphi$) in the open and ($\phi$) closed string sector. 
The closed-\cite{OV} and open-string\cite{mar} vertex operators 
are given by:
\begin{eqnarray}
V_o|_{\theta=0} &=& e^{i(k\cdot{\bar x}+{\bar k}\cdot x)},\cr
& \cr 
V_c|_{\theta = {\bar\theta}=0} &=& e^{i(k\cdot{\bar x}+{\bar
  k}\cdot x)}, \cr
& \cr
V_c^{\rm int} &=&
\biggl(ik\cdot\partial{\bar x}-i{\bar k}\partial x
-k\cdot{\bar\psi}_R{\bar k}\cdot\psi_R\biggr)
\biggl(ik\cdot{\bar\partial}{\bar x}-i{\bar k}\cdot{\bar\partial} x
-k\cdot{\bar\psi}_L{\bar k}\cdot\psi_L\biggr)\times \cr
& \cr
&\times& e^{i(k\cdot {\bar x}+{\bar k}\cdot x)}, \cr
& \cr
V_o^{\rm int} &=& \biggl(ik\cdot\partial_\tau{\bar x}
-i{\bar  k}\cdot{\partial}_\tau x
-k\cdot({\bar\psi}_L+{\bar\psi}_R){\bar
  k}\cdot(\psi_L+\psi_R)\biggr)
e^{i(k\cdot{\bar x}+{\bar k}\cdot x)},
\label{eq:8}
\end{eqnarray}
where $V^{\rm int}_{c,o}$ are the closed- and open-string vertex
operators that have been integrated w.r.t. their fermionic
supercoordinates. Following \cite{ademo}, $\theta_L$ is set equal
to $\theta_R$ for $V^{\rm int}_o$. Also, the bosonic component $x^i,\ 
x^{\bar i}$ denote $(x,y)$ 
and $({\bar x}, {\bar y})$  which originate from 
(anti-)chiral and (anti-)twisted chiral fields of $N=2$. 

The two-point function for both bosons and fermions appearing in
(\ref{eq:8}) can be written 
together using the superspace 2-point function in $N=1$ notation
of \cite{ITY}
(with $\alpha^\prime={1\over{2\pi}}$):
\begin{eqnarray}
\label{eq:6}
& & 
\langle X^i(Z_1,{\bar Z}_1) X^{\bar j}(Z_2,{\bar Z}_2)\rangle
=-g^{i{\bar j}}
ln[(z_1-z_2-\theta_1^L\theta_2^L)({\bar z}_1-{\bar
  z}_2-\theta_1^R\theta_2^R)]\nonumber\\
& & +(g^{i{\bar j}}-2G^{i{\bar j}})ln[(z_1-{\bar
  z}_2-\theta_1^L\theta_2^R)
({\bar z}_1-z_2-\theta_1^R\theta_2^L)]\nonumber\\
& & -2\Theta^{i{\bar j}}
ln\biggl[{z_1-{\bar z}_2-\theta_1^L\theta_2^R\over{\bar z}_1
-z_2-\theta_1^R\theta_2^L}\biggr].\nonumber\\
& & 
\end{eqnarray}
The indices $i, {\bar j}$ run over $1,2$ and ${\bar 1}, {\bar 2}$
respectively. The open string metric $G^{i{\bar j}}$ and the
noncommutativity parameter $\Theta^{i{\bar j}}$ can be expressed in 
terms of $g^{i{\bar j}}$ and $B^{i{\bar j}}$ as in \cite{SW}. For our case, 
$g^{i{\bar j}}$ denotes the flat closed string metric and 
$B^{i{\bar j}}$, constant antisymmetric background of `magnetic' type. 
Now, using the above results, we calculate various string amplitudes
in the open and mixed sectors. In the closed sector the results of 
\cite{OV,mar} are still valid as closed strings have no boundary,
and hence are insensitive to the addition of boundary terms to the
world-sheet action. In the open- and mixed-string sectors, 
the super-M\"{o}bius transformations allow two complex  
fermionic supercoordinates to be set to zero, and three real bosonic
coordinates to be fixed to any arbitrary value.

\subsect{Open Sector:}

The 3-point function using obvious notations is given by:
\begin{eqnarray}
\label{eq:9}
& & A_{ooo}(B\neq0)=\langle V_o|_{\theta=0}(0) V_o^{\rm int}(1)
V_o(\infty)|_{\theta=0}\rangle
\nonumber\\
& & = e^{{i\over2}({\bar k_1}\Theta k_2-{\bar k}_2\Theta
  k_1)}A_{ooo}(B=0),
\end{eqnarray}
where
$A_{ooo}(B=0)=c_{12}\equiv k_1 G^{-1}{\bar k}_2 -k_2G^{-1}{\bar
  k}_1$. 
Now, as in \cite{mar}, one has to
impose Bose symmetry on $A_{ooo}$ in 
(\ref{eq:9}). Unlike \cite{mar}, for the noncommutative case,
one can have an ``isoscalar'' as well as an ``isovector'' component
of the amplitude:
\begin{equation}
\label{eq:isoscvec1}
A_{ooo}=A_{ooo}^{\rm S}+A_{ooo}^{\rm AS\ abc},
\end{equation}
where $A_{ooo}^{\rm S}$ is the isoscalar part of the 
amplitude that is symmetric under the interchange of the
momentum labels of particles 1 and 2, and
$A_{ooo}^{\rm AS\ abc}$
is the isovector part of the amplitude that is
antisymmetric under the interchange of the momentum and group 
labels separately, but is symmetric under the simultaneous interchange
of both types of labels. Denoting
$K_{12}\equiv{1\over2}({\bar k_1}\Theta k_2-{\bar k}_2\Theta k_1)$
one sees that 
\begin{equation}
\label{eq:isoscvec2}
A_{ooo}^{\rm S}=c_{12}\sin(K_{12}),\>\>\>\>
A_{ooo}^{\rm AS\ abc}=c_{12}\cos(K_{12})f^{abc}.
\end{equation}
Similarly, the 4-point function is given by:
\begin{eqnarray}
\label{eq:10}
& & A_{oooo}(B\neq0)=\int_0^1\langle V_0|_{\theta=0}(0) V^{\rm
  int}_o(x) V^{\rm int}_o(1)V_o|_{\theta=0}(\infty)\rangle\nonumber\\
& & = e^{i(K_{12}+K_{23}+K_{13})}A_{oooo}(B=0),
\end{eqnarray}
where $A_{oooo}(B=0)=F{\Gamma(1-2s)\Gamma(1-2t)\over\Gamma(2u)}$ as in 
\cite{mar}. The $\Theta$-dependent phase factor in
(\ref{eq:10}) matches with the phase factor in equation (2.11) of \cite{SW}.
The null kinematic factor $F$ is that of \cite{OV}
with the difference that the  open-string metric is used for 
contracting momenta in $s, t$, $u$ as well as $c_{ab}$'s.

Moreover, since in the purely open string
sector, the $\Theta$ dependence of the amplitude enters via a phase
factor, one sees that the above result for the
noncommutative 3- and 4-point functions readily generalizes to the
noncommutative $n$-point function, implying that, like the claim for
the commutative $N=2$ theory, all $n$-point functions with $n\geq4$
also vanish. Hence, the noncommutative $N=2$ theory is ``topological'' 
in the closed- and open-string sectors.

We now analyze in some detail the implications of the above
modifications to the field theory of open string scalars.
Using (\ref{eq:isoscvec1}) and (\ref{eq:isoscvec2}),  one can
evaluate the field theory (FT) amplitude 
$A_{oooo\ FT}$ which consists of contributions from 
two 3-point functions ($A_{ooo\ FT}$'s) as well as a contact vertex 
$V_{oooo\ FT}$ (whose form is determined from the requirement that 
$A_{oooo\ FT}$ like $A_{oooo}$ of string theory, vanishes). 
One can verify that for $A_{ooo\ FT}$ corresponding to 
$A^S_{ooo}$ in (\ref{eq:isoscvec2}), there are no poles in 
$A_{oooo\ FT}$. This can be seen by adding contributions to the 
4-point amplitudes from $s$, $t$ and $u$-channels as:
\bea
\label{eq:Ampl}
A^S_{oooo FT} &=& A \sin(K_{12}) \sin(K_{34}) + B \sin(K_{23})
\sin(K_{41})\cr
\cr
&+& C \sin(K_{31}) \sin (K_{24}),
\eea
where 
\begin{equation}
\label{eq:ABCdefs}
A={c_{12}c_{34}\over s},\ B={c_{23}c_{41}\over t}=u-A,\
C={c_{31}c_{24}\over u}=t+A. 
\end{equation}
Then by eliminating $k_4$ using momentum conservation, it is noticed
that pole part of the amplitude above cancels for $A_{ooo}^{S}$ in 
equation (\ref{eq:Ampl}). In other words, $sin(K_{ab})$ in
 $A^S_{oooo FT}$ acts as a structure constant. To generalize this
 result further, one can consider a more general 3-point function 
\begin{equation}
\label{eq:oood}
A^{S\ abc}_{ooo}=c_{12}\sin(K_{12})d^{abc},
\end{equation}
 with $d^{abc}$ being
symmetric structure constants. Vanishing of poles in $A^S_{oooo\ FT}$ 
then implies a strong condition on $d^{abc}$'s
leading to multiple copies of abelian noncommutative FT's 
mentioned in (\ref{eq:openaction}) below.

For $A_{ooo}^{AS}$, on the other hand, we get
a constraint on $f^{abc}$:
\begin{eqnarray}
\label{eq:ASconstr}
& & \cos(K_{12})\sin(K_{31})\sin(K_{32})f^{abx}f^{xcd}
-\cos(K_{23})\sin(K_{21})\sin(K_{31})f^{bcx}f^{xda}\nonumber\\
& & +\cos(K_{31})\sin(K_{21})\sin(K_{23}) f^{cax}f^{xbd}=0.
\end{eqnarray}
One sees that the above constraint can not be satisfied by any 
classical group. Perhaps it may be satisfied for some quantum 
group. One now observes that 
$U(N)$ gauge groups can be obtained from 3-point string vertex,
eqn.(\ref{eq:9}), by considering mixed (isoscalar-isovector)
vertices.
In particular, for $U(2)$, after imposing Bose
symmetry on two of the external legs in eqn.(\ref{eq:9}), the 
corresponding isoscalar-isovector field theory vertex is given as: 
\begin{equation}
A^{a b}_{Mixed} = c_{12}\sin(K_{12})\delta^{ab}.
\end{equation}
Then it can once again be shown that the poles in the isovector 4-point 
amplitude, obtained by sewing together two 3-point vertices with 
isoscalar and isovector internal states, cancel. Higher rank groups 
can also be incorporated by including 3-point vertex in 
eqn.(\ref{eq:oood}) (See for example \cite{lech}). 

We now write down the FT corresponding to $A^S_{ooo\ FT}$ mentioned 
before, as well as the contact vertex appearing in equation (\ref{eq:Ampl}),
after using equation (\ref{eq:ABCdefs}) whose explicit form is
\begin{equation}
\label{eq:contactV}
V^{int}_{oooo} = u \sin(K_{23}) \sin(K_{41}) + t \sin(K_{31})
\sin(K_{24}).
\end{equation}
One then obtains 
the field theory action corresponding to $A^S_{ooo}$ and
$V^{int}_{oooo}$ up to terms quartic in $\varphi$: 
\begin{eqnarray}
\label{eq:openaction}
& & 
{\cal L}_{FT}=G^{{\bar i}j}
\biggl[{1\over2}{\bar\partial}_{\bar i}\varphi * \partial_j\varphi
+{i\over3}[{\bar\partial}_{\bar i}\varphi,\partial_j\varphi]_*
*\varphi-{1\over12}{\bar\partial}_{\bar i}\varphi*
[[\partial_j\varphi,\varphi]_*,\varphi]_*\biggr],\nonumber\\
& & 
\end{eqnarray}
where $[\xi,\eta]_*\equiv \xi*\eta-\eta*\xi$. The $*$ product is defined
(in momentum space) as:
\begin{equation}
\label{eq:Moydef}
e^{ik_1\cdot x}*e^{ik_2\cdot x}=e^{iK_{12}}e^{i(k_1+k_2)\cdot x}.
\end{equation}
with $k\cdot x \equiv k\cdot\bar{x} + \bar{k}\cdot x$. 
A generalization of the above noncommutative abelian field theory action
to $U(N)$ case is straightforward. 
The Moyal deformation of self dual Yang Mills (and gravity) were also
considered in \cite{PLEBA}. We now study the mixed sector of 
the noncommutative $N$=2 theory.

\subsect{Mixed sector}

(a) $A_{ooc}$: We now  show, in $N=2$ context, 
that a mixed  amplitude with 
two open and one closed string vertices generates a field
theory with a generalized *-product\cite{GAROUSI,MEHEN} 
(the $\Theta=0$ limit of which reduces to the result of
\cite{mar}). 

Following
\cite{mar}, for the purpose of setting the limits of
integration, it is convenient to fix the bosonic coordinate of one
of the two $V_o$'s (to 0) and that of $V_c$ to $z=x+iy$. Then 
\begin{eqnarray}
\label{eq:11}
& & A_{ooc}=\int_{-\infty}^\infty db \langle V_o|_{\theta=0}(b)
V^{\rm int}_o(z=x+iy)
V_o|_{\theta=0}(\tau\rightarrow\infty)\rangle\nonumber\\
& & = e^{{i\over2}({\bar k}_b\Theta k_\tau-{\bar k}_\tau\Theta k_b)}\times
\nonumber\\ 
& & \times {y\over{4\pi^2}} \int_{-\infty}^\infty db \biggl(c_{b\tau}^2-
(k_\tau\Theta{\bar k}_b+k_b\Theta {\bar k}_\tau)^2\biggr)
{e^{-{1\over{2\pi}}(k_b\Theta{\bar k}_\tau-k_\tau
\Theta{\bar  k}_b)ln
\biggl({b-(x-iy)\over{b-(x+iy)}}\biggr)}\over{([b-x]^2 +
y^2)}}\nonumber\\
& & = 
{e^{{i\over2}({\bar k}_b\Theta k_\tau-{\bar k}_\tau\Theta k_b)}\over{4\pi^2}}
\biggl(c_{b\tau}^2-
(k_\tau\Theta{\bar k}_b+k_b\Theta {\bar k}_\tau)^2\biggr)
\pi e^{-i{(k_b\Theta{\bar k}_\tau-k_\tau
\Theta{\bar  k}_b)\over2}}{sin\biggl(
{k_b\Theta{\bar k}_\tau-k_\tau\Theta{\bar k}_b\over2}
\biggr)\over{(k_b\Theta{\bar k}_\tau-k_\tau\Theta{\bar k}_b)}}\nonumber\\
& & ={\pi\over2} c^L_{b\tau}c^R_{b\tau}{sin(K_{b\tau})\over{K_{b\tau}}},
\end{eqnarray}
where 
$c^{L,R}_{ab}\equiv{1\over{2\pi}}\biggl(k_aG^{-1}{\bar k}_b-
k_bG^{-1}{\bar k}_a \mp k_a\Theta{\bar k}_b
\pm k_b\Theta{\bar k}_a\biggr)$, the upper and lower signs
corresponding to $L$ and $R$ respectively. The $b$ integral above was
done using Mathematica and is also given in appendix A of \cite{GARMYRS}.
Before commenting on the topological nature of this amplitude, we
now  write down the corresponding interaction term in the 
FT action which is
given in terms of a generalized *-product\cite{GAROUSI,MEHEN}: 
\begin{equation}
\label{eq:mixedaction}
{\cal L}_{ooc} = \phi
(\partial_i\partial_{j}
\varphi
*^\prime
{\bar\partial}_{\bar i}{\bar\partial}_{{\bar j}}\varphi
-\partial_i{\bar\partial}_{{\bar i}}
\varphi *^\prime
\partial_{j}
{\bar\partial}_{{\bar j}}\varphi)
(G^{-1}-\Theta)^{i{\bar j}}(G^{-1}+\Theta)^{j{\bar i}},
\end{equation}
where we have made use of 
\begin{equation}
\label{eq:Pr*def}
e^{ik_1\cdot x} *^\prime e^{ik_2\cdot x}=
{\sin(K_{12})\over{K_{12}}}e^{i(k_1+k_2)\cdot x}.
\end{equation}
in arriving at (\ref{eq:mixedaction}) from (\ref{eq:11}). 

Now, to interpret
$A_{ooc}$ as a topological theory in the sense of \cite{OV}, one now
sees that by expanding ${\sin(K_{12})\over{K_{12}}}$
in a power series in $\Theta$, an infinite number of
($\Theta$ or equivalently $B$-dependent) terms are generated 
at the 3-point level in the mixed sector. As the
radius of convergence of the
${\sin x\over x}$ expansion is infinite, this implies that {\it after
expansion},
$A_{ooc}$ can be interpreted as an infinite series of
local interactions between the closed and open string scalars. 

(b) $A_{oooc}$:
Like
\cite{mar}, we set the bosonic coordinates of the three $V_o$'s
at $0, 1, \infty$ and the fermionic coordinates of the first and third
$V_o$'s to zero. We will hence require to integrate over 
the bosonic coordinates of $V_c^{\rm int}(z,{\bar z})$. Defining
$t^{L,R}\equiv{1\over{2\pi}} 
\biggl( k_1G^{-1}{\bar k}_4+k_4G^{-1}{\bar k}_1
\mp k_1\Theta{\bar k}_4\mp k_4\Theta{\bar k}_1\biggr)$, 
$u^{L,R}
\equiv{1\over{2\pi}} \biggl( k_1G^{-1}{\bar k}_3+k_3G^{-1}{\bar k}_1
\mp k_1\Theta{\bar k}_3\mp k_3\Theta{\bar k}_1\biggr)$,
one gets:
\begin{eqnarray}
\label{eq:13}
& & A_{oooc}=\int\int_{\rm UHP} dz d{\bar z}
\langle V_o|_{\theta=0}(0)V^{\rm int}_o(1)V_o|_{\theta=0}(\infty) 
V^{\rm int}_c(z,{\bar z})\rangle\nonumber\\
& & \sim \int\int_{\rm UHP} dz d{\bar z}
z^{t^L}{\bar z}^{t^R}(1-z)^{u^L}(1-{\bar z})^{u^R}\nonumber\\
& & \times
\Biggl(\biggl[-{c^L_{14}\over z}+{c^L_{24}\over(1-z)}\biggr]\biggl[
-{c^R_{14}\over {\bar z}}+{c^R_{24}\over(1-{\bar
    z})}\biggr]\biggl[-2c_{12}
+{c^L_{24}\over(1-z)}+{c^R_{24}\over(1-{\bar z})}\biggr]\nonumber\\
& & +\biggl[-{c^L_{14}\over z}+{c^L_{24}\over(1-z)}\biggr]\biggl[{-u^R
+(u^R)^2-(c^R_{24})^2\over(1-{\bar z})^2}\biggr] \nonumber\\
& &  +\biggl[-{c^R_{14}\over {\bar z}}+{c^R_{24}\over(1-{\bar z})}
\biggr]\biggl[{-u^L
+(u^L)^2-(c^L_{24})^2
\over(1-z)^2}\biggr]\Biggr),
\end{eqnarray}
which gives the $A_{oooc}$ amplitude of \cite{mar} for $\Theta=0$.
However, because 
of the lack of $z\rightarrow{\bar z}$ symmetry in the presence 
of $B$, unlike \cite{mar}, one can not enlarge the domain of
integration from the upper half complex plane to the entire complex
plane. The evaluation of the above integral and the vanishing of
4-point amplitude in $B \rightarrow\infty$ limit has been discussed 
in\cite{kamal3}. 

\sect{Summary}

In this chapter, we have discussed various tree level amplitudes of  
string theory with $N=2$ superconformal symmetry on the worldsheet
with and without the presence of constant antisymmetric tensor field.

In section (4.1), we have discussed $N=2$ open and closed string theories 
in $(2,2)$ dimensional spacetime. These theories
contain a single massless scalar in their spectrum.
The closed string has the interpretation of 
self-dual gravity. The open string, on the otherhand, has the interpretation 
of the self-dual Yang-Mills in the K\"ahler background of the closed
sector. But the resulting spacetime is no longer self-dual. Because
of the topological nature of the theory, in the sense of vanishing 
$n$-point function $(n \ge 4)$, the usual constraint on string theories
coming from unitarity and factorization are weakened. In particular,
the open string can be defined for any gauge group. 

In section (4.2), we have discussed $N=2$ strings in the presence of
antisymmetric tensor background which is magnetic type. Because
of the presence of an antisymmetric tensor field, the coordinates
become noncommutative and one gets a noncommutative 
field theory. The expected topological nature of the open string is shown to
impose nontrivial constraints on the corresponding noncommutative
field theory. The $n$-point functions of the theory are local, or 
vanish identically in the open sector, even for the finite 
noncommutativity. In the mixed sector, we have calculated a
3-point amplitude and have shown that the corresponding field theory
is written in terms of a generalized *-product.
We have also analyzed a 4-point function $A_{oooc}$ in
$\Theta\rightarrow \infty$ limit.


\chapter{OVERVIEW and CONCLUSION}
\markright{Chapter 5. OVERVIEW and CONCLUSION}
In this thesis, we have studied various nonperturbative and
noncommutative aspects of string theory. Construction of the nonthreshold 
bound of various branes, their supersymmetric properties and the open
string construction in various string backgrounds have 
been the topics of discussion of the present thesis. We have also studied a 
particular example of the
mixed boundary condition on $D$-branes in the context of 
$N=2$ strings, and analyzed various tree level scattering amplitudes.

In chapter (2), we have explicitly constructed nontrivial bound 
states of $D$-branes starting with charged macroscopic strings.
These string solutions are generated from the neutral ones 
by a solution generating technique. The charged macroscopic string
solutions have been useful in the past for establishing 
several duality symmetries of string theory. The $D$-brane bound states
presented in this thesis have been generated from the charged macroscopic 
string solutions
by the application of SL(2,Z) transformation and $T$-duality symmetry
of string theory.
These bound state solutions carry a general set of string charges as well
as non-zero momenta. They
are also shown to be $1/2$ supersymmetric objects. It would be nice 
to understand the conformal field theory description of these 
$D$-brane bound states.

In chapter (3),  we have presented the
supergravity solutions of various $D$-branes and their bound states
in an exact string background, known as pp-wave background supported
by $NS-NS$ and $R-R$ flux. These backgrounds, in general,
can be obtained from $AdS_p \times S^q$ type of geometry and in some
cases provide exact string theory backgrounds. The novelty  
of pp-wave background is that string theory simplifies a lot due to the
presence of light cone gauge. Our study mainly concentrated on the 
pp-wave background arising out of $AdS_3\times S^3$ type of geometry.
We have analyzed the supersymmetry 
of the branes by solving type IIB Killing spinor equations explicitly. We 
have also presented the open string construction of the intersecting
branes in pp-wave background. Our search for supergravity brane 
solutions in other pp-waves also includes the ones in 
little string theories. The gauge theory duals of these branes
can be found out by defining the appropriate `conformal operators'
in the field theory corresponding to the supergravity modes
by using the pp-wave/CFT correspondence. 

In chapter (4) (following the pioneering work of Seiberg and Witten),
we have studied an example of open strings
ending on $D$-branes with mixed boundary condition. 
We have been able to reproduce the `topological' nature of the 
corresponding field theory in this specific example and have
also shown the appearance of a `generalized star product' in the
noncommutative field theory. 

At the moment, string theory seems like the most promising quantum theory
of gravity. Not only does string theory combines gravity with
other forces, it also is rich enough to explain, in principle, all the 
symmetries found in nature.
It avoids the ultraviolet infinities that arise in trying
to quantize gravity. At low energies, string theory is described by
a supersymmetric field theory. There is no running coupling constant 
in the theory. Instead there is a dilaton field, whose VEV acts as the gauge 
and gravitational coupling constants. Supersymmetry appears to be 
inevitable for the consistency of string theory. The discovery of
various string dualities have played an important role in enlarging
our understanding of string theory. String theory also predicts the
presence of $D$-branes, which are proved to be very important 
in understanding various dualities.
These D-branes have played a key role in understanding the 
gauge theory - string
theory duality, which goes by the name of AdS/CFT duality. Formulation
of string theory in various nontrivial backgrounds with flux has also been
the topic of intense discussion all along. One of these backgrounds is 
known as plane wave (pp-wave) background. PP-wave backgrounds in the presence
of fluxes drew lots of attention in the
recent past. These backgrounds have the property that string theory  
is exactly solvable in the light cone gauge. 
The pp-wave/CFT duality is an attempt to understand 
gauge-string duality beyond supergravity approximation. 
At present, string theory in pp-wave space-time appears to be more 
promising than many other curved backgrounds.  

Despite its phenomenal theoretical success, string theory has
certain problems. One of the outstanding problems of this theory 
is to find the correct vacuum which describes the physical universe,
namely one has to find out the one which can reproduce the standard
model with symmetry $SU(3)\bigotimes SU(2)\bigotimes U(1)$. Recently,
the proposal of moduli stabilization of string theory by turning on
additional fluxes has shed new light on this problem. The other 
problem is the supersymmetry breaking in string theory. 
In order to explain the physics of the observed universe, there must
be supersymmetry breaking at some scale. Upto now there is no concrete 
method of supersymmetry breaking in string theory has been proposed.
The next question which string theory has not answered yet is the 
value of cosmological constant, which is known to be vanishingly
small. The key of all these problems lie in the non-perturbative
formulation of string theory. The best hope, so far, has been with
the M-theory, an eleven dimensional theory. From 11-dimensional 
perspective, all the superstring theories can be viewed as  different
vacua of the same theory. But very little, up to now, has been known
for the M-theory, apart from it's low energy limit, the
11-dimensional supergravity. Unless we know how to quantize the 
membranes, hardly anything about the spectrum of the theory 
could be speculated. Apart from all the theoretically unanswered 
questions, string theory lacks experimental verification, 
because of the fact that the present day accelerators can't probe a 
distance of $10^{-33}$cm, the characteristic length of a string.
Therefore, a long way has to be covered before we accept string theory
as the theory all fundamental interactions including gravity.


\begin{thebibliography}{99}
\bibitem{polyakov} A. Polyakov, ``Quantum Geometry of Bosonic
    Strings'', Phys.Lett. {\bf B103} (1981) 207; {\it ``Quantum
    Geometry of Fermionic Strings''}, Phys.Lett. {\bf B103} (1981)
  211.


\bibitem{WITTEND} E. Witten, ``String Theory Dynamics In Various 
Dimensions'', Nucl. Phys. {\bf B443} (1995) 85, 
[hep-th/9503124].

\bibitem{DABH} A. Dabholkar, ``Ten Dimensional Heterotic String 
as a Soliton'', Phys. Lett. {\bf B357} (1995) 307, 
[hep-th/9506160].

\bibitem{HULLOPEN} C. Hull, ``String-String Duality in Ten Dimensions'', 
Phys. Lett. {\bf B357} (1995) 545, 
[hep-th/9506194]. 

\bibitem{POLCWIT} J. Polchinski and E. Witten, 
``Evidence for Heterotic - Type I String Duality'', 
Nucl. Phys. {\bf B460} (1996) 525, [hep-th/9510169].

\bibitem{HULLTOWN} C. Hull and P. Townsend, 
``Unity of Superstring Dualities'', 
Nucl. Phys. {\bf B438} (1995) 109, [hep-th/9410167].

\bibitem{DUFFSS}
M. Duff, ``Strong/Weak Coupling Duality from the Dual String'',
Nucl. Phys. {\bf B442} (1995) 47, [hep-th/9501030];

M. Duff and R. Khuri, ``Four-Dimensional String/String Duality'',
Nucl. Phys. {\bf B411} (1994) 473,
[hep-th/9305142].

\bibitem{SSSD} A. Sen, ``STRING STRING DUALITY CONJECTURE IN 
SIX DIMENSIONS AND CHARGED SOLITONIC
STRINGS'', Nucl. Phys. {\bf B450} (1995) 103. [hep-th/9504027].

\bibitem{HARSTRSSD} J. Harvey and A. Strominger, 
``The Heterotic String is a Soliton'',
Nucl. Phys. {\bf B449} (1995) 535, [hep-th/9504047].

\bibitem{poratti} A. Giveon, M. Porrati and E. Rabinovici,
``Target Space Duality in String Theory'', 
Phys. Rep. {\bf 244} (1994) 77, [hep-th/9401139].


\bibitem{sen} A. Sen, ``Strong-Weak Coupling Duality in Four
    Dimensional String Theory'', Int.J.Mod.Phys. {\bf A9} (1994)
  3707,hep-th/9402002. 

\bibitem{schwarzi}  J. H. Schwarz,  ``The Second Superstring
    Revolution'', hep-th/9607067. 

\bibitem{polchinski} J. Polchinski, ``String Duality--A
    Colloquium'',  Rev. Mod. Phys. {\bf 68} (1996) 1245, hep-th/9607050.

\bibitem{dine} M. Dine, ``String Theory Dualities'', hep-th/9609051.

\bibitem{sagnotti87} A. Sagnotti, ``Open Strings and their
    Symmetry Group'',  Cargese Summer Inst. (1987) 0521, hep-th/0208020.

\bibitem{ps89} G. Pradisi, A. Sagnotti, ``Open String
    Orbifolds'', Phys. Lett. {\bf B216} (1989) 59.
 
\bibitem{bps92} M. Bianchi, G. Pradisi, A. Sagnotti, ``Toroidal
    Compactification and Symmetry breaking in Open String theories'',
  Nucl. Phys. {\bf B376} (1992) 365.

\bibitem{dine89} M. Dine, P. Y. Huet, N. Seiberg, ``Large and Small
radius in String theory'', Nucl. Phys. {\bf B322} (1989)301. 

\bibitem{pol89} J. Dai, R.G. Leigh, J. Polchinski, ``New
    Connections Between String theories'', Mod. Phys. Lett. {\bf A4}
(1989) 2073. 
\bibitem{lei89} R.G. Leigh, ``Dirac-Born-Infeld action from Dirichlet
  Sigma model'', Mod. Phys. Lett. {\bf A4} (1989) 2767. 

\bibitem{pol94}  J. Polchinski, ``Combinatorics of Boundaries in
    String Theory'', Phys.Rev. {\bf D50} (1994) 6041, hep-th/9407031.

\bibitem{narain86} K.S. Narain, ``New Heterotic string theories
    in uncompactified dimensions < 10'',  Phys.Lett.{\bf B169} (1986) 
  41. 
\bibitem{narain87} K. S. Narain, M. H. Sarmadi, E. Witten, `` A
  note on Toroidal Compactification of Heterotic string theory'',
Nucl. Phys. {\bf B279} (1987) 369. 

\bibitem{gins87} P. Ginsparg, ``Comments on Toroidal
    Compactification of Heterotic Superstrings'', Phys. Rev. {\bf
    D35} (1987) 648.

\bibitem{harmin} J. A. Harvey, J. A. Minahan, ``Open Strings on
    Orbifolds'', Phys. Lett. {\bf B188} (1987) 44.

\bibitem{ishi} N. Ishibashi, T. Onogi, `` Open String Model
    Building'', Nucl. Phys. {\bf B318} (1989) 239.

\bibitem{sagnotti} A. Sagnotti, ``Closed strings and their Open
    string Descendants'',  Phys. Rept. {\bf 184} (1989) 167.

\bibitem{horava} P. Horava, ``Strings on Worldsheet Orbifolds'',
 Nucl. Phys. {\bf B327} (1989) 461.

\bibitem{polchinskid}  J. Polchinski, `` Dirichlet-Branes and
    Ramond-Ramond Charges'', Phys.Rev.Lett. 75 (1995) 4724,
  hep-th/9510017.

\bibitem{pol-tasi} J. Polchinski, ``TASI Lectures on D-Branes'', 
hep-th/9611050.

\bibitem{JCH} J. Polchinski, S. Chaudhuri, and C. V. Johnson,
``Notes on D-Branes'', hep-th/9602052.

\bibitem{john-d} C. V. Johnson, ``D-Brane Primer'', hep-th/0007170.

\bibitem{Open-Sag} C. Angelantonj, A. Sagnotti,
``Open Strings'', Phys. Rept. {\bf 371} (2002) 1; 
Erratum-ibid. 376 (2003) 339, hep-th/0204089.

\bibitem{bachasd} C. Bachas, ``(Half) a Lecture on D-branes'',
hep-th/9701019.

\bibitem{myersd}  J. C. Breckenridge, G. Michaud, R. C. Myers,
``More D-brane bound states'', Phys.Rev. {\bf D55} (1997) 6438,
 hep-th/9611174.

\bibitem{douglas} A. Connes, M. R. Douglas, A. Schwarz, 
``Noncommutative Geometry and Matrix Theory: Compactification on
  Tori'', JHEP {\bf 9802} (1998) 003, hep-th/9711162.

\bibitem{douglas2}  M. R. Douglas, C. Hull, ``D-branes and the
    Noncommutative Torus'', hep-th/9711165.

\bibitem{seiwit}  N. Seiberg, E. Witten, ``String Theory and
    Noncommutative Geometry'', JHEP {\bf 9909} (1999) 032,
  hep-th/9908142.

\bibitem{'thooft} G. 't Hooft, ``A Planar Diagram Theory For
    Strong Interactions'', Nucl.Phys. {\bf B72} (1974) 461.
 
\bibitem{malda97}J. M. Maldacena, ``The Large N Limit of
    Superconformal Field Theories and Supergravity'', 
Adv.Theor.Math.Phys. {\bf 2} (1998) 231; 
Int. J. Theor. Phys. {\bf 38} (1999) 1113, hep-th/9711200.

\bibitem{gubser}  S.S. Gubser, I.R. Klebanov, A.M. Polyakov,
    ``Gauge Theory Correlators from Non-Critical String Theory'',
Phys. Lett. {\bf B428} (1998) 105, hep-th/9802109.

\bibitem{ewitten} E. Witten, ``Anti De Sitter Space And
    Holography'', Adv.Theor.Math.Phys. {\bf 2} (1998) 253, hep-th/9802150.

\bibitem{freedman}  E. D'Hoker, D. Z. Freedman, W. Skiba, 
``Field Theory Tests for Correlators in the AdS/CFT
  Correspondence'', Phys.Rev. {\bf D59} (1999) 045008, hep-th/9807098. 

\bibitem{BMN} D. Berenstein, J. Maldacena, H. Nastase,
`` Strings in flat space and pp waves from ${\cal N}=4$ Super
  Yang Mills'', JHEP {\bf 0204} (2002) 013, hep-th/0202021.

\bibitem{gsw} M. Green, J. H. Schwarz, E. Witten, ``Superstring Theory
(Vol-I), Introduction'', Cambridge Press. 
\end{thebibliography}

\begin{thebibliography}{99}

\bibitem{witten} E. Witten, ``Bound States Of Strings And
    $p$-Branes'', Nucl.Phys. {\bf B460} (1996) 335,   
 hep-th/9510135.

\bibitem{mli} M. Li, ``Boundary States of D-Branes and
    Dy-Strings'', Nucl. Phys. {\bf B460} (1996) 351, hep-th/9510161. 

\bibitem{mrd} M. R. Douglas, ``Branes within Branes'',
``Cargese 1997, Strings, branes and dualities'' 267, hep-th/9512077.

\bibitem{green} M. B. Green, N.D. Lambert, G. Papadopoulos,
P.K. Townsend, ``Dyonic p-branes from self-dual (p+1)-branes'', 
Phys.Lett. {\bf B384} (1996) 86, hep-th/9605146.

\bibitem{russo-tsey} J. G. Russo, A.A. Tseytlin, ``Waves, boosted 
branes and BPS states in M-theory'', Nucl.Phys. {\bf B490} (1997) 
121, hep-th/9611047.

\bibitem{costa} M. S. Costa, G. Papadopoulos, ``Superstring
    dualities and p-brane bound states'', Nucl.Phys. {\bf B510}
(1998) 217, hep-th/9612204.

\bibitem{araf} H. Arfaei, M. M. Sheikh-Jabbari, ``Mixed Boundary 
Conditions and Brane-String Bound States'', 
Nucl.Phys. {\bf B526} (1998) 278, hep-th/9709054.

\bibitem{jabb} M. M. Sheikh-Jabbari, ``More on Mixed Boundary 
Conditions and D-branes Bound States'', Phys.Lett. {\bf B425} (1998) 48, 
hep-th/9712199.

\bibitem{kamal}  A. Kumar, R. R. Nayak, K. L. Panigrahi, ``Bound 
States of String Networks and D-branes'',
Phys. Rev. Lett. {\bf 88} (2002) 121601,  hep-th/0108174.

\bibitem{malda-russo} J. M. Maldacena, J. G. Russo, ``Large N Limit 
of Non-Commutative Gauge Theories'', JHEP {\bf 9909} (1999)
  025, hep-th/9908134.

\bibitem{hashi} A. Hashimoto, K. Hashimoto, ``Monopoles and 
Dyons in Non-Commutative Geometry'', JHEP {\bf 9901} (1999) 005, 
  hep-th/9909202.

\bibitem{yoz} M. Alishahiha, Y. Oz, M.M. Sheikh-Jabbari,
    ``Supergravity and Large N Noncommutative Field Theories'', JHEP {\bf
    9911} (1999) 007, hep-th/9909215.

\bibitem{kamal1} A. Kumar, S. Mukherji, K. L. Panigrahi, ``D-brane 
Bound States from Charged Macroscopic Strings'', 
JHEP {\bf 0205} (2002) 050, hep-th/0112219.
  
\bibitem{sen92} A. Sen, ``Macroscopic Charged Heterotic
    String'', Nucl.Phys.{\bf B388}(1992) 457,
  hep-th/9206016.

\bibitem{kumar} A. Kumar, ``Charged Macroscopic type II Strings 
and their Networks'', JHEP {\bf 9912} (1999) 001, hep-th/9911090. 

\bibitem{rcmyers} J.C. Breckenridge, G. Michaud, R.C. Myers,
`` More D-brane bound states'',
 Phys.Rev.{\bf D55} (1997) 6438, hep-th/9611174. 

\bibitem{dabh1} A. Dabholkar, J. A. Harvey, ``Nonrenormalization 
Of The Superstring Tension'', Phys. Rev. Lett.{\bf 63}
(1989) 478.  

\bibitem{dabh2} A.Dabholkar, G. W. Gibbons, J. A. Harvey, F. Ruiz
  Ruiz, ``Superstrings And Solitons'', Nucl. Phys. {\bf B 340} (1990) 33.
 
\bibitem{wit} E. Witten, ``Cosmic Superstrings'',
Phys. Lett. {\bf B153} (1985) 243. 

\bibitem{jmalda} see for example, J. Maldacena, ``Black Holes in
    String Theory, hep-th/9607235 and {\it references therein}.

\bibitem{sen95} A. Sen, ``String String Duality Conjecture in 
Six Dimensions and Charged Solitonic Strings'', Nucl.Phys. {\bf B450} 
(1995) 103, [hep-th/9504027].

\bibitem{senijmp} A. Sen, ``Strong-Weak Coupling Duality in 
Four Dimensional String Theory'', Int. J. Mod. Phys {\bf A9} (1994) 3707,
hep-th/9402002. 

\bibitem{chull} E. Bergshoeff, C. M. Hull, T. Ortin, 
``Duality in the Type--II Superstring Effective Action'', Nucl.Phys.{\bf B451}
(1995) 547,\\ 
hep-th/9504081.

\bibitem{lu} J. X. Lu, ``ADM masses for black strings and p-branes'',
Phys.Lett. {\bf B313} (1993) 29, hep-th/9304159.

\bibitem{jhschwarz} J.H. Schwarz, ``An SL(2,Z) Multiplet of Type 
IIB Superstrings'', Phys. Lett. {\bf B360} (1995) 13, hep-th/9508143.
 
\bibitem{roy} J. X. Lu, S. Roy, ``U-duality p-Branes in Diverse 
Dimensions'', Nucl.Phys. {\bf B538} (1999) 149, 
hep-th/9805180.

\bibitem{townsend} C. Hull and P. K. Townsend, ``Unity of 
Superstring Dualities'', Nucl. Phys. {\bf B438}
  (1995) 109, [hep-th/9410167].

\bibitem{papado} J. M. Izquierdo, N. D. Lambert, G. Papadopoulos,
  P. K. Townsend, ``Dyonic Membranes'', Nucl. Phys. {\bf B460} 
(1996) 560, [hep-th/9508177].

\bibitem{mcvetic} M. S. Costa, M. Cvetic, ``Non-threshold D-brane bound 
states and black holes with non-zero entropy'', Phys. Rev. {\bf D56} (1997)
  4834, [hep-th/9703204].

\bibitem{ashokesen} A. Sen, ``Extremal Black holes and Elementary 
String States'', Mod. Phys. Lett.{\bf A10} (1995) 495,
hep-th/9504147.

\bibitem{wald}  A. Dabholkar, J. P. Gauntlett, J. A. Harvey,
 D. Waldram, ``Strings as Solitons \& Black Holes as Strings''
Nucl.Phys.{\bf B474} (1996) 85, hep-th/9511053.

\bibitem{callan-malda} C. G. Callan, Jr., J. M. Maldacena,
`` D-brane Approach to Black Hole Quantum Mechanics''
Nucl.Phys.{\bf B472} (1996) 591, hep-th/9602043.

\bibitem{divecchia} P. Di Vecchia and A Liccardo, ``D branes in 
string theory, II'', hep-th/9912275 {\it and references therein}.

\bibitem{lerda} P. Di Vecchia, M. Frau, A. Lerda and A. Liccardo 
``(F,Dp) bound states from the boundary state'',
Nucl. Phys. B565 (2000) 397, hep-th/9906214  
\end{thebibliography}

\begin{thebibliography}{99}

\bibitem{penrose} R. Penrose, ``Any space-time has a plane wave as a
  limit'',in Differential geometry and relativity, pp.271-275,
  Reidel, Dordrecht, (1976).

\bibitem{tsyt1} 
G. T. Horowitz and  A. A. Tseytlin, 
``A New Class of Exact Solutions in String Theory",
 Phys. Rev {\bf D 51} (1995) 2896, hep-th/9409021.

\bibitem{guven} R. Gueven, ``Plane Wave Limits and T-Duality", 
Phys.Lett. {\bf B482} (2000) 255, hep-th/0005061.

\bibitem{metsaev} R.R.Metsaev, ``Light cone gauge formulation 
of IIB supergravity in $AdS_5 \times S^5$ background and 
AdS/CFT correspondence", Phys.Lett. {\bf B468} (1999) 65, hep-th/9908114.

R.R.Metsaev, ``Type IIB Green-Schwarz
superstring in plane wave Ramond-Ramond background", Nucl. Phys. {\bf
B625}, (2002) 70, hep-th/0112044.

R.R. Metsaev, A.A. Tseytlin, ``Exactly solvable model of
superstring in Ramond-Ramond plane wave background'',
Phys.Rev. {\bf D65} (2002) 126004, hep-th/0202109.

\bibitem{blau} M.Blau, J.Figuero-O'Farrill, C. Hull and 
G. Papadopoulos, `` A new maximally supersymmetric background of
IIB superstring theory,'' JHEP {\bf 0201}, 047 (2000), hep-th/0110242

M.Blau, J.Figuero-O'Farrill and G. Papadopoulos, `` 
Penrose limits and maximal supersymmetry,'' Class. Quant. Grav.
{\bf 19, L87}(2000), hep-th/0201081.

M.Blau, J.Figuero-O'Farrill and G. Papadopoulos,
``Penrose limits, supergravity and brane dynamics,''
Class. Quant. Grav. {\bf 19} (2002) 4753, hep-th/0202111.

\bibitem{tsyt2} A. A. Tseytlin, ``On limits of superstring in
$AdS(5)\times S^5$", hep-th/0201112.

\bibitem{malda} D. Berenstein, J. Maldacena, H. Nastase, ``Strings 
in flat space and pp waves from ${\cal N}=4$ Super Yang Mills",
JHEP{\bf 0204} (2002) 013, hep-th/0202021.

\bibitem{mukhi} N.~Itzhaki, I.~R.~Klebanov and S.~Mukhi,
``PP wave limit and enhanced supersymmetry in gauge theories,''
JHEP {\bf 0203}, 048 (2002), hep-th/0202153.

\bibitem{GO} J.~Gomis and H.~Ooguri, ``Penrose limit of N = 1 
gauge theories", Nucl. Phys. {\bf B635} (2002) 106, hep-th/0202157.

\bibitem{tseytlin} J.G. Russo, A.A. Tseytlin, ``On solvable 
models of type IIB superstring in NS-NS and R-R plane wave
backgrounds," JHEP {\bf 0204} (2002) 021, hep-th/0202179.

\bibitem{sonn} L.~A. Pando~Zayas and J.~Sonnenschein,
``On Penrose limits and gauge theories,'' hep-th/0202186.

\bibitem{mohsen} M. Alishahiha and M. M. Sheikh-Jabbari,
``The PP-wave limits of orbifolded $AdS_5 \times S^5$,''
hep-th/0203018.

\bibitem{chu} C. S. Chu, P. M. Ho, ``Noncommutative D-brane and 
Open String in pp-wave Background with B-field,''
Nucl.Phys. {\bf B636} (2002) 141, hep-th/0203186.

\bibitem{lee} P.~Lee and J.~W.~Park,
``Open strings in PP-wave background from defect conformal field
theory'', hep-th/0203257.

\bibitem{Bala} V.~Balasubramanian, M.~X.~Huang, T.~S.~Levi and A.~Naqvi,
``Open strings from N = 4 super Yang-Mills,'' hep-th/0204196.

\bibitem{Taka} H.~Takayanagi and T.~Takayanagi,
``Open strings in exactly solvable model of curved space-time and
PP-wave limit,'' hep-th/0204234.

\bibitem{klim} D. Amati, C. Klimcik, ``Nonperturbative 
Computation Of The Weyl Anomaly For A Class Of Nontrivial 
Backgrounds'', Phys. Lett. {\bf B219} (1989) 443.

\bibitem{steif}G. T. Horowitz, A. R. Steif, ``Space-Time 
Singularities In String Theory'', Phys. Rev. Lett.{\bf 64} 
(1990) 260.

\bibitem{met0} R.R.Metsaev, ``Light cone gauge formulation 
of IIB supergravity in $AdS_5 \times S^5$ background and AdS/CFT
    correspondence'', Phys.Lett. {\bf B468} (1999) 65, 
hep-th/9908114.

\bibitem{met1}R.R.Metsaev, ``Type IIB Green-Schwarz superstring
    in plane wave Ramond-Ramond background'', Nucl. Phys. {\bf B625}, 
   (2002) 70, hep-th/0112044.

\bibitem{met2} R.R. Metsaev, A.A. Tseytlin, ``Exactly solvable
    model of superstring in Ramond-Ramond plane wave background'', 
    Phys. Rev. {\bf D65} (2002) 126004, hep-th/0202109.

\bibitem{meessen} P. Meessen, ``A Small Note on PP-Wave Vacua 
in 6 and 5 Dimensions'', Phys.Rev.{\bf D65}(2002) 087501,
hep-th/0111031.

\bibitem{pope}  M. Cvetic, H. Lu, C.N. Pope, ``Penrose Limits, 
PP-Waves and Deformed M2-branes," hep-th/0203082.

\bibitem{hull}  J. P. Gauntlett, C. M. Hull, ``pp-waves in
    11-dimensions with extra supersymmetry", JHEP {\bf 0206}, (2002)
  013, hep-th/0203255.

\bibitem{bkp} A. Biswas. A. Kumar and K. L. Panigrahi, ``p-p'
    Branes in PP-wave Background'',
Phys. Rev. {\bf D66} (2002) 126002, hep-th/0208042.

\bibitem{ks} K. L. Panigrahi and Sanjay, ``D-branes in pp-wave 
spacetime with nonconstant NS-NS flux'', Phys. Lett {\bf B561} (2003)
184, hep-th/0303182. 

\bibitem{dabh} A. Dabholkar, S. Parvizi, ``Dp branes in pp-wave
background", Nucl. Phys. {\bf B641} (2002) 223, hep-th/0203231.

\bibitem{akumar} A. Kumar, R. R. Nayak, Sanjay, ``D-brane solutions
in pp-wave background'', Phys. Lett. {\bf B541}(2002) 183, hep-th/0204025.

\bibitem{sken} K.~Skenderis and M.~Taylor,
``Branes in AdS and pp-wave spacetimes,'' JHEP {\bf 0206} (2002) 
025, hep-th/0204054.

K. Skenderis, M. Taylor, ``Open strings in the plane wave 
background I: Quantization and symmetries'',
Nucl. Phys. {\bf B665} (2003) 3, hep-th/0211011.

K. Skenderis, M. Taylor, ``Open strings in the plane wave 
background II: Superalgebras and Spectra'',
JHEP {\bf 0307} (2003) 006, hep-th/0212184.

\bibitem{alishah} M. Alishahiha and A. Kumar, ``D-Brane Solutions
from New Isometries of pp-Waves", Phys.Lett. {\bf B542} (2002)
130, hep-th/0205134, 

\bibitem{bain} P. Bain, P. Meessen and M. Zamaklar,
``Supergravity solutions for D-branes in Hpp wave backgrounds,''
Class. Quant. Grav. {\bf 20} (2003) 913, hep-th/0205106.

\bibitem{singh} H. Singh, ``M5-branes with $3/8$ Supersymmetry 
in PP-wave background,'' Phys. Lett. {\bf B543} (2002) 147, hep-th/0205020.
 
\bibitem{pal} S. S. Pal, ``Solution to worldvolume action of D3 
brane in pp-wave background", Mod. Phys. Lett. {\bf A17} (2002)1735,
hep-th/0205303.

\bibitem{michi} Y. Michishita, ``D-branes in NSNS and RR pp-wave 
backgrounds and S-duality'', JHEP {\bf 0210} (2002) 048,
hep-th/0206131.

\bibitem{OPS} N. Ohta, K. L. Panigrahi and Sanjay, 
``Intersecting branes in pp-wave spacetime'',
Nucl. Phys. {\bf B 674} 306, 2003, hep-th/0306186. 

\bibitem{NP} R.~R.~Nayak and K.~L.~Panigrahi,
``More D-branes in plane wave spacetime,''
Phys. Lett. {\bf B575} (2003) 325 [arXiv:hep-th/0310219].

\bibitem{HNP} S.F. Hassan, R. R. Nayak and K. L. Panigrahi, 
''D-branes in the NS5 Near-horizon pp-Wave Background'',
hep-th/0312224.

\bibitem{NPS} R. R. Nayak, K. L. Panigrahi, and S. Siwach,
``Brane Solutions with/without Rotation in PP-wave Spacetime'',
Nucl. Phys. {B698} (2004) 149, hep-th/0405124.
 
\bibitem{rashmi} R. R. Nayak, ``D-Branes at angle in pp-wave 
Background'', Phys. Rev. {\bf D67} (2003) 086006, hep-th/0210230.

\bibitem{maoz} J. Maldacena and L. Maoz, ``Strings on pp-waves and
  massive two dimensional field theories'', JHEP {\bf 0212} (2002) 046,
 hep-th/0207284.

\bibitem{russo} J. G. Russo and A. A. Tseytlin, ``A class of
    exact pp-wave string models with interacting light-cone 
gauge actions'', JHEP {\bf 0209} (2002) 035, hep-th/0208114.


\bibitem{hikida} Y. Hikida and S. Yamaguchi, ``D-branes in 
PP-Waves and Massive Theories on Worldsheet with Boundary'', 
JHEP {\bf 0301} (2003) 072, hep-th/0210262.
 
\bibitem{kim} N. Kim, ``Remarks on type IIB pp waves with 
Ramond-Ramond fluxes and massive two dimensional nonlinear 
sigma models'', Phys. Rev. {\bf D67} (2003) 046005, hep-th/0212017.

\bibitem{bonelli}  G. Bonelli, ``On Type II strings in exact 
superconformal non-constant RR backgrounds'', JHEP {\bf 0301} 
(2003) 065, hep-th/0301089. 

\bibitem{cvetic} M. Cvetic, H. Lu, C.N. Pope, ``M-theory PP-waves, 
Penrose Limits and Supernumerary Supersymmetries'', Nucl. Phys. {\bf B644} 
(2002) 65, hep-th/0203229.

M. Cvetic, H. Lu, C.N. Pope, K.S. Stelle, ``Linearly-realised 
Worldsheet Supersymmetry in pp-wave Background'', hep-th/0209193.

\bibitem{bachas} C. Bachas, M. Petropoulos, ``Anti-de-Sitter
D-branes'', JHEP {\bf 0102} (2001) 025, hep-th/0012234.

\bibitem{malda-moore} J. Maldacena, G. Moore, N. Seiberg,
``Geometrical interpretation of D-branes in gauged WZW models'', 
JHEP {\bf 0107} (2001) 046, hep-th/0105038. 

\bibitem{ponsot} B. Ponsot, V. Schomerus, J. Teschner, ``Branes
    in the Euclidean $AdS_3$'', JHEP {\bf 0202} (2002) 016,
  hep-th/0112198.

\bibitem{billo} M. Billo', I. Pesando, ``Boundary States for GS 
superstrings in an Hpp wave background'', Phys. Lett. {\bf B536} 
(2002) 121, hep-th/0203028.

\bibitem{alis-jab} M. Alishahiha, M. M. Sheikh-Jabbari, ``Strings
in PP-Waves and Worldsheet Deconstruction'', Phys. Lett. {\bf B538} 
(2002) 180, hep-th/0204174.

\bibitem{faya1} A. Fayyazuddin, D. J. Smith, ``Localized 
intersections of M5-branes and four-dimensional superconformal 
field theories'', JHEP {\bf 9904} (1999) 030, hep-th/9902210.

\bibitem{faya2}  A. Fayyazuddin, D. J. Smith, ``Warped 
AdS near-horizon geometry of completely localized intersections 
of M5-branes'', JHEP {\bf 0010} (2000) 023, hep-th/0006060.   

\bibitem{lupo} H. Lu, C.N. Pope,``Interacting Intersections'',
Int.J.Mod.Phys. A13 (1998) 4425, hep-th/9710155.

\bibitem{boe} J. de Boer, A. Pasquinucci, and K. Skenderis 
``ADS / CFT Dualities Involving Large 2-D N=4 Superconformal
  symmetry'', Adv. Theor. Math. Phys.{\bf 3} (1999) 577,
hep-th/9904073.  
\bibitem{papa} G. Papadopoulos, J. G. Russo, A. A. Tseytlin,
    ``Curved branes from string dualities'', 
    Class.Quant.Grav. 17 (2000) 1713, hep-th/9911253.

\bibitem{duff} M. J. Duff, Ramzi R. Khuri, J. X. Lu, ``String
    Solitons", Phys.Rept. {\bf 259} (1995) 213, hep-th/9412184.

\bibitem{oz} M. Alishahiha, H. Ita, Y. Oz, ``Graviton Scattering
    on D6 Branes with B Fields", JHEP {\bf 0006} (2000) 002,
  hep-th/0004011.

\bibitem{myers} J. C. Breckenridge, G. Michaud, R. C. Myers, 
``More D-brane bound states,'' Phys. Rev. {\bf D55} (1997) 6438,
hep-th/9611174.

\bibitem{kamal2} A. Kumar, S. Mukherji and K. L. Panigrahi,
    ``D-brane bound states from Charged Macroscopic Strings", 
JHEP{\bf 05} (20020 050, hep-th/0112219.

\bibitem{tseyt6} A.A. Tseytlin, ``Harmonic superpositions of
    M-branes'', Nucl.Phys. {\bf B475} (1996) 149, hep-th/9604035.

\bibitem{schwarz} J. H. Schwarz, ``Covariant Field Equations of 
Chiral N=2 D = 10 Supergravity", Nucl. Phys.{\bf B226} (1983) 269. 

\bibitem{fawad} S.F. Hassan, `` T-Duality, Space-time Spinors
and R-R Fields in Curved Backgrounds,'' Nucl.Phys. {\bf B568}
(2000) 145, hep-th/9907152.
\bibitem{rangamani} V. E. Hubeny, M. Rangamani and  E. Verlinde, 
``Penrose Limits and Non-local theories,'' hep-th/0205258.

\bibitem{sakai} Y. Oz, T. Sakai, ``Penrose Limit and Six-Dimensional 
Gauge Theories," hep-th/0207223.

\bibitem{kumar1} M. Alishahiha, A. Kumar, ``PP-waves from Nonlocal
    Theories,", hep-th/0207257.

\bibitem{berkooz} M. Berkooz, M. Rozali and N. Seiberg, ``Matrix
    Description of M-theory on $T^4$ and $T^5$", Phys.Lett. {\bf B408}
 (1997) 105, hep-th/9704089.

\bibitem{seiberg} N. Seiberg, ``Matrix Description of M-theory on
    $T^5$ and $T^5/Z_2$", Phys.Lett. {\bf B408} (1997) 98,
  hep-th/9705221.
\end{thebibliography}

\begin{thebibliography}{99}
\bibitem{bruce} D.J. Bruce, D.B. Fairlie, R.G. Yates, 
``Dual Models with a Color Symmetry'',
Nucl. Phys. {\bf B108} (1976) 310. 

\bibitem{ademo} M. Ademollo, L. Brink, A. D'Adda, R. D'Auria, E. Napolitano, 
S. Sciuto, E. Del Giudice, P. Di Vecchia, S. Ferrara, F. Gliozzi,
R. Musto, R. Pettorino, J.H. Schwarz, `` Dual String With U(1) Color 
Symmetry'', Nucl.Phys. B{\bf 111} (1976) 77.

\bibitem{OV} H. Ooguri and C. Vafa, `` Selfduality And N=2 
String Magic'', Mod. Phys. Lett. A{\bf 5} (1990) 1389.

H. Ooguri and C. Vafa, ``Geometry Of N=2 Strings'',  
Nucl.Phys. B {\bf 361} (1991) 469-518. 

\bibitem{OV1}H. Ooguri and C. Vafa, ``N=2 Heterotic Strings'', 
Nucl.Phys. B{\bf 367} (1991) 83. 

\bibitem{pleb} J.F. Plebanski, ``Some Solutions Of Complex 
Einstein Equations'', J. Math. Phys. {\bf 16} (1975) 2395.

\bibitem{mar} N. Marcus, ``The N=2 open string'',
Nucl.Phys. B{\bf 387} (1992) 263, hep-th/9207024. 

N. Marcus, ``A tour through N=2 strings'',
Presented at International Workshop on String Theory, 
Quantum Gravity and the Unification of Fundamental
Interactions, Rome, Italy, 21-26 Sep 1992{\it Rome String Theory
  Wkshp.} (1992) 391, hep-th/9211059.

\bibitem{MART}  E. Martinec, ``M-theory and N=2 Strings'',
Talk given 
at the NATO Advanced Study Institute on Strings, Branes and Dualities, 
Cargese, France, 26 May - 14 Jun 1997. 
In *Cargese 1997, Strings, branes and dualities* 241, [hep-th/9710122]. 

\bibitem{BERK} N. Berkovits, `` Super-Poincare Invariant 
Superstring Field Theory'', Nucl. Phys. B{\bf 450} (1995) 90,
[hep-th/9503099].

N. Berkovits and W. Siegel, ``Covariant field theory for self-dual 
strings'', Nucl. Phys. B{\bf 505} (1997) 139, [hep-th/9703154].

\bibitem{kamal3} A. Kumar, A. Misra, K. L. Panigrahi, 
``Noncommutative N=2 Strings'', JHEP {\bf 0102} 
(2001) 037, hep-th/0011206.
 
\bibitem{DOUG} A. Connes, M.R. Douglas and A. Schwarz, 
``Noncommutative Geometry and Matrix Theory: Compactification on Tori'',
JHEP {\bf 9802} 003, (1998), [hep-th/9711162].   

\bibitem{SW}N. Seiberg and E. Witten, ``String Theory and 
Noncommutative Geometry'', JHEP {\bf 9909} 032, (1999), [hep-th/9908142].

\bibitem{SOLITONS}  R. Gopakumar, S. Minwalla and A. Strominger,
`` Noncommutative Solitons'', JHEP {\bf 0005} 020,(2000), 
[hep-th/0003160].

\bibitem{OMNCSOL} R. Gopakumar, S. Minwalla, N. Seiberg and
A. Strominger,  ``OM Theory in Diverse Dimensions'', 
JHEP {\bf 0008} 008, (2000),[hep-th/0006062].

\bibitem{ketov} S. V. Ketov, ``The OSp(32|1) versus OSp(8|2) 
supersymmetric M-brane action from self-dual (2,2) strings'', 
Mod.Phys.Lett. A{\bf 11} (1996) 2369,[hep-th/9609004.

\bibitem{HULL}S.J. Gates, Jr., C.M. Hull and M. Rocek,
``wisted Multiplets And New Supersymmetric Nonlinear Sigma Models'',
Nucl.Phys. B{\bf 248} (1984) 157.

\bibitem{GAROUSI} M. R. Garousi, ``Non-commutative world-volume 
interactions on D-brane and Dirac-Born-Infeld action'',
Nucl.Phys. B{\bf 579} (2000) 209, [hep-th/9909214].

M.R. Garousi, R.C. Myers, ``World-Volume Potentials on D-branes'',
JHEP {\bf 0011} (2000) 032, hep-th/0011147.

\bibitem{MEHEN} H. Liu and J. Michelson, ``*-TREK: The One-Loop N=4 
Noncommutative SYM Action'', Nucl.Phys. {\bf B 614} (2001) 279-304, 
hep-th/0008205.

F. Ardalan and N. Sadooghi, ``Anomaly and Nonplanar Diagrams 
in Noncommutative Gauge Theories'', 
Int. J. Mod. Phys. {\bf A17} (2002) 123, hep-th/0009233;

T. Mehen and M.B. Wise, ``Generalized *-Products, Wilson Lines 
and the Solution of the Seiberg-Witten Equations'',
JHEP {\bf 0012} (2000) 008, hep-th/0010204.

S. Das and S. Trivedi, ``Supergravity couplings to Noncommutative Branes, 
Open Wilson Lines and Generalised Star Products'',
JHEP {\bf 0102} (2001) 046, hep-th/0011131.

\bibitem{KAW} H. Kawai, D.C. Lewellen, S.H.H. Tye, 
``A Relation Between Tree Amplitudes Of Closed And 
Open Strings'', Nucl.Phys. B{\bf 269} (1986) 1. 

\bibitem{CH}C.S. Chu and P.M. Ho, ``Noncommutative Open String and
  D-brane'', Nucl.Phys. B{\bf 550} (1999) 151, hep-th/9812219.

\bibitem{ITY}B.Chen, H. Itoyama, T. Matsuo and K. Murakami,
``Worldsheet and Spacetime Properties of p-p' System with 
B Field and Noncommutative Geometry'', Nucl. Phys. {\bf B593} (2001)
505, hep-th/0005283.

\bibitem{lech} O. Lechtenfeld, A.D. Popov, B. Spendig, {\it ``Open N=2 
strings in a B-field background and noncommutative self-dual 
Yang-Mills''}, Phys.Lett. {\bf B507} (2001) 317, hep-th/0012200.

\bibitem{PLEBA}  H. Garcia-Compean, J. F. Plebanski,  M. Przanowski,
``Geometry Associated with Self-dual Yang-Mills and the Chiral 
Model Approaches to Self-dual Gravity'', Acta Phys. Polon. 
B{\bf 29} (1998) 549.    

\bibitem{GARMYRS} M. R. Garousi, R. Myers, ``World-Volume 
Potentials on D-branes'', JHEP {\bf 0011} (2000) 032, hep-th/0010122.

\end{thebibliography}
\end{document}